%% file: main.tex
\documentclass[12pt,letterpaper]{article}

\usepackage{mathtools}
\usepackage{arxiv_preprint}
\usepackage{array}
\hypersetup{hypertexnames=false}

\newlength{\panelht}

\input{notations}

\title{Rethinking Data Value: Asymmetric Data Shapley for Structure-Aware Valuation in AI Model Markets}

\author{%
Xi Zheng\textsuperscript{1}\quad
Yinghui Huang\textsuperscript{2}\quad
Xiangyu Chang\textsuperscript{2}\\[0.35em]
Ruoxi Jia\textsuperscript{3}\quad
Yong Tan\textsuperscript{1}\\[0.65em]
{\small\textsuperscript{1}Michael G. Foster School of Business, University of Washington}\\
{\small\textsuperscript{2}School of Management, Xi'an Jiaotong University}\\
{\small\textsuperscript{3}Bradley Department of Electrical and Computer Engineering, Virginia Tech}%
}
\date{}
\hypersetup{
  pdftitle={Rethinking Data Value: Asymmetric Data Shapley for Structure-Aware Valuation in AI Model Markets},
  pdfauthor={Xi Zheng; Yinghui Huang; Xiangyu Chang; Ruoxi Jia; Yong Tan}
}

\begin{document}

\MISQmaketitle

\begin{abstract}
Modern AI model markets assemble models from heterogeneous pipeline contributions, including raw data, derivative data, and data-related services, whose value can depend on upstream inputs and inherited model states.
Conventional valuation approaches focus on data contributions without explicitly accounting for these production relationships.
They can therefore misattribute value by evaluating contributions outside their upstream production context.
We formalize this pipeline-context valuation mismatch and develop Asymmetric Data Shapley (\texttt{ADS}), a unified framework for structure-aware value attribution.
\texttt{ADS} translates documented production relationships into ordered valuation groups and evaluates contributions among peers conditional on their upstream production context.
We provide an axiomatic characterization of the resulting allocation and quantify allocation uncertainty when evidence supports multiple valuation structures.
To support valuation at larger scales or with costly utility evaluations, we develop a Monte Carlo estimator (\texttt{MC-ADS}) and a surrogate-assisted estimator (\texttt{SA-MC-ADS}).
Experiments show that ignoring production dependencies systematically shifts value toward redundant derivative data and away from useful upstream data and services.
Moreover, experiments on derivative data selection and on dataset procurement during multi-stage large language model fine-tuning show that decisions guided by \texttt{ADS} improve downstream model performance relative to the respective benchmarks.
\end{abstract}

\MISQkeywords{Data Valuation; AI Model Markets; Shapley Value; Structured AI Pipelines; Data Provenance; Data-Related Services}

\MISQBeginMainText

\section{Introduction}\label{sec:intro}
\input{S1_Introduction}

\section{Related Work}\label{sec:related_work}
\input{S2_Background_and_Related_Work}

\section{Pipeline-Context Valuation Mismatch in Structured AI Pipelines}
\label{sec:counterintuition}
\input{S3_Limitations_of_Data_Shapley}

\section{Asymmetric Data Shapley for Structured AI Pipelines}
\label{sec:asymmetric data shapley}
\input{S4_Asymmetric_Data_Shapley}

\section{Operationalizing ADS in AI Model Markets}
\label{sec:computation}
\input{S5_Efficient_Computation_of_Asymmetric_Data_Shapley}

\section{Applications}\label{sec:experiments}
\input{S6_Applications}

\section{Extension of ADS Under Uncertain Pipeline Structure}
\label{sec:robustness_strategy}
\input{S7_Extension}

\section{Concluding Remarks}
\label{sec:conclusion}
\input{S8_Concluding_Remarks}

\MISQPrintReferences

\input{Online_Appendices}

\end{document}

%% file: notations.tex
\newcommand{\ds}[2]{\ensuremath{\phi(#1;\,v,#2)}}

\newcommand{\ads}[2]{\ensuremath{\phi^\sigma(#1;\,v,#2)}}

\newcommand{\pre}{\operatorname{Pre}}

%% file: S1_Introduction.tex
\begin{figure}[htbp]
    \centering
    \includegraphics[width=1.0\textwidth]{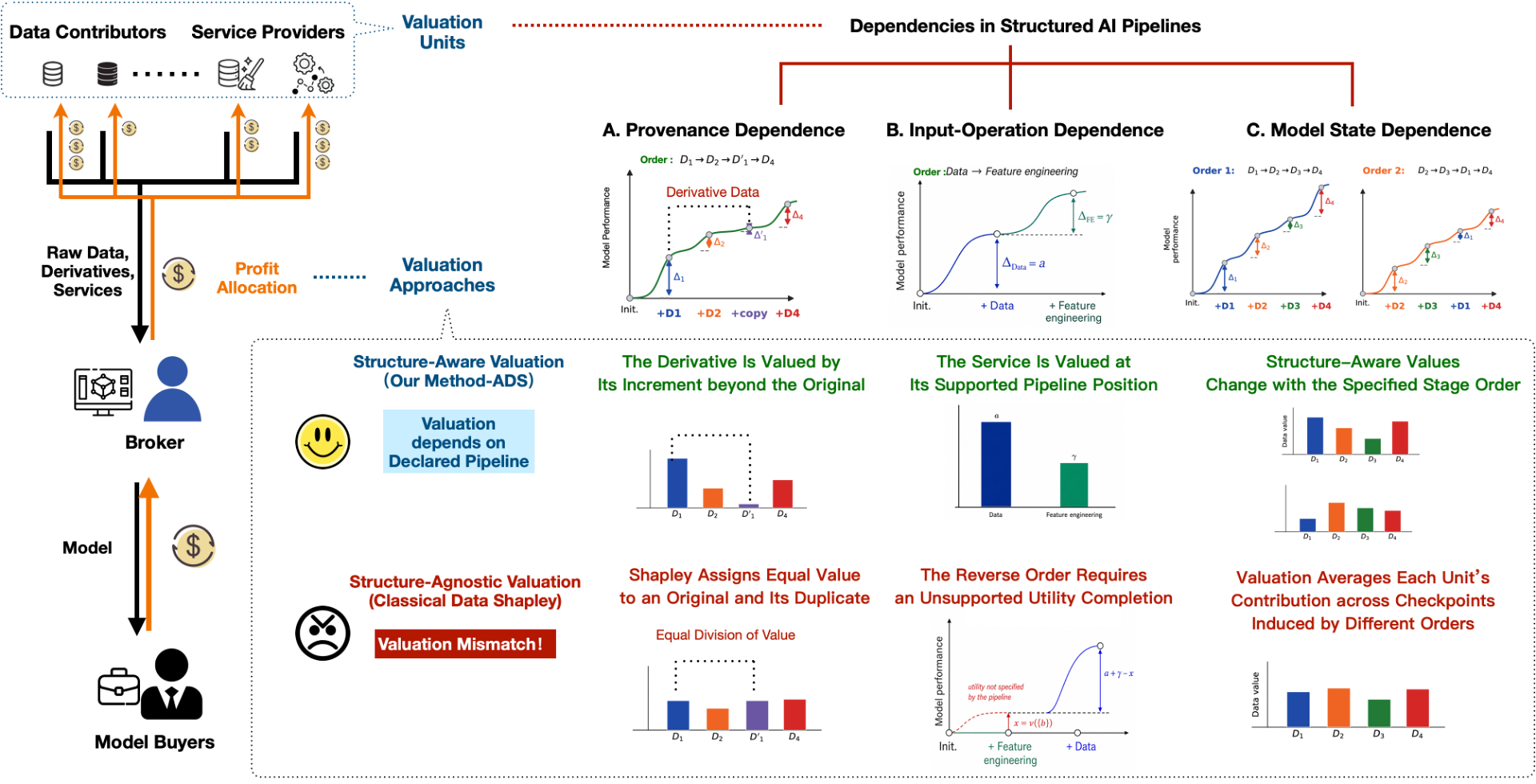}
    \caption{Motivating example of pipeline-context-conditioned valuation in AI model markets.
    The left panel illustrates four roles in an AI model market. 
    The right panels illustrate representative structured AI production settings, and contrast classical structure-agnostic valuation with the proposed \texttt{ADS} valuation. }
    \label{fig:data_market_compensation_pipeline}
\end{figure}

Producing valuable AI models involves coordinating data and complementary capabilities across organizations~\citep{bi2024understanding,pujol2024commoning,vandenbroek2021machine}. AI model production can draw on heterogeneous data sources and requires specialized expertise across data preparation, model development, and production deployment~\citep{polyzotis2018data,xin2021production}. At the same time, downstream firms may demand trained model instances or prediction services without possessing all the data and production capabilities needed to build them internally~\citep{chen2019towards,liu2021dealer}. 
These conditions motivate AI model markets that involve four analytically distinct roles: data contributors supply proprietary data assets, service providers transform these assets or support their use in model production, brokers organize the production process, and model buyers purchase the resulting model instances or prediction services~\citep{chen2019towards,liu2021dealer,pei2023data} (as depicted in Figure~\ref{fig:data_market_compensation_pipeline}). These roles need not map onto separate organizations; the same entity may, for example, operate the market as a broker while also contributing data or production services. Such markets therefore coordinate a production pipeline in which raw data, derivative data, and data-related services jointly generate model value. 
For the broker coordinating this process, a central valuation challenge is to attribute the resulting model value across heterogeneous contributions supplied by different organizations~\citep{chen2019towards,liu2021dealer,xia2023equitable}.

This value-attribution problem is challenging because modern AI models are produced through structured, multi-stage pipelines rather than through a single training step over a fixed dataset.
Contemporary AI workflows encompass raw-data collection, preprocessing, data augmentation, model architecture design, large-scale training, post-training alignment, evaluation, deployment, and iterative updating~\citep{baylor2017tfx,polyzotis2018data,xin2021production}. 
In these pipelines, model performance is shaped by a sequence of interdependent transformations that generate, refine, and reuse intermediate data artifacts and model states~\citep{derakhshan2022materialization}. 
Critically, raw data, derivative data, and data-related services enter the pipeline at different points and may rely on upstream inputs or inherited model states. Their marginal contributions are therefore conditioned by the
production context in which they enter or operate.

In a structured AI production pipeline, the valuation rule must satisfy two requirements.
First, it must accommodate heterogeneous contributions, including raw data, derivative data, and data-related services. 
Second, it must evaluate each contribution relative to the production context in which it enters or operates, including its upstream inputs, pipeline position, and, when relevant, inherited model state. 
Existing data-valuation methods only partially address the first requirement because they are formulated primarily for data examples or data sources rather than heterogeneous valuation units that also include data-related services~\citep{ghorbani2019data,kwon2022beta,wang2023data}. 
The second requirement exposes a distinct limitation of structure-agnostic valuation. 
Classical Data Shapley (\texttt{DS}), for example, averages marginal contributions over all possible orders~\citep{ghorbani2019data}. 
It therefore assigns positive weight to predecessor contexts that can conflict with documented production dependencies, such as placing a derivative before its source data, a service before the inputs it transforms, or a late-stage dataset before the model checkpoint at which it actually enters. 
Existing order-aware extensions address parts of this problem by restricting causal feature orderings or conditioning raw-data valuation on federated-learning rounds~\citep{frye2020asymmetric,wang2020principled}. 
What remains missing is a unified valuation framework that accommodates heterogeneous contributions using documented production relationships. 
These observations lead to our central research question: \textit{How should AI model markets attribute model value among heterogeneous contributions whose marginal value depends on their documented production context?}

This paper proposes Asymmetric Data Shapley (\texttt{ADS}), a pipeline-context-conditioned valuation framework for structured AI pipelines. \texttt{ADS} values raw data, derivative data, and data-related services within a unified framework. \texttt{ADS} uses documented production relationships to evaluate each contribution conditional on the production context in which it enters the pipeline.   Within comparable pipeline positions, \texttt{ADS} retains Shapley-style marginal-contribution comparisons among peer contributions. In doing so,  \texttt{ADS} resolves the mismatch between structure-agnostic valuation methods and the sequential, interdependent nature of modern AI model production pipelines.

This paper makes three contributions. 
First, we identify and formalize the pipeline-context valuation mismatch that arises in structured AI production. We show that provenance, input--operation, and model-state dependencies jointly determine the predecessor contexts that are relevant for value attribution. This distinction clarifies when unrestricted marginal-contribution averaging becomes misaligned with valuation.
Second, we develop \texttt{ADS}, a unified framework for valuing raw data, derivative data, and data-related services at their documented positions in the pipeline. \texttt{ADS} translates production relationships into ordered valuation groups and applies Shapley comparisons among peers conditional on upstream contributions. We provide an axiomatic characterization of the allocation rule and establish group efficiency and completion invariance. 
These properties ensure that contributions entering at the same point in the pipeline jointly receive exactly the utility they add, and that allocations do not depend on utilities assigned to coalitions outside the pipeline-supported domain. 
We further extend the \texttt{ADS} framework to characterize allocation uncertainty when available evidence supports multiple valuation structures.
Third, we operationalize \texttt{ADS} and examine its consequences for value allocation and downstream decisions. 
We develop a Monte Carlo \texttt{ADS} estimator (\texttt{MC-ADS}) and a surrogate-assisted Monte Carlo \texttt{ADS} estimator (\texttt{SA-MC-ADS}), which can improve estimation precision at a given high-fidelity evaluation budget.
Both estimators come with finite-sample guarantees, making \texttt{ADS} applicable when exact evaluation is impractical. 
Experiments show that ignoring production dependencies shifts value toward redundant derivatives and away from useful upstream data and services, and that decisions guided by \texttt{ADS} improve final performance in dataset procurement for large language model (LLM) fine-tuning and in derivative-data selection.

The remainder of the paper is organized as follows. Section~\ref{sec:related_work} reviews the relevant literature. Section~\ref{sec:counterintuition} examines how unrestricted permutation averaging can create pipeline-context valuation mismatch. Section~\ref{sec:asymmetric data shapley} develops \texttt{ADS} for fixed production structures and establishes its principal allocation properties. Section~\ref{sec:computation} presents \texttt{MC-ADS} and \texttt{SA-MC-ADS} for scalable estimation. Section~\ref{sec:experiments} evaluates the allocation and procurement implications of \texttt{ADS}. Section~\ref{sec:robustness_strategy} extends the framework to uncertain production structures, and Section~\ref{sec:conclusion} concludes the paper.

%% file: S2_Background_and_Related_Work.tex
This study relates to two streams of literature. The first examines AI model markets and pipeline-based AI production, which motivate value attribution that reflects model production processes. The second examines Shapley-based data valuation, which provides the foundation for marginal-contribution-based attribution. We review both streams and position \texttt{ADS} at their intersection.

\subsection{Market Valuation and Pipeline-Based AI Production}\label{subsec:pipelines}

A growing literature studies market mechanisms for allocating value in data-driven and model-driven settings. 
Early work in database systems treats data as an information good and develops arbitrage-free pricing schemes for query-based access~\citep{balazinska2011data,koutris2012query,linkifer2014arbitrage}. 
These studies price data access rather than contributions to downstream model performance, whereas more recent machine learning (ML) market studies link valuation directly to predictive performance.
One line of work prices data according to its marginal contribution to model accuracy, often via Shapley-style rules in two-sided marketplaces~\citep{agarwal2019marketplace,ohrimenko2019collaborative}. 
Another line of work prices trained model instances or prediction services, developing accuracy-tiered or arbitrage-free pricing curves and combining them with privacy-aware or instrumental valuation~\citep{liu2021dealer,han2023optimal}. 
Subsequent work extends these ideas to the right to be forgotten~\citep{xia2023equitable}, human-annotated data for LLMs~\citep{zhang2025fairshare}, and generative models~\citep{mahmood2024pricing}.
From the buyer's perspective, \citet{liu2023cost} examine which features to acquire to improve predictive performance under a limited budget. Their CREATE approach accounts for feature-specific costs and bundled purchases in linear and logistic regression.

Complementing these market-design efforts, research on pipeline-based AI production documents the structural complexity of real-world systems.
Modern AI applications are organized as structured, multi-stage pipelines rather than single-step training procedures~\citep{xin2021production}. 
These pipelines are often iterative and path-dependent, as datasets, features, intermediate artifacts, and model states are repeatedly transformed, reused, and refined through evolving computation graphs~\citep{xin2021production}.
System performance is consequently sensitive to upstream decisions and cross-stage dependencies~\citep{sculley2015hidden}. 
Related organizational research shows that changing data uses can require substantial work to restore data integrity~\citep{pine2026myth}. 
Importantly, contemporary ML infrastructure makes these processes increasingly observable. 
Systems such as TensorFlow Extended (TFX) and ML Metadata (MLMD) capture detailed provenance graphs linking executions and artifacts~\citep{xin2021production}, while dataset documentation practices emphasize recording composition and preprocessing histories~\citep{gebru2021datasheets}.
Research on data repurposing likewise calls for documenting data transformations to support transparency, replicability, and further repurposing~\citep{parsons2026repurposing}. 
The structure of modern AI pipelines is therefore not only complex but also auditable, in principle, through explicit execution traces.

These literatures leave a gap between value attribution and production structure. Market-valuation studies provide performance-based pricing and revenue-sharing rules but generally abstract away documented production dependencies, whereas pipeline research makes those dependencies increasingly observable without specifying how they should enter value attribution. This gap is especially salient when raw data, derivative data, and data-related services contribute through interdependent production relationships.
Our work addresses this gap by developing a pipeline-context-conditioned valuation framework that evaluates each contribution in the production context defined by its documented dependencies.

\subsection{Shapley-Based Valuation Methods}\label{subsec:shapley_methods}

Shapley-based valuation provides a foundation for allocating model value among multiple contributors. 
The Shapley value is a cooperative game-theoretic solution concept that assigns credit by averaging each player's marginal contribution over all possible permutations~\citep{shapley1953value,ghorbani2019data}. 
Its axiomatic properties, including efficiency, symmetry, linearity, and nullity, make it attractive for fair value attribution. 
Subsequent theoretical work has generalized the classical framework to semivalues~\citep{dubey1981value}, weighted and quasi-values~\citep{kalai1987weighted,nowak1995axiomatizations}, and probabilistic random-order values~\citep{weber1988probabilistic}. 
These generalizations offer greater flexibility in weighting marginal contributions, with random-order values also allowing nonuniform or restricted distributions over player orderings.

In machine learning and information systems research, Shapley-based methods have been applied primarily along two lines.
One line attributes model predictions~\citep{lundberg2017unified,lundberg2020local} and predictive performance~\citep{covert2020understanding} to input features.
XPER decomposes a trained model's performance relative to a benchmark without retraining for each coalition~\citep{hue2026measuring}. Performance attribution also informs privacy protection, where useful prediction provides data utility but inference of confidential attributes creates disclosure risk. Separately attributing these outcomes to features guides selective masking that balances utility and risk~\citep{qu2026shapley}.

The other line focuses on data valuation, with Data Shapley quantifying the contribution of training examples to model performance~\citep{ghorbani2019data}.
Subsequent work has improved computational scalability through methods such as KNN-Shapley~\citep{jia2019knn}, enhanced robustness through methods such as Beta Shapley~\citep{kwon2022beta} and Data Banzhaf~\citep{wang2023data}, and extended data valuation to distributional settings~\citep{ghorbani2020distributional}.
These refinements primarily concern how marginal contributions are computed or weighted, or how reference distributions are defined.
Related attribution research in online advertising accounts for consumer context by valuing state--action pairs relative to a no-ad counterfactual~\citep{singal2022shapley}.
An open question for structured AI production is how documented dependencies among contributions should determine the admissible predecessor contexts used for marginal evaluation.

Weighted, precedence-constrained, and probabilistic random-order values provide a theoretical foundation for incorporating priorities through nonuniform or restricted distributions over player orderings~\citep{kalai1987weighted,faigle1992shapley,weber1988probabilistic}.
Recent extensions combine precedence constraints with contributor-specific weights and accommodate weighted or cyclic priority relations~\citep{lee2026priority,lee2026generalized}.
These ordering choices can reflect domain knowledge, as Asymmetric Shapley Values (ASVs) incorporate known causal precedence in feature attribution~\citep{frye2020asymmetric}.
In federated learning, \citet{wang2020principled} evaluate data contributions relative to the model state inherited from earlier training rounds.
In vertical federated learning, FedValue's master-conditioned Shapley rule fixes the task party first in every ordering, valuing external data relative to the task party's internal data~\citep{han2026data}.
For AI model markets, a complementary challenge is to identify separately valued contributions and specify economically meaningful marginal comparisons based on documented production dependencies.

The applications reviewed above primarily value features or data contributions.
Joint attribution across raw data, derivative data, and data-related services in structured AI production pipelines requires an explicit specification of valuation units, pipeline-supported predecessor contexts, and the coalition utilities needed for marginal comparisons.
Our approach treats these heterogeneous contributions as separate valuation units and uses documented provenance, input--operation, and model-state dependencies to organize them into ordered valuation groups.
These groups determine the supported predecessor contexts and the coalition utilities required for valuation.
On the resulting utility domain, \texttt{ADS} applies the equal-weight specialization of the weighted Shapley value~\citep{kalai1987weighted,nowak1995axiomatizations}.
We characterize this allocation axiomatically and establish that each group receives the utility it adds to its upstream prefix and that individual allocations are invariant to utilities assigned to unsupported coalitions.

%% file: S3_Limitations_of_Data_Shapley.tex
Valuing heterogeneous contributions in structured AI pipelines requires specifying the predecessor contexts in which their marginal contributions are evaluated. 
Ignoring production dependencies can cause contributions to be evaluated in contexts that never arise in the documented pipeline, creating pipeline-context valuation mismatch. 
We develop this argument by first establishing a valuation benchmark and then examining three forms of production dependence.
Section~\ref{subsec:data-shapley} introduces classical Data Shapley as the benchmark, Section~\ref{subsec:input-service-dependence} examines provenance and input--operation dependencies, and Section~\ref{subsec:sequential} examines model-state dependence in staged training. 
These analyses motivate the framework developed in Section~\ref{sec:asymmetric data shapley}.

\subsection{Classical Data Shapley as a Structure-Agnostic Baseline}
\label{subsec:data-shapley}

Classical Data Shapley (\texttt{DS}) provides a structure-agnostic benchmark when raw-data contributions are treated as peers without production dependencies.
Let \(\mathcal D^{\mathrm{raw}}:=\{D_1,\ldots,D_n\}\) denote the raw-data bundles supplied by \(n\) contributors, where each \(D_i\) comprises multiple training data examples.
For a coalition \(S\subseteq\mathcal D^{\mathrm{raw}}\), let \(D_S\)
denote the pooled training data obtained by concatenating the bundles in
\(S\), retaining repeated training examples.
For each nonempty coalition \(S\), let \(\widehat h_S:=\mathsf{Train}(D_S)\) denote the model obtained by applying a prespecified learning procedure \(\mathsf{Train}\) to \(D_S\), and define its utility as \(v(S):=V(\widehat h_S)\), where \(V\) is a prespecified measure of model performance or market utility.
Let \(v(\varnothing)\) be the utility of the prespecified baseline model.
The learning and evaluation protocol is held fixed across coalitions, so that
\(
v:2^{\mathcal D^{\mathrm{raw}}}\rightarrow\mathbb R
\)
defines the utility function for the \texttt{DS} game
\(
(\mathcal D^{\mathrm{raw}},v).
\)
\begin{definition}[Marginal Contribution]\label{def:ds_marginal}
For \(D_i \in \mathcal D^{\mathrm{raw}}\) and a coalition \(S \subseteq \mathcal D^{\mathrm{raw}} \setminus \{D_i\}\), define the marginal contribution of \(D_i\) to predecessor coalition \(S\) as 
$\Delta(D_i \mid S) := v(S \cup \{D_i\}) - v(S)$. 
\end{definition}
Let \(\Pi(\mathcal D^{\mathrm{raw}})\) denote the set of all permutations of the raw-data bundles. For \(\pi \in \Pi(\mathcal D^{\mathrm{raw}})\), define
\(
\pre_{\pi}(D_i) 
:=
\{D_j \in \mathcal D^{\mathrm{raw}} : D_j \text{ precedes } D_i \text{ in } \pi\}.
\)
Then \texttt{DS} averages each bundle's marginal contributions across predecessor coalitions.
Specifically, the \texttt{DS} value of \(D_i\) is
\begin{equation}\label{eq:ds_marginal}
\ds{D_i}{\mathcal D^{\mathrm{raw}}}
=
\frac{1}{n!}
\sum_{\pi \in \Pi(\mathcal D^{\mathrm{raw}})}
\Delta\!\left(D_i \mid \pre_{\pi}(D_i)\right).
\end{equation}

For the fixed game
\(
(\mathcal D^{\mathrm{raw}},v),
\)
the classical Shapley value is the unique allocation satisfying efficiency, linearity, nullity, and symmetry.
These properties make \texttt{DS} attractive for market valuation.
Efficiency allocates exactly the total incremental utility
\(
v(\mathcal D^{\mathrm{raw}})-v(\varnothing)
\),
thereby providing budget balance in terms of \(v\).
Nullity assigns zero value to a bundle that never changes utility.
Symmetry assigns equal value to bundles with identical marginal-contribution profiles.
Linearity preserves additive combinations of valuation objectives.
These properties explain how \texttt{DS} allocates value once the utility game is specified. 

Because \texttt{DS} averages over all permutations, every subset of
\(\mathcal D^{\mathrm{raw}}\setminus\{D_i\}\) can appear as the predecessor coalition of \(D_i\) with positive probability.
\texttt{DS} is therefore structure-agnostic in the sense that these predecessor contexts used to evaluate \(D_i\)'s marginal contribution do not depend on production-based dependencies among the data bundles. This is well aligned with valuation settings in which the raw-data bundles are treated as unordered peers, but can become less appropriate when value is generated through a structured production pipeline in which upstream and downstream units are linked by production dependencies.
Sections~\ref{subsec:input-service-dependence} and~\ref{subsec:sequential}
show how provenance, input--operation, and model-state dependence make this distinction consequential.

\subsection{Provenance and Input--Operation Dependencies}
\label{subsec:input-service-dependence}

Structured AI production pipelines often combine contributor-supplied raw data,
derivative data such as augmented or synthetic data, and data-related
services such as deduplication~\citep{heinrich2025different}, label auditing, and feature engineering.
These components create two forms of production dependence relevant to
valuation.
Provenance dependence links a derivative-data unit to the upstream data and
transformations that produced it.
Input--operation dependence links a service to the data or intermediate state
on which it operates.
We examine these two dependencies in turn and show how each can create a mismatch between the marginal comparisons induced by unrestricted Shapley ordering and the production relationships that govern how individual units contribute to value.

We begin with provenance dependence. Data augmentation and synthetic-data generation are widely used to expand training datasets and improve generalization \citep{shorten2019survey}. These techniques often create derivative training data from existing data examples.
Because the information in a derivative may overlap with its upstream source,
its marginal value can depend on whether that source is already present when
the derivative is evaluated.
Example~\ref{ex:MD} and Lemma~\ref{lemma:example1_redundancy} formalize this provenance dependence in the simple case of exact duplication. 
The proof of Lemma~\ref{lemma:example1_redundancy} is provided in Appendix~\ref{sec:proofs}.

\begin{example}[Exact Duplication]
\label{ex:MD}

Let
\(
\mathcal D^{\mathrm{raw}}=\{D_1,\ldots,D_n\}
\)
be contributor-supplied raw-data bundles.
A derivative-data provider creates
\(
\mathcal D^{\mathrm{aug}}
=
\{\widetilde D_1,\ldots,\widetilde D_n\},
\)
where each \(\widetilde D_i\) is a separately indexed copy of \(D_i\)
containing exactly the same training examples.
Let
\(
\mathcal D^{\mathrm{dup}}
:=
\mathcal D^{\mathrm{raw}}
\dot\cup
\mathcal D^{\mathrm{aug}}.
\)
A full-permutation Shapley valuation over \(
\mathcal D^{\mathrm{dup}}\) treats the \(2n\) raw and derivative
bundles as distinct players. 
Lemma~\ref{lemma:example1_redundancy} characterizes the resulting allocation
under exact duplication.

\end{example}

\begin{lemma}[\texttt{DS} under exact duplication]
\label{lemma:example1_redundancy}

Consider the game
\(
(\mathcal D^{\mathrm{dup}},v)
\)
defined in Example~\ref{ex:MD}.
For every \(i=1,\ldots,n\) and every
\(
S\subseteq
\mathcal D^{\mathrm{dup}}\setminus\{D_i,\widetilde D_i\},
\)
replacing \(D_i\) with its exact copy \(\widetilde D_i\) makes the two coalitions instantiate identical training data, so that
\(
v(S\cup\{D_i\})
=
v(S\cup\{\widetilde D_i\}).
\)
Shapley symmetry therefore gives
\(
\phi(D_i;v,\mathcal D^{\mathrm{dup}})
=
\phi(\widetilde D_i;v,\mathcal D^{\mathrm{dup}}),
\text{for every } i=1,\ldots,n.
\)
Moreover, the total value assigned to the raw-data bundles equals that assigned to the derivative-data bundles, with each group accounting for one half of the total surplus:
\begin{equation}
\label{eq:dup-tier-split}
\sum_{i=1}^{n}
\phi(D_i;v,\mathcal D^{\mathrm{dup}})
=
\sum_{i=1}^{n}
\phi(\widetilde D_i;v,\mathcal D^{\mathrm{dup}})
=
\frac{1}{2}
\Bigl[
v(\mathcal D^{\mathrm{dup}})
-
v(\varnothing)
\Bigr].
\end{equation}
If the evaluated model utility is invariant to uniform duplication of data, then
\(
v(\mathcal D^{\mathrm{dup}})
=
v(\mathcal D^{\mathrm{raw}}).
\)

\end{lemma}

Lemma~\ref{lemma:example1_redundancy} highlights the market consequence of ignoring provenance when derivative data reproduce information already supplied upstream.
Even when the duplicate data add no incremental utility at their documented downstream position, \texttt{DS} assigns equal value to the original and duplicated data. 
This can shift compensation from original contributors toward redundant derivative data, potentially weakening incentives to supply original data or to create derivative data that provide genuinely incremental information
\citep{agarwal2019marketplace,ohrimenko2019collaborative}.
Beyond these incentive effects, in copyright-sensitive settings, such reallocation may divert compensation from rightsholders whose data were duplicated
\citep{wang2024economic}.

We next consider input--operation dependence.
Data-related services pose a different valuation challenge because their
contributions depend on the specific data or intermediate output on which they operate.
Their value must therefore be evaluated in the operating
context that gives them economic meaning.
Outside that context, the documented pipeline may not determine what the
service would do or what utility it would create, 
leaving the resulting value
attribution sensitive to analyst-imposed assumptions.
Example~\ref{ex:broker_service} illustrates this ambiguity with a label-audit
service and shows how it affects the allocation of value between data and
service contributions.

\begin{example}[Label Audit Service]
\label{ex:broker_service}
Consider a pipeline in which a raw-data bundle \(D\) is first supplied and a label-audit service \(b\) is then applied to it.
The documented pipeline specifies $v(\varnothing)=0$, $v(\{D\})=a$, and $v(\{D,b\})=a+\gamma$, so the data add $a$ and the audit adds $\gamma$.
It does not, however, determine the utility of applying the audit without the data it is intended to audit.
Any valuation rule that requires this off-workflow case must therefore specify that utility by assumption, say $x:=v(\{b\})$.
Under \texttt{DS},
\[
\phi(D;v,\{D,b\}) = a+\frac{\gamma-x}{2},
\qquad
\phi(b;v,\{D,b\}) = \frac{\gamma+x}{2}.
\]
The allocation thus changes with \(x\), even though the documented pipeline is unchanged.
A natural assumption, \(x=0\), under which the audit alone creates no value, yields \((a+\gamma/2,\ \gamma/2)\) and shifts half of the audit's documented increment to the data.
Recovering the documented increments \((a,\gamma)\) requires \(x=\gamma\), a choice that nothing in the documented pipeline identifies.
\end{example}

Together, Examples~\ref{ex:MD} and~\ref{ex:broker_service} show that ignoring production relationships distorts the valuation of both derivative data and data-related services. 
\texttt{DS} can shift compensation from original contributors to redundant derivative data. 
It also makes a service's allocated value depend on an off-workflow assumption, which weakens the transparency of value attribution. 
Even under the natural assumption that the service alone creates no value, \texttt{DS} credits part of the service's contribution to its input data. 
In both examples, the distortion arises because \texttt{DS} evaluates contributions in predecessor contexts inconsistent with the documented pipeline structure.

\subsection{Model-State Dependence}
\label{subsec:sequential}

Section~\ref{subsec:input-service-dependence} examined how provenance and input--operation relationships shape marginal contributions within a production stage. 
Staged training adds a dependence across stages because later contributions are applied to a model state that earlier stages have already updated. 
We refer to this as model-state dependence. 
As a result, the same data or service can create different incremental value at different stages, even when its content or operation is unchanged.

\begin{example}[Identical Data at Different Model Checkpoints]
\label{ex:sequential_llm}
Consider a two-stage training pipeline in which each stage continues training from the model checkpoint produced by the previous stage.
Let \(\theta_0\) denote the prespecified baseline model, and let \(\mathsf{Train}_k(\theta;D)\) denote the stage-\(k\) training operation that updates checkpoint \(\theta\) using data \(D\).
At stage~1, a contributor supplies a data bundle \(D_1\), yielding \(\theta_1=\mathsf{Train}_1(\theta_0;D_1)\).
At stage~2, another contributor supplies a separate bundle \(D_2\) containing exactly the same training examples as \(D_1\), yielding \(\theta_2=\mathsf{Train}_2(\theta_1;D_2)\).
The documented pipeline therefore specifies \(v(\varnothing)=V(\theta_0)\), \(v(\{D_1\})=V(\theta_1)\), and \(v(\{D_1,D_2\})=V(\theta_2)\).
Suppose \(v(\{D_1\})-v(\varnothing)=5\) and \(v(\{D_1,D_2\})-v(\{D_1\})=1\).
Although the two bundles contain identical training examples, \(D_1\) adds five units of utility at the baseline model \(\theta_0\), whereas \(D_2\) adds one unit at the updated checkpoint \(\theta_1\).
Unlike Example~\ref{ex:MD}, where the original and its copy enter the same pooled training run, \(D_1\) and \(D_2\) enter at different checkpoints, so the difference in their marginal contributions reflects only the model states they inherit.
\end{example}

Example~\ref{ex:sequential_llm} shows that identical data can make different marginal contributions when they enter at different model states.
Earlier training may absorb information contained in a later submission, reducing its incremental value, or create capabilities that make the later submission more useful.
A valuation rule that ignores model-state dependence can therefore misallocate value by evaluating a contribution against a model state it never encounters in the documented pipeline.
Under \texttt{DS}, for instance, \(D_2\) is also evaluated at the baseline model \(\theta_0\), which requires the off-workflow utility \(v(\{D_2\})\).
If stage-2 training on \(\theta_0\) would also add five units, \texttt{DS} assigns three units to each bundle despite their documented increments of five and one.

More generally, staged pipelines combine this cross-stage dependence with the provenance and input--operation relationships examined in Section~\ref{subsec:input-service-dependence}.
A valuation framework for structured AI production must therefore account for both the model state each contribution inherits and the within-stage relationships that shape its marginal contribution.
Section~\ref{sec:asymmetric data shapley} develops such a framework.

%% file: S4_Asymmetric_Data_Shapley.tex
Section~\ref{sec:counterintuition} shows that provenance, input--operation, and model-state dependence determine the meaningful predecessor contexts for each contribution, and that ignoring them can misallocate value. 
This section develops \texttt{ADS}, which evaluates each contribution only in pipeline-supported predecessor contexts. 
Section~\ref{subsec:ads-units-groups} defines valuation units and maps documented production relationships into ordered valuation groups. 
Sections~\ref{subsec:ads-definition} and~\ref{subsec:ads-properties} then define the \texttt{ADS} rule, characterize it axiomatically, and derive its allocation properties.

\subsection{From Pipeline Structure to Valuation Groups}
\label{subsec:ads-units-groups}
To translate a documented AI pipeline into a valuation structure, the
marketplace makes two linked design choices.
It first identifies the pipeline contributions to be treated as separate valuation units.
It then uses documented production relationships to determine which of these units can be treated as peers and which are subject to production precedence.
The resulting ordered valuation groups determine the predecessor contexts used by \texttt{ADS}.

The first design choice concerns the valuation units.
A valuation unit is a pipeline contribution that is valued separately and for which inclusion or exclusion at its documented pipeline position defines a meaningful counterfactual.
Let the finite set of valuation units be
\(
\mathcal U
=
\mathcal D^{\mathrm{raw}}
\cup
\mathcal D^{\mathrm{aug}}
\cup
\mathcal B^{\mathrm{svc}},
\)
where \(\mathcal D^{\mathrm{raw}}\) contains original data bundles,
\(\mathcal D^{\mathrm{aug}}\) contains derivative data, and \(\mathcal B^{\mathrm{svc}}\) contains service modules.
The three categories are disjoint as valuation units, although their underlying data content may overlap.
For example, a derivative-data unit may reproduce information contained in an upstream raw-data bundle while remaining a distinct valuation unit.
The boundary of a valuation unit is a design choice: an
augmentation activity may be represented by its derivative output, by
the service that produces that output, or by separate units for both
when their counterfactual effects can be defined without double counting.
\texttt{ADS} allocates value to valuation units, and a participant associated with multiple units receives the sum of their values.

Documented production evidence then determines how the valuation units are grouped.
The marketplace may use provenance links, transformation recipes, service logs, timestamps, checkpoint histories, and other auditable execution records to identify production precedence.
Units are placed in the same valuation group when they share the same required predecessors and no documented production relationship requires one to precede another.
When documented production relationships require some units to be included in the predecessor context of others, the required predecessors are assigned to earlier valuation groups.
For the fixed-structure setting studied in Sections~\ref{sec:asymmetric data shapley}--\ref{sec:experiments},
this mapping produces a single declared valuation structure specified
before the resulting allocation is observed.
Section~\ref{sec:robustness_strategy} relaxes this assumption when the available evidence, such as a partial precedence order, supports multiple plausible structures.

\begin{definition}[\bf Ordered valuation structure]
\label{definition:order_partition}

An ordered valuation structure on \(\mathcal U\) is an ordered partition
\(
\sigma=(G_1,\ldots,G_Q),
\)
where the valuation groups are nonempty, pairwise disjoint, and cover
\(\mathcal U\):
\[
G_q\neq\varnothing
\quad\text{for }q=1,\ldots,Q,
\qquad
G_q\cap G_r=\varnothing
\quad\text{for }q\neq r,
\qquad
\mathcal U=\bigcup_{q=1}^{Q}G_q.
\]
Units within \(G_q\) are treated as peers with respect to production
precedence, while every group \(G_r\) with \(r<q\) precedes \(G_q\).
Define the pipeline prefix through group \(G_q\) by
\[
U_0:=\varnothing,
\qquad
U_q:=\bigcup_{r=1}^{q}G_r,
\qquad
q=1,\ldots,Q.
\]
\end{definition}

Given the ordered valuation groups, we next specify the predecessor contexts used to evaluate each unit.
For \(u\in G_q\), all units in the preceding pipeline prefix
\(U_{q-1}\) are held in place, while only the presence of peer units in
\(G_q\) is allowed to vary.
Accordingly, define the pipeline-supported predecessor contexts of
\(u\) as
\[
\mathcal C_\sigma(u)
:=
\left\{
U_{q-1}\cup S
:
S \subseteq G_q\setminus\{u\}
\right\}.
\]
This construction maps the three production dependencies in Section~\ref{sec:counterintuition} into pipeline-supported predecessor contexts. 
For provenance dependence in Example~\ref{ex:MD}, let $\sigma_1=(\mathcal{D}^{\mathrm{raw}},\mathcal{D}^{\mathrm{aug}})$. For $\widetilde D_i\in\mathcal{D}^{\mathrm{aug}}$, $\mathcal{C}_{\sigma_1}(\widetilde D_i)=\{\mathcal{D}^{\mathrm{raw}}\cup S: S\subseteq\mathcal{D}^{\mathrm{aug}}\setminus\{\widetilde D_i\}\}$, so each derivative is evaluated with all upstream raw-data bundles in place and only its peer derivatives varying. 
For input--operation dependence in Example~\ref{ex:broker_service}, let $\sigma_2=(\{D\},\{b\})$. Then $\mathcal{C}_{\sigma_2}(b)=\{\{D\}\}$, so the service is evaluated only with its input data in place, and the off-workflow utility $v(\{b\})$ is never needed. 
For model-state dependence in Example~\ref{ex:sequential_llm}, let $\sigma_3=(\{D_1\},\{D_2\})$. Then $\mathcal{C}_{\sigma_3}(D_2)=\{\{D_1\}\}$, so $D_2$ is evaluated only at the checkpoint $\theta_1$ it inherits, and $v(\{D_2\})$ is likewise never needed.

The supported predecessor contexts also determine which coalition utilities are needed to define the valuation rule.
We therefore define the utility domain required by \texttt{ADS} as
\[
\mathcal F_\sigma
:=
\bigcup_{q=1}^{Q}
\left\{
U_{q-1}\cup S
:
S \subseteq G_q
\right\}
=
\bigcup_{u\in\mathcal U}
\bigcup_{P\in\mathcal C_\sigma(u)}
\left\{
P,\,
P\cup\{u\}
\right\}.
\]
The first expression builds \(\mathcal F_\sigma\) from the ordered valuation groups.
The second shows that \(\mathcal F_\sigma\) contains exactly the predecessor coalition \(P\) and the coalition \(P\cup\{u\}\) for every supported marginal comparison, so no utility outside \(\mathcal F_\sigma\) is required to define \texttt{ADS}.
The marketplace specifies $v:\mathcal{F}_\sigma\to\mathbb{R}$, where $v(P)$ is the prespecified utility, such as out-of-sample model performance, obtained when the valuation units in $P$ are incorporated according to the declared pipeline structure.
Since \(\varnothing,\mathcal U\in\mathcal F_\sigma\), the total utility increment \(v(\mathcal U)-v(\varnothing)\) is well defined.

\subsection{\texttt{ADS} Valuation Rule}
\label{subsec:ads-definition}

Section~\ref{subsec:ads-units-groups} defines the supported predecessor contexts \(\mathcal C_\sigma(u)\) for each valuation unit and the utility
domain \(\mathcal F_\sigma\) containing the coalitions needed to evaluate
the supported marginal contributions. We now define how \texttt{ADS}
weights these marginal contributions. The construction preserves the
declared order across valuation groups while retaining Shapley-style
random ordering among peer units within each group.

Let \(\Pi(S)\) denote the set of permutations of a finite set \(S\).
We define the set of pipeline-consistent permutations as
\[
\Pi_\sigma(\mathcal U)
:=
\left\{
\pi_1\oplus\cdots\oplus\pi_Q
:
\pi_q\in\Pi(G_q),\ q=1,\ldots,Q
\right\},
\]
where $\oplus$ denotes sequence concatenation, so that $|\Pi_\sigma(\mathcal U)|=\prod_{q=1}^{Q}|G_q|!$. 
For $\pi\in\Pi(\mathcal U)$, let $\mathrm{Pre}_\pi(u)$ denote the units that precede $u$ in $\pi$. 
For $u\in G_q$ and $\pi\in\Pi_\sigma(\mathcal U)$, $\mathrm{Pre}_\pi(u)=U_{q-1}\cup S$ for some $S\subseteq G_q\setminus\{u\}$, so $\mathrm{Pre}_\pi(u)\in\mathcal C_\sigma(u)$. 
Following Definition~\ref{def:ds_marginal}, for $P\in\mathcal C_\sigma(u)$, let $\Delta(u\mid P):=v(P\cup\{u\})-v(P)$, which is well defined because $P,P\cup\{u\}\in\mathcal F_\sigma$.

\begin{definition}[\bf Asymmetric Data Shapley]
\label{definition:ads}
For $u\in\mathcal U$, define its \texttt{ADS} value under the ordered valuation structure $\sigma$ as
\begin{equation}\label{eq:asv_permutation}
\phi^\sigma(u;v,\mathcal U):=\frac{1}{|\Pi_\sigma(\mathcal U)|}\sum_{\pi\in\Pi_\sigma(\mathcal U)}\Delta\bigl(u\mid \mathrm{Pre}_\pi(u)\bigr).
\end{equation}
\end{definition}
Eq.~\eqref{eq:asv_permutation} is the
equal-weight specialization of the weighted Shapley value~\citep{nowak1995axiomatizations}, as derived in Appendix~\ref{sec:proofs}.

\begin{proposition}[Predecessor-context and induced-game representations of \texttt{ADS}]
\label{propo:asv_subset}

For \(u\in G_q\) and \(P\subseteq\mathcal U\setminus\{u\}\),
uniform sampling from \(\Pi_\sigma(\mathcal U)\) induces the
predecessor-context distribution
\begin{equation}
\label{eq:asv_prob}
p^\sigma_u(P)
:=
\Pr_{\pi\sim\operatorname{Unif}(\Pi_\sigma(\mathcal U))}
\!\left(
\pre_{\pi}(u)=P
\right)
=
\begin{cases}
\displaystyle
\frac{1}{|G_q|}
\binom{|G_q|-1}{|P\setminus U_{q-1}|}^{-1},
&
P\in\mathcal C_\sigma(u),
\\[8pt]
0,
&
P\notin\mathcal C_\sigma(u).
\end{cases}
\end{equation}
Hence, the support of the predecessor-context distribution is
\(
\mathcal C_\sigma(u)
\),
and
\begin{equation}\label{eq:asv_subset}
\ads{u}{\mathcal U}
=
\frac{1}{|G_q|}
\sum_{P\in\mathcal C_\sigma(u)}
\binom{|G_q|-1}{|P\setminus U_{q-1}|}^{-1}
\Delta(u\mid P).
\end{equation}

Equivalently, define the induced game
\(v_q:2^{G_q}\to\mathbb R\) by
\(
v_q(S)
:=
v(U_{q-1}\cup S)-v(U_{q-1}),
\)
where $S\subseteq G_q$.
Then
\(
\ads{u}{\mathcal U}
=
\phi(u;v_q,G_q),
\)
where \(\phi(\cdot;v_q,G_q)\) is the classical Shapley value of the
induced game \((G_q,v_q)\).
\end{proposition}
Proposition~\ref{propo:asv_subset} makes explicit how \texttt{ADS} incorporates pipeline structure into the Shapley averaging developed in Section~\ref{subsec:data-shapley}.
Whereas classical \texttt{DS} allows every subset of the other valuation units to serve as a predecessor coalition of the focal unit, \texttt{ADS} restricts the predecessor contexts of \(u\) to \(\mathcal C_\sigma(u)\) according to the declared structure \(\sigma\).
Within this set of pipeline-supported contexts, \texttt{ADS} applies the canonical Shapley weights induced by the relative orderings of peer units in \(G_q\).
Equivalently, once the pipeline prefix \(U_{q-1}\) is fixed, the allocation among units in \(G_q\) is the classical Shapley allocation of the induced game \((G_q,v_q)\).

In staged-training settings, Eq.~\eqref{eq:asv_subset} can be simplified when the inherited model checkpoint summarizes all valuation-relevant effects of the preceding pipeline prefix. 
Let $\theta_{q-1}$ denote the checkpoint inherited before $G_q$, and let $\tilde v_q(S;\theta_{q-1})$ denote the utility obtained by incorporating $S\subseteq G_q$ into this checkpoint. 
We formalize this requirement as the state-sufficiency condition
\begin{equation}\label{eq:state_sufficiency}
v_q(S)=\tilde v_q(S;\theta_{q-1})-\tilde v_q(\varnothing;\theta_{q-1}),\qquad S\subseteq G_q. \end{equation}
In Example~\ref{ex:sequential_llm}, this condition holds by construction because $\theta_2=\mathsf{Train}_2(\theta_1;D_2)$ depends on stage~1 only through $\theta_1$. 
If current-stage execution requires information from $U_{q-1}$ not encoded in $\theta_{q-1}$, such as upstream data, provenance information, or service outputs, the condition may fail, and \texttt{ADS} is evaluated directly through Eq.~\eqref{eq:asv_subset}.
For \(u\in G_q\) and \(S \subseteq G_q\setminus\{u\}\), define the corresponding state-conditioned marginal contribution as
\begin{equation}
\label{eq:state_marginal}
\Delta_q(u\mid S;\theta_{q-1})
:=
\widetilde v_q(S\cup\{u\};\theta_{q-1})
-
\widetilde v_q(S;\theta_{q-1}).
\end{equation}
Every \(P\in\mathcal C_\sigma(u)\) can be written uniquely as
\(
P=U_{q-1}\cup S
\) with $S \subseteq G_q \setminus \{u\}$.
Under the state-sufficiency condition in
Eq.~\eqref{eq:state_sufficiency}, the corresponding marginal
contributions satisfy
\begin{equation}
\label{eq:state_marginal_equivalence}
\Delta\!\left(u\mid U_{q-1}\cup S\right)
=
\Delta_q(u\mid S;\theta_{q-1}).
\end{equation}
Thus, substituting Eq.~\eqref{eq:state_marginal_equivalence} into Eq.~\eqref{eq:asv_subset} yields
\begin{equation}
\label{eq:asv_seq}
\ads{u}{\mathcal U}
=
\frac{1}{|G_q|}
\sum_{S \subseteq G_q\setminus\{u\}}
\binom{|G_q|-1}{|S|}^{-1}
\Delta_q(u\mid S;\theta_{q-1}),
\qquad
u\in G_q.
\end{equation}
Eq.~\eqref{eq:asv_seq} is a state-conditioned representation of the \texttt{ADS} allocation. 
It evaluates each unit in $G_q$ relative to the checkpoint inherited from the preceding pipeline prefix, thereby capturing the model-state dependence in Section~\ref{subsec:sequential}. 
As a special case, when valuation groups correspond to federated-learning rounds and each contributor's participation in a round is treated as a separate valuation unit, Eq.~\eqref{eq:asv_seq} reduces to the round-conditioned federated Shapley value of \citet{wang2020principled}. 
\texttt{ADS} differs from this construction in three respects. 
First, its valuation groups are constructed from documented provenance, input--operation, and model-state dependencies rather than fixed by training rounds.
Second, it values raw data, derivative data, and data-related services within a unified valuation framework, rather than data contributors alone.
Third, it does not require state sufficiency, because Eqs.~\eqref{eq:asv_permutation} and~\eqref{eq:asv_subset} remain valid when later units need information not encoded in the inherited checkpoint, such as the upstream data on which a service operates.

\subsection{Axiomatic Characterization and Allocation Properties}
\label{subsec:ads-properties}
We now characterize \texttt{ADS} axiomatically on the utility domain $\mathcal F_\sigma$ and derive its allocation properties.
We focus on efficiency, linearity, nullity, and symmetry, which parallel the classical Shapley axioms but are adapted to the utility domain and predecessor contexts induced by the pipeline structure. Specifically, efficiency and linearity are defined over utility functions \(v:\mathcal F_\sigma\to\mathbb R\), whereas nullity and symmetry are imposed only with respect to pipeline-supported predecessor contexts $\mathcal{C}_\sigma(\cdot)$. Let
\(\psi^\sigma(u;v,\mathcal U)\) denote an arbitrary candidate allocation rule under \(\sigma\). 
We state the four axioms for this candidate rule and then show that they uniquely characterize \texttt{ADS} for a fixed $\sigma$.

\begin{axiom}[Pipeline-conditioned efficiency]
\label{axiom:ads_efficiency}
For every \(v:\mathcal F_\sigma\to\mathbb R\),
\[
\sum_{u\in\mathcal U}\psi^\sigma(u;v,\mathcal U)
=
v(\mathcal U)-v(\varnothing).
\]
\end{axiom}

\begin{axiom}[Pipeline-conditioned linearity]
\label{axiom:ads_linearity}
For \(v_1,v_2:\mathcal F_\sigma\to\mathbb R\),
\(\alpha,\beta\in\mathbb R\), and \(u\in\mathcal U\),
\[
\psi^\sigma(u;\alpha v_1+\beta v_2,\mathcal U)
=
\alpha\psi^\sigma(u;v_1,\mathcal U)
+
\beta\psi^\sigma(u;v_2,\mathcal U).
\]
\end{axiom}

\begin{axiom}[Pipeline-conditioned nullity]
\label{axiom:ads_nullity}
For \(u\in G_q\), if
\[
v(P\cup\{u\})=v(P)
\qquad
\text{for every }P\in\mathcal C_\sigma(u),
\]
then
\(\psi^\sigma(u;v,\mathcal U)=0\).
\end{axiom}

\begin{axiom}[Within-group symmetry]
\label{axiom:ads_symmetry}
For distinct \(u,u'\in G_q\), if
\[
v(P\cup\{u\})=v(P\cup\{u'\})
\qquad
\text{for every }
P\in\mathcal C_\sigma(u)\cap\mathcal C_\sigma(u'),
\]
then
\(
\psi^\sigma(u;v,\mathcal U)
=
\psi^\sigma(u';v,\mathcal U).
\)
\end{axiom}

\begin{theorem}[Axiomatic characterization of \texttt{ADS}]
\label{proposition:ads_characterization}
For a fixed ordered valuation structure $\sigma$, an allocation rule $\psi^\sigma$ satisfies Axioms~1--4 if and only if $\psi^\sigma(u;v,\mathcal U)=\phi^\sigma(u;v,\mathcal U)$ for every $v:\mathcal F_\sigma\to\mathbb R$ and every $u\in\mathcal U$.
\end{theorem}

Theorem~\ref{proposition:ads_characterization} follows from the induced-game representation in Proposition~\ref{propo:asv_subset}.
Pipeline-conditioned linearity and nullity imply that the allocation of \(u\in G_q\) depends only on the induced game \(v_q\).
Applying pipeline-conditioned efficiency to a utility in which only \(G_q\) creates value, so that all units outside \(G_q\) are null, shows that the allocations within \(G_q\) sum to \(v_q(G_q)\).
Together with linearity, nullity, and within-group symmetry, the classical Shapley characterization~\citep{shapley1953value} therefore gives \(\psi^\sigma(u;v,\mathcal U)=\phi(u;v_q,G_q)=\phi^\sigma(u;v,\mathcal U)\).
The detailed proof is provided in Appendix~\ref{sec:proofs}, which also proves Propositions~\ref{propo:asv_subset}--\ref{proposition:ads_completion_invariance}.

\begin{proposition}[Reduction to classical Shapley valuation]
\label{proposition:sv_as_special_asv}
If \(\sigma=(\mathcal U)\) consists of a single valuation group, then
\(
\Pi_\sigma(\mathcal U)=\Pi(\mathcal U)\), \(\mathcal F_\sigma=2^{\mathcal U},
\)
and \texttt{ADS} reduces to the classical Shapley value of
\((\mathcal U,v)\). When
\(\mathcal U=\mathcal D^{\mathrm{raw}}\), \texttt{ADS} coincides with classical \texttt{DS}, and Eq.~\eqref{eq:asv_permutation} reduces to Eq.~\eqref{eq:ds_marginal}.
\end{proposition}

Proposition~\ref{proposition:sv_as_special_asv} connects \texttt{ADS} to the structure-agnostic setting developed in Section~\ref{subsec:data-shapley}.
When the production process imposes no precedence among valuation units, the declared structure contains a single peer group, and the pipeline-supported predecessor contexts coincide with all subsets of the other units.
This case applies, for example, to simple data markets in which raw-data bundles are pooled in a single training step and no services are separately valued.
When documented provenance, input--operation, or model-state dependencies instead impose production precedence, the valuation units are assigned to different ordered groups and their predecessor contexts are restricted accordingly.

\begin{proposition}[Group efficiency]
\label{propo:efficiency_asv_intra_group}
For every \(q=1,\ldots,Q\),
\[
\sum_{u\in G_q}\ads{u}{\mathcal U}
=
v(U_q)-v(U_{q-1}).
\]
\end{proposition}

Group efficiency assigns each valuation group exactly the utility increment it creates relative to its preceding pipeline prefix \(U_{q-1}\), and these increments telescope to \(v(\mathcal U)-v(\varnothing)\) across groups.
In Example~\ref{ex:MD}, if uniform duplication leaves model utility unchanged, the derivative group receives zero aggregate value, while the raw-data group receives \(v(\mathcal D^{\mathrm{raw}})-v(\varnothing)\).
Because the raw-data group is the first valuation group, each raw-data bundle receives the same value as in the game involving raw data alone, whereas \texttt{DS} assigns the raw-data group only half of the total surplus (Lemma~\ref{lemma:example1_redundancy}).
Adding redundant derivative data therefore does not dilute the value attributed to original contributors, which discourages duplicative submissions while preserving incentives to supply original data and genuinely value-adding transformations.

\begin{proposition}[Completion invariance]
\label{proposition:ads_completion_invariance}
Fix an ordered valuation structure \(\sigma\) and a pipeline-supported
utility \(v:\mathcal F_\sigma\to\mathbb R\). 
Let $\bar v_1,\bar v_2:2^{\mathcal U}\to\mathbb R$ be any two completions of $v$, that is, full-domain utility functions satisfying $\bar v_1(P)=\bar v_2(P)=v(P)$ for every $P\in\mathcal F_\sigma$.
Then, for every \(u\in\mathcal U\),
\(
\phi^\sigma(u;\bar v_1,\mathcal U)
=
\phi^\sigma(u;\bar v_2,\mathcal U)
=
\phi^\sigma(u;v,\mathcal U).
\)
\end{proposition}

Proposition~\ref{proposition:ads_completion_invariance} shows that the \texttt{ADS} allocation depends only on utilities in $\mathcal F_\sigma$. 
This matters when the documented pipeline does not determine the utility of some coalitions. 
In Example~\ref{ex:broker_service}, the pipeline specifies $v(\varnothing)$, $v(\{D\})$, and $v(\{D,b\})$ but not $v(\{b\})$, and the \texttt{DS} allocation varies with the imputed value $x=v(\{b\})$. 
Under $\sigma=(\{D\},\{b\})$, $\{b\}\notin\mathcal F_\sigma$, so \texttt{ADS} assigns $a$ to the data and $\gamma$ to the audit for every $x$. 
Likewise, in Example~\ref{ex:sequential_llm}, \texttt{ADS} assigns five units of utility to $D_1$ and one to $D_2$ regardless of the off-workflow utility $v(\{D_2\})$. 
Completion invariance therefore ensures that allocations depend only on utilities defined by the documented pipeline rather than on analyst-imposed values for off-workflow coalitions, improving the transparency and auditability of value allocation.

%% file: S5_Efficient_Computation_of_Asymmetric_Data_Shapley.tex
Section~\ref{sec:asymmetric data shapley} defines \texttt{ADS} for a fixed
ordered valuation structure. We now develop two estimators for approximating \texttt{ADS} when exact computation is costly.
\texttt{MC-ADS} replaces exact within-group averaging with Monte Carlo
sampling, while \texttt{SA-MC-ADS} combines low-fidelity surrogate evaluations with high-fidelity residual corrections.

\subsection{Monte Carlo Estimation}
\label{subsec:mc_ads}

\subsubsection{Estimator and Implementation}
\label{subsubsec:mcads-estimator}

For a fixed structure
\(\sigma=(G_1,\ldots,G_Q)\),
Proposition~\ref{propo:asv_subset} decomposes exact \texttt{ADS}
computation into a Shapley calculation within each valuation
group.
For a non-singleton group \(G_q\), exact computation still requires
enumerating all \(|G_q|!\) peer orderings in permutation form or all
\(2^{|G_q|}\) within-group coalitions in the equivalent subset form.
\texttt{MC-ADS} replaces this exact within-group averaging with uniform
sampling of peer orderings.

Let \(R_{\mathrm{MC}}\in\mathbb{N}_+\) denote the common Monte Carlo sample size, defined as the number of within-group permutation draws used for each
non-singleton valuation group.
For each group \(G_q\), draw
\(\pi_q^{(1)},\ldots,\pi_q^{(R_q)}\)
independently from
\(\operatorname{Unif}(\Pi(G_q))\), where
\(R_q=1\) for singleton groups
and \(R_q=R_{\mathrm{MC}}\) otherwise.
The same sampled permutations are used to estimate the values of all
units in \(G_q\).
For \(\pi_q\in\Pi(G_q)\), let
\(\pre_{\pi_q}(u)\) denote the peer units in \(G_q\)
that precede \(u\) in \(\pi_q\).
Because \texttt{ADS} averages $\Delta(u\mid U_{q-1}\cup\mathrm{Pre}_{\pi_q}(u))$ over uniformly drawn peer orderings (Proposition~\ref{propo:asv_subset}), the \texttt{MC-ADS} estimator for $u\in G_q$ is
\begin{equation}
\label{eq:mc_ads_estimator}
\widehat{\phi}_{\mathrm{MC}}^{\sigma}(u;v,\mathcal U)
:=
\frac{1}{R_q}
\sum_{b=1}^{R_q}
\Delta\!\left(
u
\mid
U_{q-1}\cup
\pre_{\pi_q^{(b)}}(u)
\right).
\end{equation}
Singleton groups are therefore evaluated exactly, while non-singleton groups are approximated by averaging pipeline-supported marginal contributions across sampled peer orderings.
Algorithm~\ref{alg:mc_ads} implements
Eq.~\eqref{eq:mc_ads_estimator}.
Each sampled permutation is evaluated sequentially from
\(U_{q-1}\) to \(U_q\), and successive utility differences give the
sampled marginal contributions of all units in \(G_q\).

\begin{algorithm}[htbp]
\caption{Monte Carlo Asymmetric Data Shapley (\texttt{MC-ADS})}
\label{alg:mc_ads}
\begin{algorithmic}[1]
\Require Ordered valuation structure
\(\sigma=(G_1,\ldots,G_Q)\);
pipeline-supported utility
\(v:\mathcal F_\sigma\rightarrow\mathbb R\);
Monte Carlo sample size \(R_{\mathrm{MC}}\).
\State Initialize
\(\widehat{\phi}_{\mathrm{MC}}^{\sigma}(u;v,\mathcal U)\gets 0\)
for every \(u\in\mathcal U\).
\For{\(q=1,\ldots,Q\)}
    \State Set
    \(y_{q,0}\gets v(U_{q-1})\).
    \State Set
    \(R_q\gets 1\) if \(|G_q|=1\), and
    \(R_q\gets R_{\mathrm{MC}}\) otherwise.
    \For{\(b=1,\ldots,R_q\)}
        \State Draw
        \(\pi_q^{(b)}
        =(u_{q,1}^{(b)},\ldots,u_{q,|G_q|}^{(b)})
        \sim\operatorname{Unif}(\Pi(G_q))\).
        \State Set
        \(S_{\mathrm{act}}\gets\varnothing\) and
        \(y_{\mathrm{prev}}\gets y_{q,0}\).
        \For{\(j=1,\ldots,|G_q|\)}
            \State Set
            \(u\gets u_{q,j}^{(b)}\).
            \State
            \(S_{\mathrm{act}}
            \gets
            S_{\mathrm{act}}\cup\{u\}\).
            \State
            \(y_{\mathrm{cur}}
            \gets
            v(U_{q-1}\cup S_{\mathrm{act}})\).
            \State
            \(\widehat{\phi}_{\mathrm{MC}}^{\sigma}(u;v,\mathcal U)
            \gets
            \widehat{\phi}_{\mathrm{MC}}^{\sigma}(u;v,\mathcal U)
            +
            R_q^{-1}(y_{\mathrm{cur}}-y_{\mathrm{prev}})\).
            \State
            \(y_{\mathrm{prev}}\gets y_{\mathrm{cur}}\).
        \EndFor
    \EndFor
\EndFor
\State \textbf{Output:}
\(\{\widehat{\phi}_{\mathrm{MC}}^{\sigma}(u;v,\mathcal U)
:u\in\mathcal U\}\).
\end{algorithmic}
\end{algorithm}
\FloatBarrier
When the state-sufficiency condition in
Eq.~\eqref{eq:state_sufficiency} holds,
Eq.~\eqref{eq:state_marginal_equivalence} implies
\(
\Delta(u\mid U_{q-1}\cup S)
=
\Delta_q(u\mid S;\theta_{q-1})
\).
The sampled peer orderings can then be used to compute marginal contributions directly from the inherited checkpoint \(\theta_{q-1}\):
\begin{equation}
\label{eq:mc_ads_state_estimator}
\widehat{\phi}_{\mathrm{MC}}^{\sigma}(u;v,\mathcal U)
=
\frac{1}{R_q}
\sum_{b=1}^{R_q}
\Delta_q\!\left(
u
\mid
\pre_{\pi_q^{(b)}}(u);
\theta_{q-1}
\right).
\end{equation}
Equation~\eqref{eq:mc_ads_state_estimator} is an alternative representation of the estimator in
Eq.~\eqref{eq:mc_ads_estimator} under state sufficiency.
It avoids explicitly replaying the complete prefix \(U_{q-1}\) when the model checkpoint
\(\theta_{q-1}\) fully captures the prefix information relevant to within-group valuation.
When state sufficiency does not hold,
Eq.~\eqref{eq:mc_ads_estimator} remains applicable.

\subsubsection{Finite-Sample Guarantees and Computational Cost}
\label{subsubsec:mcads-properties}

We next characterize the finite-sample accuracy, unbiasedness, group efficiency, and computational cost of \texttt{MC-ADS}.

\begin{definition}[\((\epsilon,\delta)\)-approximation]
\label{def:epsilon_delta_approximation}
For \(\epsilon>0\) and \(\delta\in(0,1)\), an estimator
\(\widehat{\phi}^{\sigma}\) of the \texttt{ADS} allocation is an
\((\epsilon,\delta)\)-approximation if
\(
\Pr\!\left(
\max_{u\in\mathcal U}
\left|
\widehat{\phi}^{\sigma}(u;v,\mathcal U)
-
\ads{u}{\mathcal U}
\right|
\leq \epsilon
\right)
\geq 1-\delta.
\)
\end{definition}

\begin{theorem}[Finite-sample properties of \texttt{MC-ADS}]
\label{thm:mc_ads_properties}
Consider the \texttt{MC-ADS} estimator in
Eq.~\eqref{eq:mc_ads_estimator}, with the same sampled permutations
used for all units within each valuation group.

(i) \textnormal{Approximation guarantee.}
Suppose there exists \(\rho>0\) such that, for every non-singleton
valuation group \(G_q\) and every \(u\in G_q\), the supported marginal
contributions
\(
\left\{
\Delta\!\left(
u\mid U_{q-1}\cup\pre_{\pi_q}(u)
\right)
:
\pi_q\in\Pi(G_q)
\right\}
\)
lie in an interval of length at most \(\rho\).
Then $\hat\phi^\sigma_{\mathrm{MC}}$ is an $(\epsilon,\delta)$-approximation whenever $R_{\mathrm{MC}}\ge\lceil\rho^2\log(2N/\delta)/(2\epsilon^2)\rceil$, where $N\coloneqq|\mathcal U|$.

\textnormal{(ii) Unbiasedness.}
For every \(u\in\mathcal U\),
\(
\mathbb E\!\left[
\widehat{\phi}_{\mathrm{MC}}^{\sigma}(u;v,\mathcal U)
\right]
=
\ads{u}{\mathcal U}.
\)

\textnormal{(iii) Finite-sample group efficiency.}
When each sampled permutation is evaluated as a complete within-group
path, for every \(q=1,\ldots,Q\) and every \(R_{\mathrm{MC}}\geq1\),
\begin{equation}
\label{eq:mc_ads_group_efficiency}
\sum_{u\in G_q}
\widehat{\phi}_{\mathrm{MC}}^{\sigma}(u;v,\mathcal U)
=
v(U_q)-v(U_{q-1}).
\end{equation}
\end{theorem}

Part~(i) quantifies the number of sampled peer orderings needed for a prespecified uniform approximation error.
Part~(ii) establishes unbiasedness relative to the exact \texttt{ADS} allocation.
Part~(iii) shows that finite sampling may affect unit-level allocations within \(G_q\) but preserves the exact group total in Proposition~\ref{propo:efficiency_asv_intra_group}.
Proofs are provided in Appendix~\ref{app:mc_accuracy}.

\begin{remark}[Computational cost of \texttt{MC-ADS}]
\label{remark:mc_ads_cost}
Algorithm~\ref{alg:mc_ads} evaluates one baseline utility
\(v(U_{q-1})\) for each group and at most \(|G_q|\) additional
pipeline-supported utilities along each sampled path.
It therefore requires at most
\(
Q+
\sum_{q=1}^{Q}R_q|G_q|
\leq
Q+R_{\mathrm{MC}}N
\)
utility evaluations.
The actual number can be smaller because prefix utilities, repeated coalition states, inherited model checkpoints, and deterministic service outputs can be cached and reused across sampled paths.
\end{remark}

\subsection{Surrogate-Assisted Estimation}
\label{subsec:surrogate-assisted-mcads}

\texttt{MC-ADS} reduces the number of peer orderings that must be
evaluated, but each sampled path may still require costly model training
or service execution.
\texttt{SA-MC-ADS} combines low-fidelity utility evaluations with paired high-fidelity evaluations while targeting the same \texttt{ADS} allocation. Its potential cost advantage depends on both residual variance reduction and the cost of evaluating the surrogate.
The construction adapts the control-variate idea used in Monte Carlo
Shapley estimation \citep{goldwasser2024stabilizing} to the pipeline-supported
valuation domain of \texttt{ADS}.

\subsubsection{Surrogate Utility and Estimator}
\label{subsubsec:surrogate-utility-estimator-main}

\texttt{SA-MC-ADS} augments the \texttt{MC-ADS} estimator in
Eq.~\eqref{eq:mc_ads_estimator} with a low-fidelity surrogate.
Let $v^\mathrm{L}:\mathcal F_\sigma\to\mathbb R$ denote a less expensive surrogate utility defined on the same pipeline-supported domain as the marketplace utility $v$, which we also refer to as the high-fidelity utility.
Examples include a \(K\)-nearest-neighbor utility for labeled data
\citep{jia2019knn}, resource-reduced training with fewer examples, fewer epochs, or reduced model capacity
\citep{wu2020practical,trofimov2023multi}, and a lightweight prediction
head applied to fixed learned representations
\citep{chen2020simple}.

For \(u\in G_q\) and \(\pi_q\in\Pi(G_q)\), define the low-fidelity
counterpart of the sampled marginal contribution in
Eq.~\eqref{eq:mc_ads_estimator} as
\[
\Delta^{\mathrm L}\!\left(
u
\mid
U_{q-1}\cup\pre_{\pi_q}(u)
\right)
:=
v^{\mathrm L}\!\left(
U_{q-1}\cup\pre_{\pi_q}(u)\cup\{u\}
\right)
-
v^{\mathrm L}\!\left(
U_{q-1}\cup\pre_{\pi_q}(u)
\right).
\]
A useful surrogate produces low-fidelity marginal contributions that
track the corresponding high-fidelity marginal contributions
\(
\Delta(
u\mid U_{q-1}\cup\pre_{\pi_q}(u))
\)
at substantially lower evaluation cost. Let \(R_{\mathrm L},R_{\mathrm H}\in\mathbb N_+\) be the common numbers of low-only and paired permutation paths for each
non-singleton valuation group. Singleton groups are evaluated exactly
by their unique within-group permutation.
For each non-singleton group \(G_q\), draw \(\widetilde{\pi}_q^{(1)},\ldots,\widetilde{\pi}_q^{(R_{\mathrm{L}})}\) for the low-only paths and \(\pi_q^{(1)},\ldots,\pi_q^{(R_{\mathrm{H}})}\) for the paired paths, all independently from \(\operatorname{Unif}(\Pi(G_q))\).
The low-only paths are evaluated using \(v^{\mathrm{L}}\) only, whereas each
paired path is evaluated using both \(v\) and \(v^{\mathrm{L}}\) on the same
permutation path. 
As in \texttt{MC-ADS}, each sampled permutation is shared across all
units in \(G_q\).
For every $u\in G_q$, the \texttt{SA-MC-ADS} estimator is
\begin{equation}
\begin{split}
\widehat{\phi}_{\mathrm{SA}}^\sigma(u;v,\mathcal U)
&:=
\frac{1}{R_{\mathrm L}}
\sum_{\ell=1}^{R_{\mathrm L}}
\Delta^{\mathrm L}\!\left(
u
\mid
U_{q-1}\cup
\pre_{\widetilde\pi_q^{(\ell)}}(u)
\right)\\
&\quad+
\frac{1}{R_{\mathrm H}}
\sum_{r=1}^{R_{\mathrm H}}
\Bigg[
\Delta\!\left(
u
\mid
U_{q-1}\cup
\pre_{\pi_q^{(r)}}(u)
\right)
-
\Delta^{\mathrm L}\!\left(
u
\mid
U_{q-1}\cup
\pre_{\pi_q^{(r)}}(u)
\right)
\Bigg].
\end{split}
\label{eq:sa-mcads-estimator-main}
\end{equation}
The first term estimates the control mean, that is, the expected low-fidelity marginal contribution, and the second term estimates the expected residual between the high- and low-fidelity marginal contributions.
When the control mean can be computed exactly, it replaces the first sample average (see Appendix~\ref{app:sa-mcads-algorithm} for this exact-control variant of \texttt{SA-MC-ADS}).

\subsubsection{Statistical and Accounting Properties}
\label{subsubsec:sa-mcads-properties-main}

\begin{theorem}[Finite-sample properties of \texttt{SA-MC-ADS}]
\label{thm:sa-mcads-main}

Consider the \texttt{SA-MC-ADS} estimator defined in
Eq.~\eqref{eq:sa-mcads-estimator-main} under the sampling scheme described
above. 

\textnormal{(i) Approximation guarantee.}
Suppose that, for every non-singleton valuation group \(G_q\) and every
\(u\in G_q\), the low-fidelity marginal contributions
\(
\{
\Delta^{\mathrm L}\!\left(
u
\mid
U_{q-1}\cup\pre_{\pi_q}(u)
\right)
:\allowbreak
\pi_q\in\Pi(G_q)
\}
\)
lie in an interval of length at most \(\rho_{\mathrm L}\), and the
residual marginal contributions
\[
\left\{
\Delta\!\left(
u
\mid
U_{q-1}\cup\pre_{\pi_q}(u)
\right)
-
\Delta^{\mathrm L}\!\left(
u
\mid
U_{q-1}\cup\pre_{\pi_q}(u)
\right)
:
\pi_q\in\Pi(G_q)
\right\}
\]
lie in an interval of length at most \(\rho_{\mathrm{res}}\).
Then $\hat\phi^\sigma_{\mathrm{SA}}$ is an $(\epsilon,\delta)$-approximation whenever
\begin{equation}\label{eq:sa-mcads-accuracy-main}
\frac{\rho_{\mathrm L}^2}{R_{\mathrm L}}+\frac{\rho_{\mathrm{res}}^2}{R_{\mathrm H}}\le\frac{2\epsilon^2}{\log(2N/\delta)}. 
\end{equation}
   
\textnormal{(ii) Unbiasedness.}
For every \(u\in\mathcal U\),
\(
\mathbb E\!\left[
\widehat{\phi}_{\mathrm{SA}}^\sigma(u;v,\mathcal U)
\right]
=
\ads{u}{\mathcal U}.
\)

\textnormal{(iii) Finite-sample group efficiency.}
When every low-only and paired permutation is evaluated as a complete within-group path, and each path assigns its successive utility increments to the corresponding units in $G_q$, for every non-singleton valuation group \(G_q\),
\begin{equation}
\label{eq:sa-mcads-group-efficiency-main}
\sum_{u\in G_q}
\widehat{\phi}_{\mathrm{SA}}^\sigma(u;v,\mathcal U)
=
v(U_q)-v(U_{q-1})
\end{equation}
for every finite \(R_{\mathrm L}\) and \(R_{\mathrm H}\).
Singleton groups are evaluated exactly and satisfy the same identity.
\end{theorem}

Part (i) gives a joint accuracy requirement on the numbers of low-only and paired permutation paths.
Part~(ii) shows that surrogate assistance introduces no systematic bias
relative to the exact \texttt{ADS} allocation.
Part (iii) shows that complete-path sampling preserves the exact group total in Proposition~\ref{propo:efficiency_asv_intra_group} regardless of surrogate quality, so the surrogate affects only estimation variance, not group-level accounting.
The proof is provided in Appendix~\ref{app:sa-mcads-proof}.

\begin{remark}[Surrogate quality and computational gain]
\label{remark:sa-mcads-variance}
Under the sampling scheme described above, for $u\in G_q$ and $\pi_q\sim\mathrm{Unif}(\Pi(G_q))$,
\begin{align}
\operatorname{Var}\!\left[
\widehat{\phi}_{\mathrm{SA}}^\sigma(u;v,\mathcal U)
\right]
&=
\frac{1}{R_{\mathrm L}}
\operatorname{Var}\!\left[
\Delta^{\mathrm L}\!\left(
u
\mid
U_{q-1}\cup\pre_{\pi_q}(u)
\right)
\right]
\nonumber\\
&\quad+
\frac{1}{R_{\mathrm H}}
\operatorname{Var}\!\left[
\Delta\!\left(
u
\mid
U_{q-1}\cup\pre_{\pi_q}(u)
\right)
-
\Delta^{\mathrm L}\!\left(
u
\mid
U_{q-1}\cup\pre_{\pi_q}(u)
\right)
\right].
\label{eq:sa-mcads-variance-main}
\end{align}
For comparison, ordinary \texttt{MC-ADS} based on
\(R_{\mathrm H}\) high-fidelity utility paths has unit-level variance
\(
\frac{1}{R_{\mathrm H}}
\operatorname{Var}\!\left[
\Delta\!\left(
u
\mid
U_{q-1}\cup\pre_{\pi_q}(u)
\right)
\right].
\)
Thus, the surrogate is useful when low-fidelity evaluations are inexpensive and the residual marginal contributions are substantially less variable than the high-fidelity marginal contributions.
\end{remark}

When low-fidelity evaluations are inexpensive, \(R_{\mathrm L}\) can be large enough to make the first variance term small.
The number of costly high-fidelity evaluations then depends on the extent to which low-fidelity contributions deviate from their high-fidelity counterparts.
A small pilot sample of paired paths can be used to estimate the residual variation and relative evaluation costs before choosing \(R_{\mathrm L}\) and \(R_{\mathrm H}\).

%% file: S6_Applications.tex
This section examines how \texttt{ADS}, our structure-aware value attribution method, affects value allocation and downstream organizational decisions.
First, we examine how conditioning attribution on documented production dependencies changes value allocation across heterogeneous contributions in structured AI pipelines, and compare \texttt{ADS} with structure-agnostic \texttt{DS}.
Second, we investigate the decision consequences of structure-aware valuation by examining sequential dataset procurement during multi-stage LLM fine-tuning and selection of derivative data under provenance dependence, comparing \texttt{ADS} with alternative valuation and selection rules.

\subsection{Value Allocation under Production Dependencies}
\label{subsec:value_allocation_dependencies}

We first examine whether conditioning valuation on documented production dependencies changes how model utility is allocated across pipeline contributions.
In particular, we study how \texttt{ADS} and structure-agnostic \texttt{DS} differ in attributing value across upstream and downstream contributions when provenance and input--operation relationships constrain the relevant predecessor contexts.
Within each market configuration, the two rules share the same valuation units, learner, and utility function and differ only in the predecessor contexts they admit.
Experiment~\ref{subsubsec:exp1a_replication_augmentation} isolates provenance dependence by comparing downstream derivative data that exactly duplicate the raw data with those that introduce new information.
Experiment~\ref{subsubsec:exp1b_broker_pipeline} then adds a data-related service to the pipeline, so that provenance and input--operation dependencies operate jointly, and uses reduced markets to separate their contributions to the allocation differences.

\subsubsection{Exact Duplication and Informative Augmentation}
\label{subsubsec:exp1a_replication_augmentation}

This experiment extends Example~\ref{ex:MD} to a controlled comparison between redundant replication and informative data transformation. We consider two matched scenarios in which downstream derivative data enter at the same documented positions. 
In the exact duplication scenario, the downstream derivative data are exact copies of the raw data and add no new information. In the augmentation scenario, the derivative data introduce new information that improves model utility.
This design tests whether structure-aware valuation can distinguish redundant replication from informative transformation.

\paragraph{Design.}
We independently generate 100 synthetic binary-classification markets from a common data-generating process with different random seeds. Each valuation unit is a data bundle containing a similar number of training observations. Each market begins with one raw-data group of three data units. The duplication and augmentation scenarios share the raw-data group, the deterministic learner, and the held-out evaluation sample. 
Starting from this common configuration, the duplication scenario sequentially adds two groups of exact copies: the first copies the raw-data group, and the second copies the first.
The augmentation scenario instead sequentially adds a minority-targeted group, which oversamples the underrepresented class, and a boundary-targeted group, which is generated from the raw-data and minority-targeted groups and concentrates data examples near the decision boundary.
In both scenarios, the second derivative group is thus produced from the first. \texttt{ADS} treats the raw-data group and the two derivative groups as three ordered valuation groups in their order of production, whereas \texttt{DS} treats all data units in a configuration as a single peer group.

The learner minimizes an averaged $\ell_2$-regularized logistic loss, under which uniform duplication leaves the fitted model and hence model utility unchanged. The utility of each coalition is the balanced accuracy on the held-out sample. For each market, we sum the allocations of the three data units within each group to obtain the group-level total. We then average these totals across the 100 generated markets and report the resulting average group-level allocations.

\begin{figure}[!htbp]
    \centering

    \makebox[\textwidth][c]{%
        \begin{subfigure}[t]{0.19\textwidth}
            \centering
            \includegraphics[width=\linewidth]
            {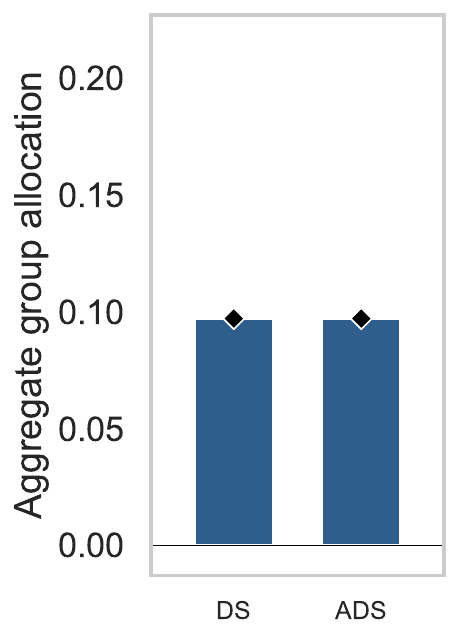}
            \caption{Original}
            \label{fig:exp1a_rep_original}
        \end{subfigure}\hspace{0.005\textwidth}%
        \begin{subfigure}[t]{0.19\textwidth}
            \centering
            \includegraphics[width=\linewidth]
            {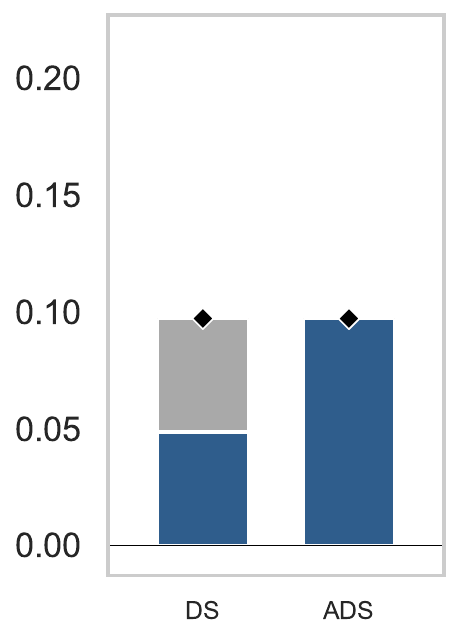}
            \caption{Duplicated once}
            \label{fig:exp1a_rep_copied_once}
        \end{subfigure}\hspace{0.005\textwidth}%
        \begin{subfigure}[t]{0.19\textwidth}
            \centering
            \includegraphics[width=\linewidth]
            {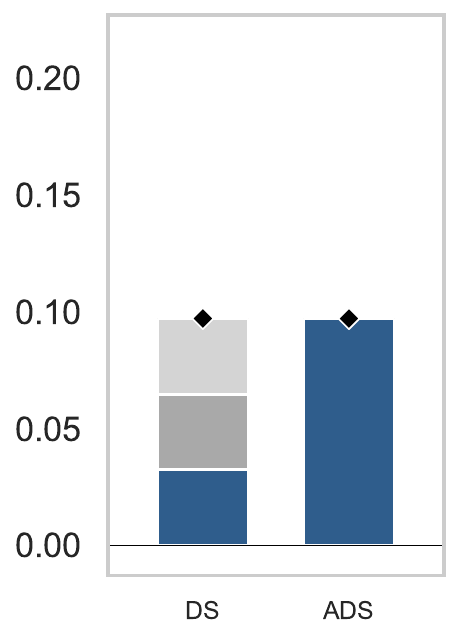}
            \caption{Duplicated twice}
            \label{fig:exp1a_rep_copied_twice}
        \end{subfigure}\hspace{0.005\textwidth}%
        \begin{subfigure}[t]{0.19\textwidth}
            \centering
            \includegraphics[width=\linewidth]
            {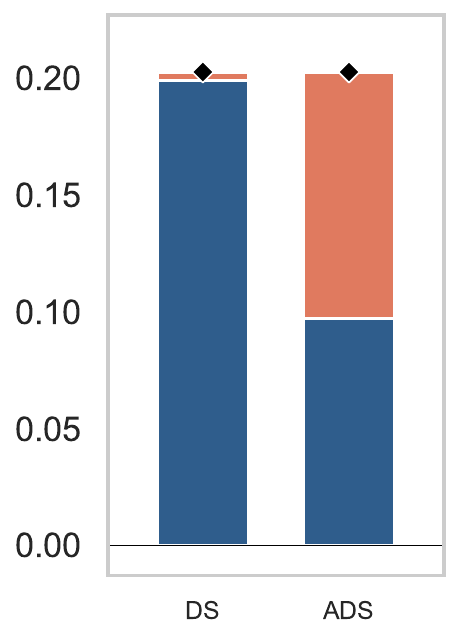}
            \caption{Augmented once}
            \label{fig:exp1a_augmented_once}
        \end{subfigure}\hspace{0.005\textwidth}%
        \begin{subfigure}[t]{0.19\textwidth}
            \centering
            \includegraphics[width=\linewidth]
            {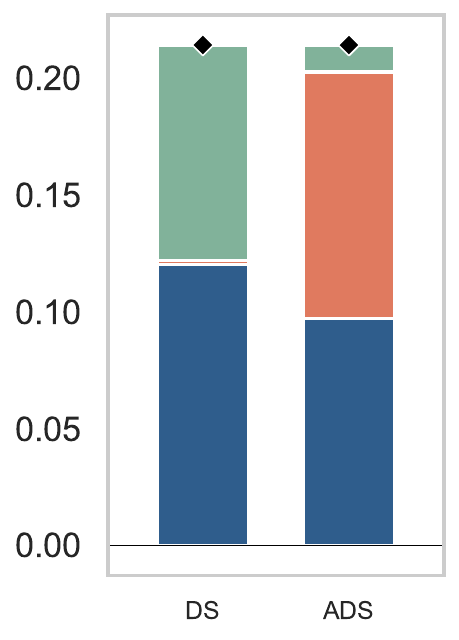}
            \caption{Augmented twice}
            \label{fig:exp1a_augmented_twice}
        \end{subfigure}%
    }

    \vspace{0.5em}
    \includegraphics[width=0.98\textwidth]
    {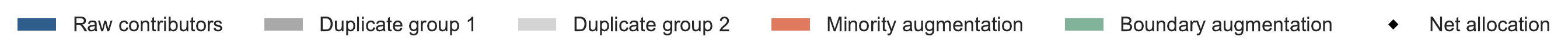}

    \caption{
    Group-level allocations under exact duplication and informative augmentation. Panel~(a) shows the common baseline containing only the raw-data group. 
    Panels~(b)--(c) show the duplication scenario after introducing the first and second duplicate groups, respectively. Panels~(d)--(e) show the corresponding augmentation scenario after introducing the minority-targeted and boundary-targeted groups. 
    Bars report average group-level allocations across 100 independently generated markets, measured in balanced accuracy above $v(\varnothing)=0.5$. Diamonds mark the net allocation, equal to total model utility above $v(\varnothing)$.
    }
    \label{fig:exp1a_exact_allocation}
\end{figure}

\paragraph{Results.}
Figure~\ref{fig:exp1a_exact_allocation} reports the average group-level allocations across the five configurations. In the duplication scenario, adding either duplicate group leaves total model utility unchanged. \texttt{ADS} therefore assigns zero group-level value to each duplicate group and preserves the allocation to the raw-data group. In contrast, \texttt{DS} redistributes the same total utility across the separately indexed but informationally identical data units. Relative to the baseline in Panel~(a), the raw-data group's average allocation falls to one-half after the first duplicate group is introduced in Panel~(b) and to one-third after the second is introduced in Panel~(c). Thus, \texttt{DS} rewards redundant copies at the expense of the original raw-data group.

The augmentation scenario exhibits a different pattern. In Panel~(d), introducing the minority-targeted group increases model utility on average, and \texttt{ADS} assigns it a positive average allocation matching its incremental contribution. However, \texttt{DS} assigns this group almost no value and instead credits most of the gain to the raw-data group. In Panel~(e), after the boundary-targeted group is introduced, model utility increases again and \texttt{ADS} assigns positive average value to both augmentation groups. Under \texttt{DS}, the minority-targeted group's average allocation remains close to zero, while the raw-data and boundary-targeted groups receive allocations exceeding the increments they add at their documented positions.

The above results show that the predecessor contexts admitted by the valuation rule materially affect group-level attribution. \texttt{DS} can both dilute original contributors' compensation by rewarding redundant copies and misattribute the value created by informative transformations. If allocations determine payments, such distortions can favor low-cost duplication and weaken incentives to invest in high-value augmentation.
In contrast, by conditioning attribution on the documented production structure, \texttt{ADS} leaves the raw-data group's allocation unaffected by downstream additions and assigns informative derivative groups their incremental contributions at the positions where they enter the pipeline. It therefore protects original contributors from copy-based dilution while rewarding derivative-data providers for useful data transformations, a property that is particularly important in copyright-sensitive markets.

\subsubsection{Heterogeneous Pipeline with Service}
\label{subsubsec:exp1b_broker_pipeline}

This experiment extends Experiment~\ref{subsubsec:exp1a_replication_augmentation} to a heterogeneous pipeline containing both data units and a data-related service.
The pipeline combines two raw-data units, a deduplication service, and a derivative-data unit, thereby introducing provenance (through the derivative-data unit) and input--operation (through the service) dependencies within the same production process.
We use this setting to examine how these two forms of dependence affect value allocation when heterogeneous contributions are valued jointly.

\paragraph{Design.}
We construct 100 independently resampled MNIST digit-classification markets. In each market, two contributors each supply one raw-data unit, and some images appear in both units as separately indexed examples, a subset of them with conflicting labels. The deduplication service pools the two raw-data units and retains a single example per detected duplicate image. The derivative-data unit is a separately indexed exact copy of the post-deduplication dataset and therefore adds no new information. The documented production order is thus the two raw-data units, which form a single group, followed by the deduplication service unit and then the redundant derivative-data unit. We use the same learner and coalition utility as in Experiment~\ref{subsubsec:exp1a_replication_augmentation}.

We compare exact \texttt{ADS} with exact \texttt{DS}. 
For \texttt{DS}, each full-market coalition is evaluated using its included raw and derivative records. If the service is present, deduplication is applied to the included raw records before training. This also defines utilities for coalitions outside the documented pipeline. 
To diagnose the resulting allocation differences, we also value three reduced markets that retain at most one form of dependence: (i) a raw-only placebo containing only the two raw-data units; (ii) a provenance diagnostic that adds the redundant derivative-data unit, with deduplication applied but not valued; and (iii) an input--operation diagnostic that instead adds the deduplication service. Because these markets contain different valuation units, they serve as diagnostics rather than an additive decomposition of the \texttt{ADS}--\texttt{DS} difference in the full market.

\begin{figure}[!htbp]
    \centering
    \captionsetup[subfigure]{font=footnotesize}
    \setlength{\panelht}{0.28\textwidth}%
    \makebox[\textwidth][c]{%
        \begin{subfigure}[t]{0.27\textwidth}
            \centering
            \includegraphics[height=\panelht]
            {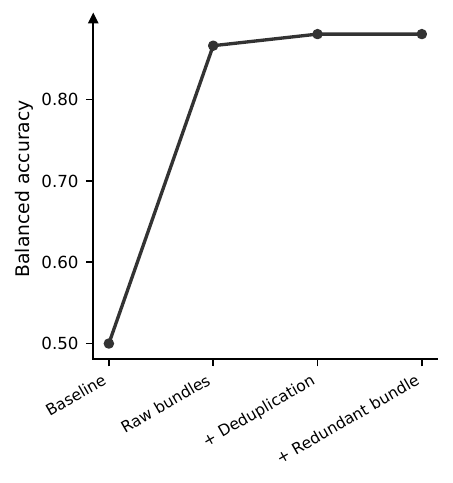}
            \caption{Pipeline utility}
            \label{fig:exp1b_panel_utility}
        \end{subfigure}\hspace{0.005\textwidth}%
        \begin{subfigure}[t]{0.19\textwidth}
            \centering
            \includegraphics[height=\panelht]
            {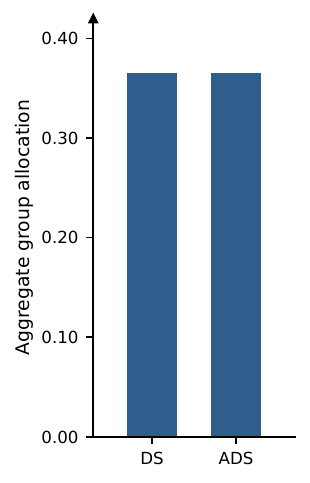}
            \caption{Placebo}
            \label{fig:exp1b_panel_placebo}
        \end{subfigure}\hspace{0.005\textwidth}%
        \begin{subfigure}[t]{0.17\textwidth}
            \centering
            \includegraphics[height=\panelht]
            {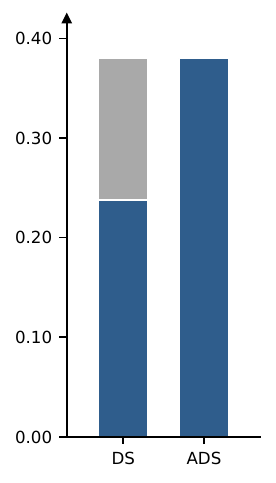}
            \caption{Provenance}
            \label{fig:exp1b_panel_provenance}
        \end{subfigure}\hspace{0.005\textwidth}%
        \begin{subfigure}[t]{0.17\textwidth}
            \centering
            \includegraphics[height=\panelht]
            {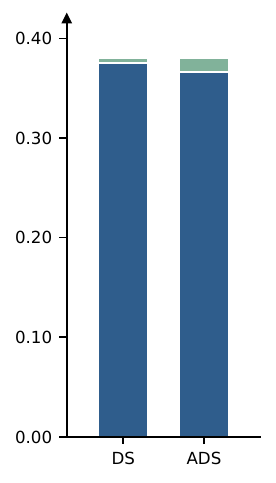}
            \caption{Input--op.}
            \label{fig:exp1b_panel_input_operation}
        \end{subfigure}\hspace{0.005\textwidth}%
        \begin{subfigure}[t]{0.17\textwidth}
            \centering
            \includegraphics[height=\panelht]
            {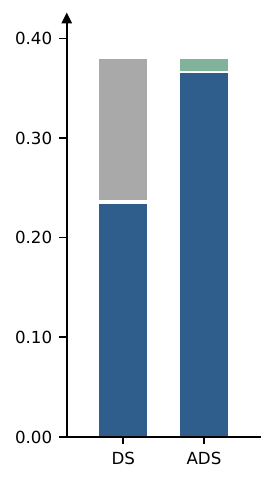}
            \caption{Full pipeline}
            \label{fig:exp1b_panel_full}
        \end{subfigure}%
    }
    \vspace{0.5em}
    \includegraphics[width=0.98\textwidth]
    {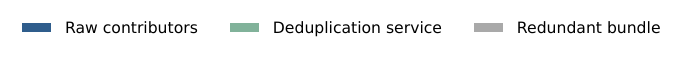}
    \caption{Value allocation in a heterogeneous pipeline. Panel~(a) traces average balanced accuracy along the documented production order: the two raw-data units, followed by the deduplication-service unit and the redundant derivative-data unit. Panels~(b)--(e) report average group-level allocations, measured in balanced accuracy above $v(\varnothing)=0.5$, under exact \texttt{DS} (left bar) and exact \texttt{ADS} (right bar) for the raw-only placebo, provenance diagnostic, input--operation diagnostic, and full pipeline, respectively. Allocations are not directly comparable across panels because the four markets contain different sets of valuation units. All reported quantities are averages across 100 independently resampled markets.}
    \label{fig:exp1b_allocation}
\end{figure}

\paragraph{Results.}
Figure~\ref{fig:exp1b_allocation} reports pipeline utility and the resulting group-level allocations. Panel~(a) traces model utility along the documented production order. The two raw-data units create most of the final utility, the deduplication service provides a further positive increment by reconciling overlapping and conflicting examples, and the redundant derivative-data unit adds no additional utility. 
Panel~(e) shows the allocations in the full heterogeneous pipeline. Under \texttt{ADS}, the raw-data group receives the utility created by its units, the service group receives the increment created by deduplication, and the redundant derivative group receives zero. \texttt{DS} produces a substantially different allocation. It assigns the redundant derivative group approximately 0.14 of the 0.38 total allocated utility, while reducing the allocation to the raw-data group by roughly one third and leaving the service group with almost no value.
Thus, a derivative group that adds no incremental utility receives substantial value under structure-agnostic attribution, while value is shifted away from the upstream data and service groups.

Panels~(b)--(d) diagnose the sources of this reallocation. In the raw-only placebo in Panel~(b), where neither provenance nor input--operation dependence is present among the valued units, \texttt{ADS} and \texttt{DS} produce the same group-level allocations. 
Panel~(c), which isolates provenance dependence, reproduces most of the full-pipeline reallocation: under \texttt{DS}, the redundant derivative-data unit receives substantial value despite adding no information beyond its upstream data.
In Panel~(d), \texttt{DS} assigns less value to the service unit than \texttt{ADS} and correspondingly more to the raw-data group. Deduplication creates value only when both raw-data units are present. However, \texttt{DS} also evaluates the service in predecessor contexts that lack one of these units, where there is nothing to deduplicate, and thus attributes part of the service's contribution to the raw data.
Finally, because \texttt{DS} depends on utilities outside $\mathcal F_\sigma$, its allocation also depends on how these utilities are defined.
Appendix~\ref{app:completion_sensitivity} shows that the \texttt{ADS} allocation is invariant across alternative completions of utilities outside $\mathcal F_\sigma$, whereas the \texttt{DS} allocation changes materially.

\subsection{Procurement and Selection Decisions under Production Dependencies}
\label{subsec:procurement_deicisions}

Section~\ref{subsec:value_allocation_dependencies} shows that conditioning value attribution on documented production structure can substantially change how model utility is allocated across contributions. 
We now examine whether these allocation differences lead organizations to make different data decisions and, ultimately, attain different levels of model performance. Experiment~\ref{subsubsec:llm_experiment} studies sequential dataset procurement, in which each purchase updates the model checkpoint used to evaluate subsequent candidate datasets, and Experiment~\ref{subsubsec:derivative_procurement} studies derivative-data selection, in which candidate derivative datasets depend on their upstream source data examples.

\subsubsection{Dataset Procurement in Multi-Stage LLM Fine-Tuning}\label{subsubsec:llm_experiment}

This experiment examines sequential dataset procurement in a multi-stage LLM fine-tuning pipeline, in which an organization repeatedly selects among candidate datasets to fine-tune an evolving LLM. 
Because each purchase updates the checkpoint inherited by the next round, the value of a candidate dataset depends on the model state created by earlier procurement decisions. We therefore examine whether \texttt{ADS} changes the sequence of datasets selected and the model performance ultimately realized, relative to leave-one-out (\texttt{LOO}) \citep{cook1977detection,ghorbani2019data} and random selection (\texttt{Random}).

\paragraph{Design.}
We consider a market-research setting in which an organization sequentially procures product-rating data over 30 rounds to fine-tune an LLM that serves as a synthetic respondent in consumer preference surveys.
In each round, three providers offer candidate datasets of 50 labeled instances each, and the budget allows the purchase of only one.
The selected dataset is used to fine-tune the incoming checkpoint, and the resulting checkpoint is carried forward to the next round.
Products are described by three categorical attributes (color, size, and material), each with four levels.
Target ratings are generated from an additive preference model of the standard conjoint form, in which each rating equals the sum of the corresponding attribute-level part-worths plus random noise.
To create systematic variation in dataset quality, we draw a separate set of off-target part-worths. In the three candidate datasets, 30\%, 60\%, and 90\% of the ratings, respectively, are generated from this off-target model, while the remaining ratings follow the target model.

We compare \texttt{ADS} with \texttt{LOO} and \texttt{Random}.
In each round, the three candidate datasets form a single valuation group, and each dataset is treated as one valuation unit.
\texttt{ADS} values a candidate by averaging its marginal contribution across all subsets of the other two candidates at the current model checkpoint. \texttt{LOO} instead measures the utility loss from removing that candidate from the full set of three candidates.
\texttt{ADS} and \texttt{LOO} purchase the candidate with the highest estimated value, whereas \texttt{Random} selects one of the three candidates uniformly at random.
We do not include \texttt{DS} as a baseline because applying it jointly across all candidate datasets over 30 rounds would require evaluating an exponential number of coalition-dependent fine-tuning paths, most of which correspond to model states that are never realized by the sequential procurement process.
All policies share the same candidate datasets and start from the same initial checkpoint, but their checkpoints subsequently diverge because of their different procurement decisions.

Procurement utility is the negative mean squared error of predicted ratings on a fixed validation set generated from the target preference model. For ex post evaluation, we obtain five LLM ratings for each of the 64 product profiles using separate decoding seeds and fit a conjoint model. \(\mathrm{AvgErr}\) averages the absolute errors of its 12 attribute-level part-worth estimates. Ground-truth part-worths and ex post results are unavailable to procurement policies. 
We conduct the experiment with Llama~3.2-3B, Llama~3.1-8B, Qwen 3-0.6B, and Qwen 3-8B and report results over 15 independent runs.

\begin{figure}[!htbp]
    \centering
\begin{subfigure}[b]{0.245\textwidth}
    \centering
    \includegraphics[width=\textwidth]{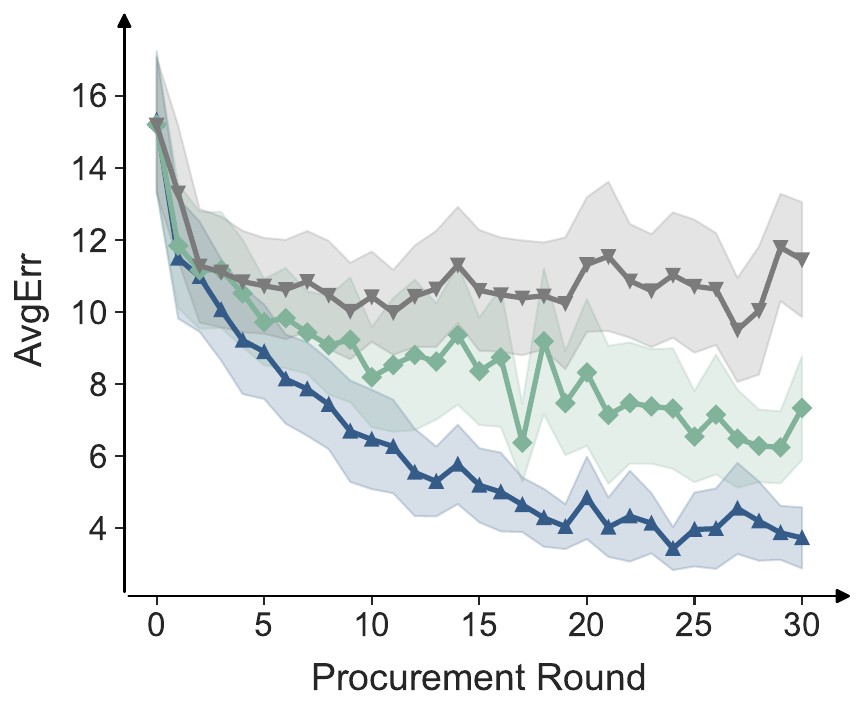}
    \caption{Llama 3.2-3B}
    \label{fig:sub1}
\end{subfigure}\hfill%
    \begin{subfigure}[b]{0.245\textwidth}
        \centering
        \includegraphics[width=\textwidth]{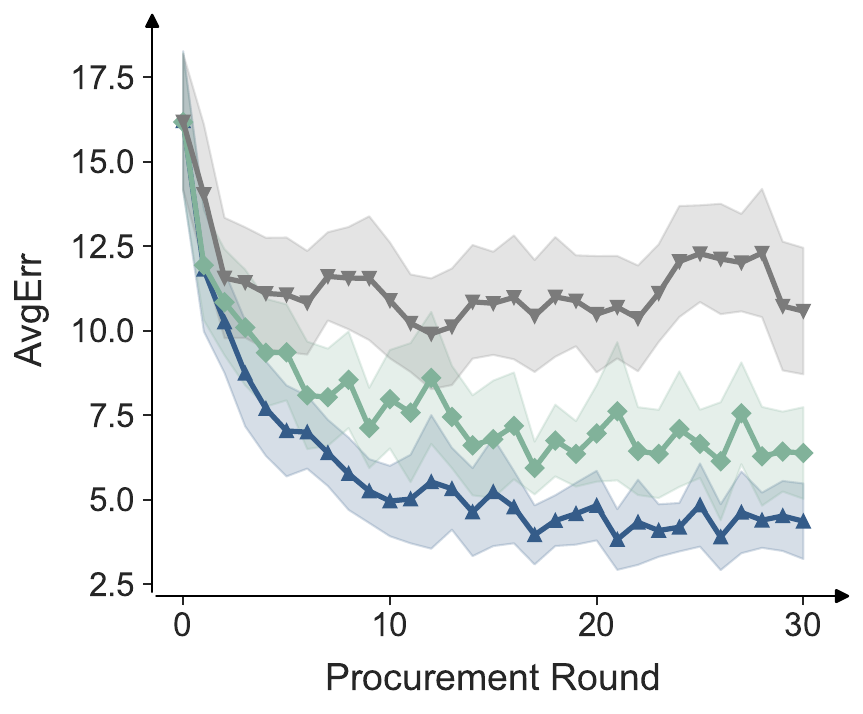}
        \caption{Llama 3.1-8B}
        \label{fig:sub2}
    \end{subfigure}\hfill%
    \begin{subfigure}[b]{0.245\textwidth}
        \centering
        \includegraphics[width=\textwidth]{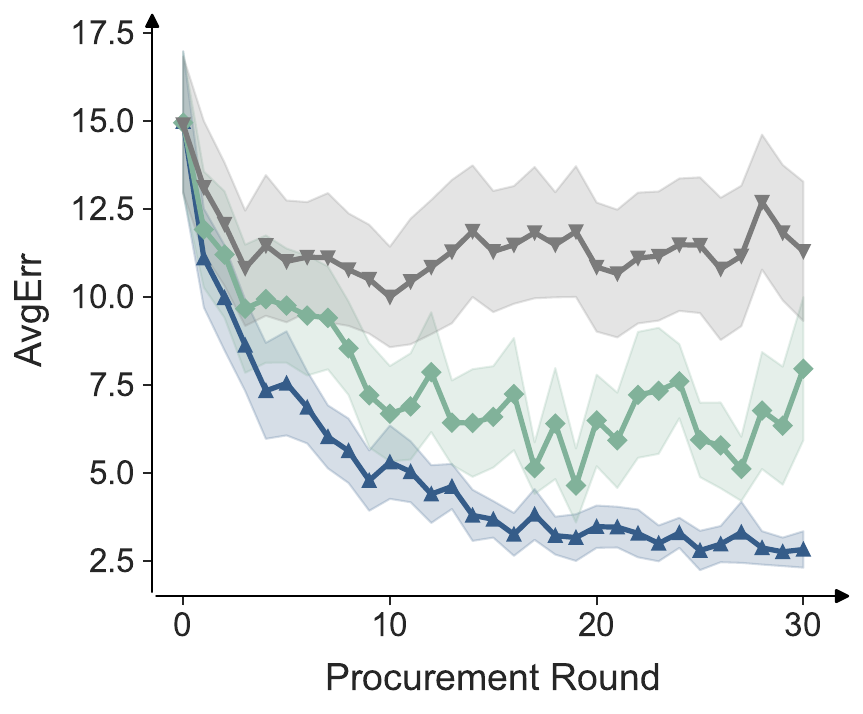}
        \caption{Qwen 3-0.6B}
        \label{fig:sub3}
    \end{subfigure}\hfill%
    \begin{subfigure}[b]{0.245\textwidth}
        \centering
        \includegraphics[width=\textwidth]{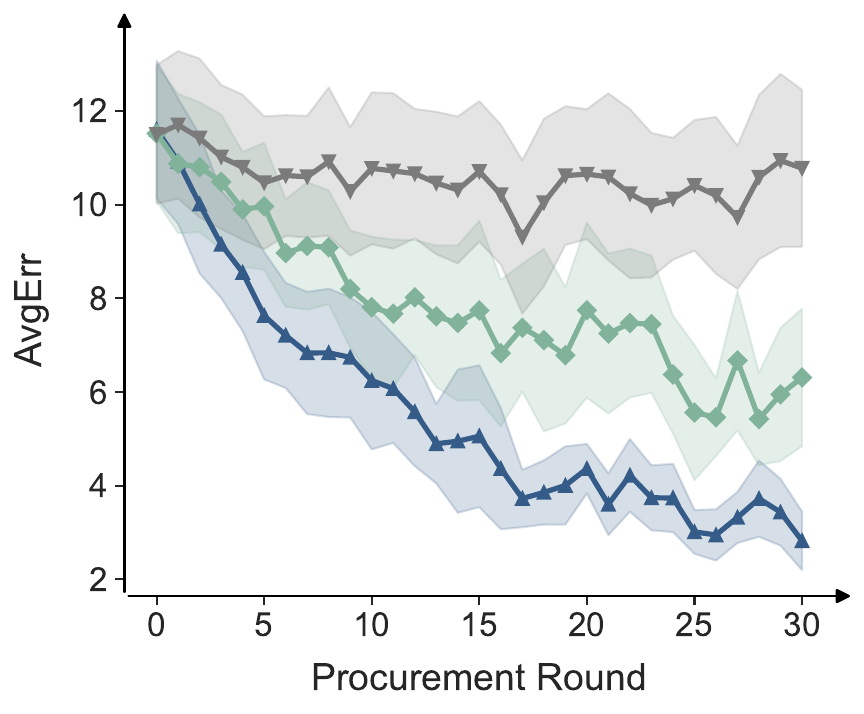}
        \caption{Qwen 3-8B}
        \label{fig:sub4}
    \end{subfigure}
    \par\vspace{4pt}
    \includegraphics[width=0.3\textwidth]{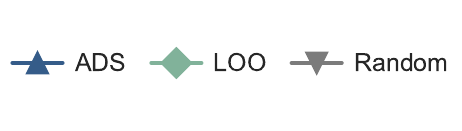}
    \caption{Dataset procurement in multi-stage fine-tuning of four base LLMs: average part-worth estimation error (\(\mathrm{AvgErr}\)) by procurement round under each policy. Results are averaged over 15 runs with 90\% confidence intervals (based on Student's \(t\)-distribution). } %
    \label{fig:llm_experiment}
\end{figure}

\begin{figure}[!htbp]
    \centering
    \begin{subfigure}{0.44\textwidth}
        \centering
        \includegraphics[width=\textwidth]{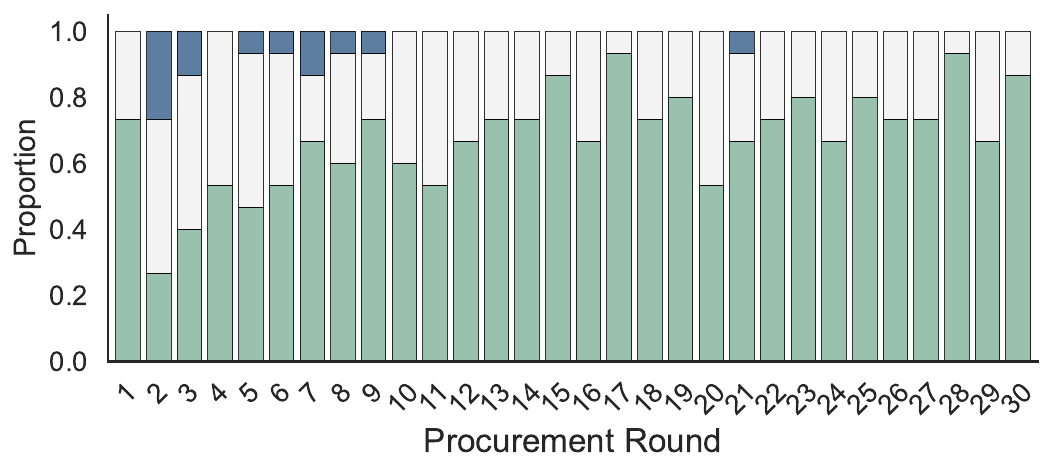}
        \caption{\texttt{LOO} for Llama 3.1-8B}
    \end{subfigure}
    \hfill
    \begin{subfigure}[b]{0.44\textwidth}
        \centering
        \includegraphics[width=\textwidth]{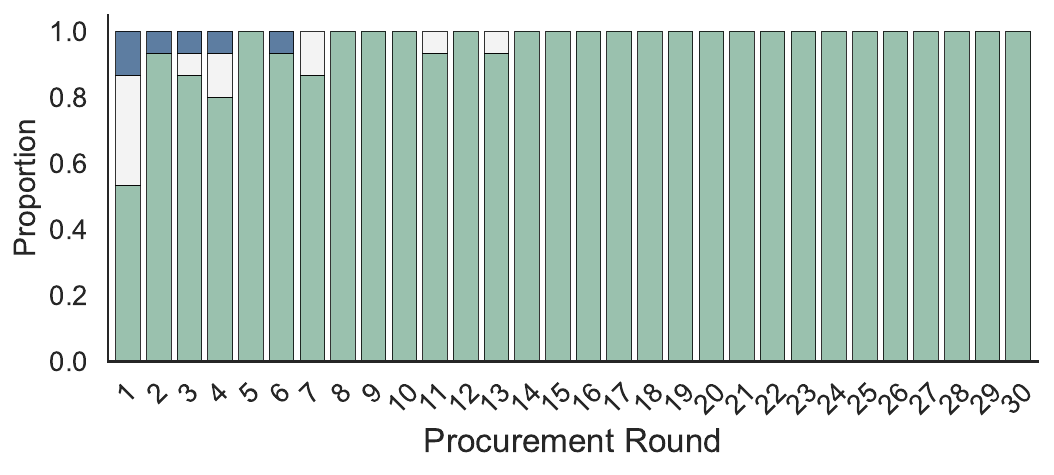}
        \caption{\texttt{ADS} for Llama 3.1-8B}
    \end{subfigure}
    
    \begin{subfigure}{0.8\textwidth}
        \centering
        \vspace{-0.2cm}
    \includegraphics[width=\textwidth]{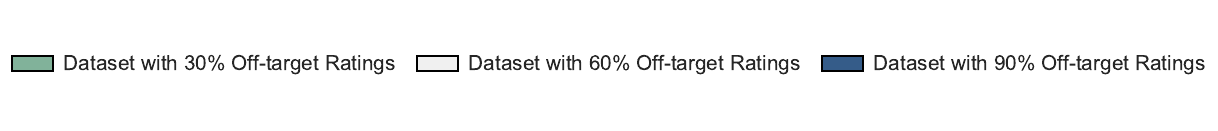}
        \caption*{}
    \end{subfigure}
    \vspace{-1.0cm}
    \caption{Share of runs selecting each candidate dataset by procurement round under \texttt{LOO} and \texttt{ADS} for Llama~3.1-8B, across 15 runs. }
    \label{fig:selection_llm_experiment}
\end{figure}

\paragraph{Results.}

Figure~\ref{fig:llm_experiment} reports the evolution of \(\mathrm{AvgErr}\) over the 30 procurement rounds.
Across all four base LLMs, \texttt{ADS} reduces part-worth estimation error more rapidly and reaches the lowest error by the end of the procurement horizon.
For Llama~3.1-8B, Figure~\ref{fig:selection_llm_experiment} shows that \texttt{ADS} selects the 30\% off-target candidate more frequently and the 90\% off-target candidate less frequently than \texttt{LOO}. This overall pattern is observed in all four base models (Appendix~\ref{app:llm_selection_additional}).

Within each round, \texttt{ADS} averages each candidate's marginal contribution over all subsets of its peers, whereas \texttt{LOO} considers only the context in which the other two candidates are present. 
Because the candidate datasets provide partially redundant information about the same underlying preference structure, the utility loss from removing one dataset while retaining the other two may be small, causing \texttt{LOO} to compress differences in their estimated values. %
Model-state dependence then amplifies these within-round differences. 
A poorer procurement decision affects not only the current round but also the checkpoint inherited by subsequent rounds. Early differences in dataset selection can therefore persist and accumulate over time.
In a path-dependent pipeline, the valuation rule therefore shapes not a single purchase but the whole trajectory of model states the organization ends up with.

\subsubsection{Selecting Derivative Data under Provenance Dependence}
\label{subsubsec:derivative_procurement}

This experiment considers an organization that already owns a raw dataset and selects additional training examples from derivative data generated from that dataset. 
The usefulness of each derivative example depends on the additional utility it provides alongside the raw data.
Because this experiment treats each derivative example as a separate valuation unit, exact valuation is prohibitively costly.
We therefore use \texttt{MC-ADS} and a Monte Carlo estimator of \texttt{DS} (\texttt{MC-DS}) to estimate the corresponding allocations and examine whether structure-aware valuation remains effective in guiding data selection under provenance dependence when implemented through Monte Carlo estimation.

\begin{figure}[!htbp]
\centering
\captionsetup[subfigure]{font=footnotesize,labelfont=bf,justification=centering}
\begin{subfigure}[t]{0.245\textwidth}
\centering
\includegraphics[width=\linewidth]{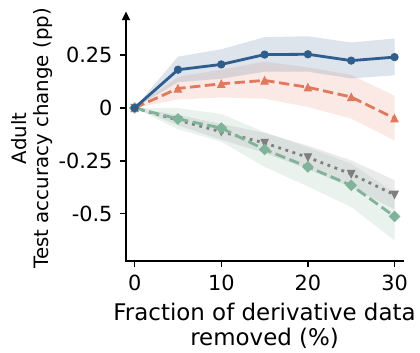}
\par\vspace{0.65em}
\includegraphics[width=\linewidth]{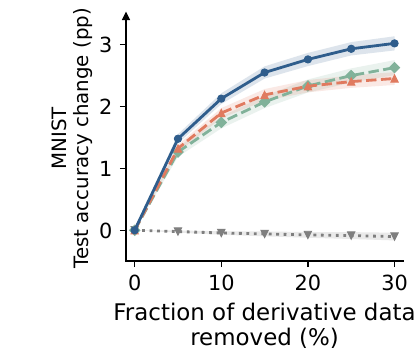}
\par\vspace{0.65em}
\includegraphics[width=\linewidth]{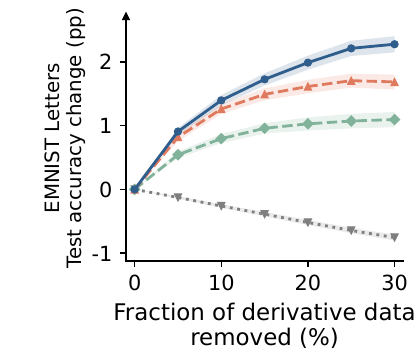}
\caption{Remove low-value data}
\label{fig:derivative_remove_low}
\end{subfigure}\hfill%
\begin{subfigure}[t]{0.245\textwidth}
\centering
\includegraphics[width=\linewidth]{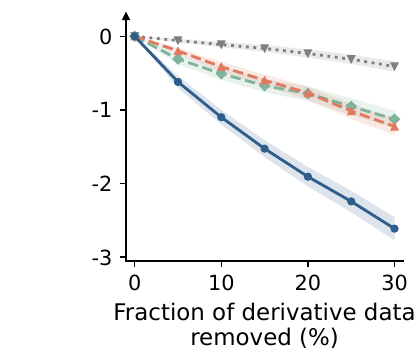}
\par\vspace{0.65em}
\includegraphics[width=\linewidth]{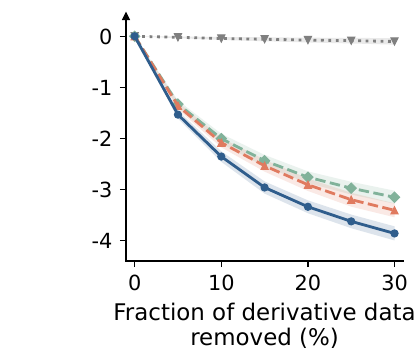}
\par\vspace{0.65em}
\includegraphics[width=\linewidth]{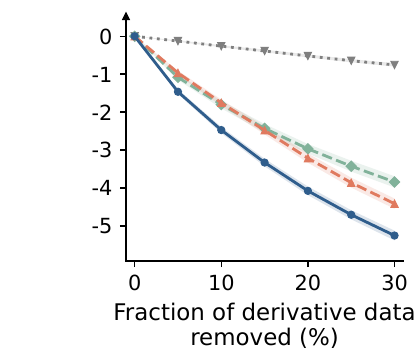}
\caption{Remove high-value data}
\label{fig:derivative_remove_high}
\end{subfigure}\hfill%
\begin{subfigure}[t]{0.245\textwidth}
\centering
\includegraphics[width=\linewidth]{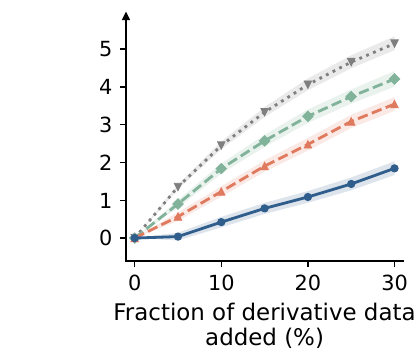}
\par\vspace{0.65em}
\includegraphics[width=\linewidth]{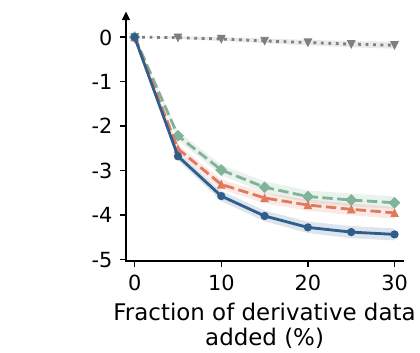}
\par\vspace{0.65em}
\includegraphics[width=\linewidth]{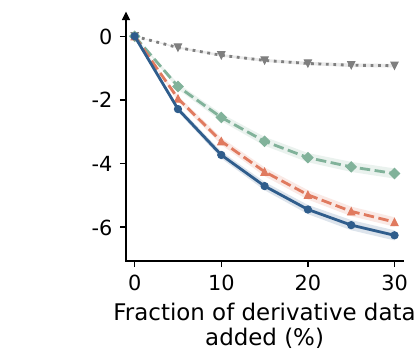}
\caption{Add low-value data}
\label{fig:derivative_add_low}
\end{subfigure}\hfill%
\begin{subfigure}[t]{0.245\textwidth}
\centering
\includegraphics[width=\linewidth]{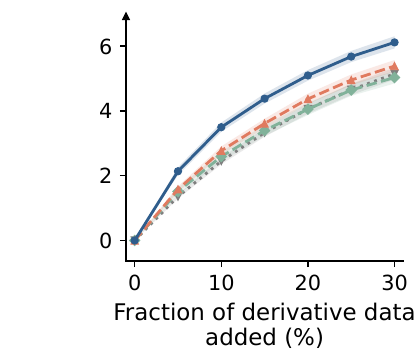}
\par\vspace{0.65em}
\includegraphics[width=\linewidth]{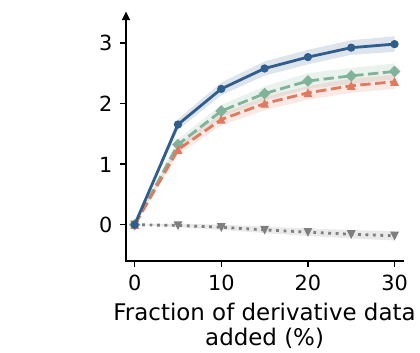}
\par\vspace{0.65em}
\includegraphics[width=\linewidth]{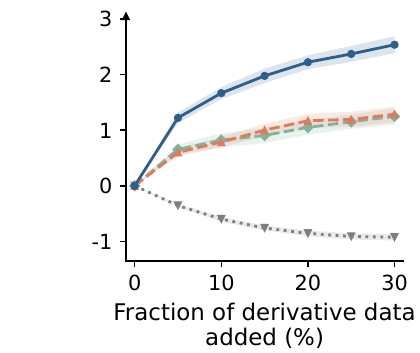}
\caption{Add high-value data}
\label{fig:derivative_add_high}
\end{subfigure}\par

\vspace{0.4em}
\includegraphics[width=0.72\textwidth]{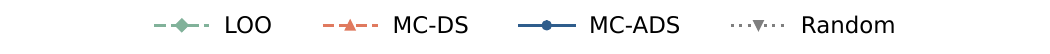}
\caption{
Changes in test classification accuracy under removal and addition of individual derivative examples. Rows correspond to Adult, MNIST, and EMNIST Letters. Columns show the removal of (a) the lowest-ranked and (b) the highest-ranked derivative examples from the complete training set, or the addition of (c) the lowest-ranked and (d) the highest-ranked derivative examples to the organization's owned raw dataset. 
The horizontal axes indicate the percentage of derivative examples removed or added.
The vertical axes report changes in test accuracy, in percentage points, relative to training on all raw and derivative examples for removal and on raw examples alone for addition. 
Curves report averages across 100 randomized markets per dataset, with 95\% bootstrap confidence intervals. For Adult, MNIST, and EMNIST Letters, raw/derivative sample sizes are 1,000/600, 500/500, and 260/260, respectively. Valuation/test sample sizes are 10,000/2,000, 10,000/2,000, and 5,200/5,200. Each Monte Carlo method uses 256, 2,048, and 256 paths per market, respectively. The random benchmark averages 20 orderings per market. Intervals use 10,000 whole-market bootstrap resamples. 
}
\label{fig:derivative_add_remove}
\end{figure}

\paragraph{Design.}
We construct 100 randomized markets for each of Adult~\citep{misc_adult_2}, MNIST~\citep{lecun1998gradient}, and EMNIST Letters~\citep{cohen2017emnist}.
Each market contains a raw dataset already owned by the organization and a pool of candidate derivative examples.
For Adult, derivative examples are generated from the raw data through minority-class oversampling, whereas for MNIST and EMNIST Letters, they are generated through affine image augmentation.
Each derivative example constitutes a separate valuation unit, and all owned raw examples remain available throughout the selection experiment.
We compare \texttt{MC-ADS}, \texttt{MC-DS}, \texttt{LOO}, and \texttt{Random} on common market data. 
\texttt{MC-ADS} treats the owned raw examples as a preceding group and the derivative examples as a single peer group. \texttt{MC-DS} ignores this structure, whereas \texttt{LOO} evaluates each derivative example only in the context of all other examples.
We use regularized logistic regression for Adult and EMNIST Letters and uniform 5-neighbor majority voting for MNIST. Utility and selection performance are accuracies on disjoint, class-balanced held-out valuation and test samples, respectively. Figure~\ref{fig:derivative_add_remove} reports sample sizes and path budgets.

We evaluate the resulting rankings through four removal and addition exercises.
Starting from the full dataset containing all raw and derivative data examples, we sequentially remove either the lowest-ranked or the highest-ranked derivative data example from the training set. 
Starting from the raw data examples, we sequentially add either the lowest-ranked or the highest-ranked derivative data example to the training set.
The number removed or added ranges from 0\% to 30\% of all derivative examples.
We report changes in test accuracy, in percentage points, relative to the complete-data baseline for removal and the raw-dataset-only baseline for addition.

\paragraph{Results.}
Figure~\ref{fig:derivative_add_remove} reports changes in test accuracy under the four selection exercises.
Across the three datasets, removing the lowest-ranked derivative data examples in Panel~(a) or adding the highest-ranked derivative data examples in Panel~(d) generally yields better average test performance under \texttt{MC-ADS} than under the other rules.
These results indicate that \texttt{MC-ADS} helps identify which derivative data examples to exclude from the complete dataset and which to add to the organization's existing raw data. 
The reverse operations provide further evidence that \texttt{MC-ADS} distinguishes useful from less useful derivative examples.
Removing the highest-ranked derivative data examples under \texttt{MC-ADS} in Panel~(b) generally causes larger performance losses, whereas adding the lowest-ranked ones in Panel~(c) produces smaller gains or larger declines.
The random benchmark also shows that using more derivative examples for training does not by itself improve prediction, except on Adult, where minority-class oversampling helps on average.
Overall, these patterns show that the selection benefits of structure-aware valuation remain observable when its allocations are estimated through Monte Carlo sampling.

These findings are consistent with the different valuation contexts used by the methods.
\texttt{MC-DS} includes contexts without upstream raw examples, so it assigns value to a derivative example that carries information the organization already possesses.
\texttt{LOO} measures the loss from deleting each derivative while retaining all other examples, so close substitutes among the remaining derivative data examples can mask its usefulness. 
By holding the owned raw data fixed and averaging marginal contributions across subsets of peer derivative data examples, \texttt{MC-ADS} aligns valuation with the context in which the selected examples will be used.
The results suggest that \texttt{MC-ADS} can help organizations select additional data that improve model performance alongside their existing data assets. Appendix~\ref{appC:computational cost} compares \texttt{MC-ADS} with \texttt{SA-MC-ADS} on these markets: the surrogate lowers Monte Carlo error at a fixed number of high-fidelity paths, but in these settings the gain may be offset by the cost of computing the surrogate.

%% file: S7_Extension.tex
The analysis in Sections~\ref{sec:counterintuition}--\ref{sec:experiments} conditions on a single ordered valuation structure \(\sigma\) that is declared before the resulting allocation is observed. 
In practice, the available information about a pipeline may establish some dependencies without uniquely determining its ordered valuation structure.
This section extends \texttt{ADS} to this setting by allowing multiple structures consistent with the available information and characterizing the resulting allocation uncertainty.

Let \(\Sigma\) be the nonempty finite set of ordered valuation structures on \(\mathcal U\) that are consistent with the available pipeline information.
For each \(\sigma\in\Sigma\), let \(\mathcal F_\sigma\) and \(\Pi_\sigma(\mathcal U)\) denote the pipeline-supported utility domain and the set of pipeline-consistent permutations defined in Section~\ref{sec:asymmetric data shapley}.
Define their unions across the admissible structures as
\begin{equation}
\label{eq:union_domains}
\mathcal F_\Sigma:=\bigcup_{\sigma\in\Sigma}\mathcal F_\sigma,\qquad \Pi_\Sigma(\mathcal U):=\bigcup_{\sigma\in\Sigma}\Pi_\sigma(\mathcal U).
\end{equation}
We assume a common utility function $v:\mathcal F_\Sigma\to\mathbb R$, so that any coalition supported by multiple admissible structures has the same utility under each of them.
For each \(\sigma\in\Sigma\), applying Eq.~\eqref{eq:asv_permutation} using \(v\) on \(\mathcal F_\sigma\) yields the corresponding fixed-structure allocation \(\phi^\sigma(u;v,\mathcal U)\).

\paragraph{Expected \texttt{ADS}.}
The set \(\Sigma\) specifies the admissible valuation structures but does not determine their probabilities.
When the available pipeline information provides a basis for specifying a probability distribution over the admissible structures, we use this distribution to average their fixed-structure \texttt{ADS} allocations.
Let \(\mu\in\Delta(\Sigma)\) denote this distribution, where \(\Delta(\Sigma)\) is the probability simplex on \(\Sigma\).
The expected \texttt{ADS} value of unit \(u\) is
\begin{equation}
\label{eq:dads}
\tilde{\phi}^{\mu}(u;v,\mathcal U):=\sum_{\sigma\in\Sigma}\mu(\sigma)\phi^\sigma(u;v,\mathcal U).
\end{equation}
Equivalently, expected \texttt{ADS} is the average marginal contribution when a structure is first drawn according to $\mu$ and a permutation is then drawn uniformly from its pipeline-consistent permutations.
Expected \texttt{ADS} thus averages marginal contributions over the predecessor contexts supported by the admissible structures, with weights induced by \(\mu\).
When \(\mu\) assigns probability one to a single structure, expected \texttt{ADS} reduces to the corresponding fixed-structure allocation.
As a probability-weighted average of fixed-structure \texttt{ADS} allocations, expected \texttt{ADS} preserves the pipeline-conditioned efficiency and therefore satisfies
$\sum_{u\in\mathcal U}\tilde{\phi}^\mu(u;v,\mathcal U)=v(\mathcal U)-v(\varnothing).$

\paragraph{Credal \texttt{ADS}.}
When the available pipeline information is insufficient to specify a single distribution over the admissible structures, we retain all distributions satisfying the specified probability constraints.
Let \(\mathcal M\subseteq\Delta(\Sigma)\) denote the set of distributions consistent with the available information, which we assume to be nonempty, compact, and convex.
If no additional probability constraints are specified, \(\mathcal M=\Delta(\Sigma)\).
Let \(\boldsymbol{\phi}^{\mu}_{\mathrm E}(v,\mathcal U):=\bigl(\tilde{\phi}^{\mu}(u;v,\mathcal U)\bigr)_{u\in\mathcal U}\) denote the expected allocation vector under \(\mu\).
The credal \texttt{ADS} allocation set collects these vectors over all distributions in \(\mathcal M\):
\begin{equation}
\label{eq:credal_ads}
\mathcal V_{\mathcal M}(v,\mathcal U):=\left\{\boldsymbol{\phi}^{\mu}_{\mathrm E}(v,\mathcal U):\mu\in\mathcal M\right\}.
\end{equation}
For each unit \(u\), define the lower and upper values over this set as
\begin{equation}
\label{eq:credal_ads_bounds}
L_{\mathcal M}(u;v,\mathcal U):=\min_{\mu\in\mathcal M}\tilde{\phi}^{\mu}(u;v,\mathcal U),\qquad H_{\mathcal M}(u;v,\mathcal U):=\max_{\mu\in\mathcal M}\tilde{\phi}^{\mu}(u;v,\mathcal U).
\end{equation}
When \(\mathcal M=\{\mu\}\), credal \texttt{ADS} reduces to the single expected allocation under \(\mu\).
Because \(\tilde{\phi}^\mu(u;v,\mathcal U)\) is linear in \(\mu\), each credal bound can be computed by linear programming whenever \(\mathcal M\) is specified by linear probability constraints, as detailed in Appendix~\ref{app:credal_bounds}.

The set \(\mathcal V_{\mathcal M}(v,\mathcal U)\) describes the jointly admissible allocations, whereas the coordinatewise bounds summarize each unit's valuation range.
These bounds need not be attained simultaneously because different units may attain their extrema under different distributions.
The widths of these intervals depend on both structural ambiguity, represented by \(\mathcal M\), and order sensitivity, reflected in how fixed-structure \texttt{ADS} allocations differ across admissible structures.
The following proposition considers two admissible structures differing only in the order of two adjacent groups.
Using group efficiency in Proposition~\ref{propo:efficiency_asv_intra_group}, it expresses the width of each affected group's total allocation range as the product of structural ambiguity and order sensitivity.

\begin{proposition}[Adjacent-order sensitivity]
\label{prop:order_sensitivity}
Let \(\sigma=(G_1,\ldots,G_Q)\), and let \(\sigma'\) differ from \(\sigma\) only by interchanging the adjacent groups \(G_q\) and \(G_{q+1}\).
Suppose \(\Sigma=\{\sigma,\sigma'\}\), and write \(U_j:=\bigcup_{k=1}^{j}G_k\), with \(U_0:=\varnothing\).
For the common utility \(v\), let \(\Phi_G^\mu:=\sum_{u\in G}\tilde{\phi}^\mu(u;v,\mathcal U)\) denote the expected total allocation to group \(G\), and define
\begin{equation}
\label{eq:conditional_interaction}
I_q:=v(U_{q-1}\cup G_q)+v(U_{q-1}\cup G_{q+1})-v(U_{q-1})-v(U_{q+1}).
\end{equation}
For every \(\mu\in\Delta(\Sigma)\), the two group totals satisfy
\begin{equation}
\label{eq:order_transfer}
\begin{aligned}
\Phi^\mu_{G_q}+\Phi^\mu_{G_{q+1}}&=v(U_{q+1})-v(U_{q-1}),\\
\Phi^\mu_{G_q}&=v(U_{q+1})-v(U_{q-1}\cup G_{q+1})+\mu(\sigma)I_q.
\end{aligned}
\end{equation}
Consequently, for each \(G\in\{G_q,G_{q+1}\}\), its total allocation range over \(\mathcal M\) has width
\begin{equation}
\label{eq:order_range_width}
\max_{\mu\in\mathcal M}\Phi_G^\mu-\min_{\mu\in\mathcal M}\Phi_G^\mu=\left(\max_{\mu\in\mathcal M}\mu(\sigma)-\min_{\mu\in\mathcal M}\mu(\sigma)\right)|I_q|.
\end{equation}
\end{proposition}

The term \(I_q\) is the difference between a group's marginal contribution when it precedes the other group and its marginal contribution when it follows that group, conditional on the common upstream prefix \(U_{q-1}\).
Reversing the order therefore transfers \(|I_q|\) between the two group totals without changing their combined allocation.
When $I_q>0$, the groups exhibit conditional substitutability: each group contributes less after the other is present and consequently receives a larger total allocation when entering first.
When \(I_q<0\), the groups exhibit conditional complementarity, and each receives a larger total allocation when entering second.
When \(I_q=0\), both group totals are invariant to their relative order, although allocations to individual units within a multiunit group may still change.

Equation~\eqref{eq:order_range_width} separates structural ambiguity, measured by the admissible range of \(\mu(\sigma)\), from order sensitivity, measured by \(|I_q|\).
Holding the utility function fixed, additional information that narrows this probability range reduces each group's allocation-range width in proportion to \(|I_q|\).
Reducing uncertainty about the distribution over orders therefore improves the precision of these group totals only when \(I_q\neq0\).
Whether obtaining additional information is worthwhile also depends on its cost and on whether the resulting precision would materially affect valuation-based decisions. Appendix~\ref{app:adjacent_order_sensitivity} provides proof and a corresponding unit-level sensitivity analysis, and Appendix~\ref{app:credal_illustration} illustrates the credal bounds when service-order information is incomplete.

%% file: S8_Concluding_Remarks.tex
This paper addresses value attribution in AI model markets where raw data, derivative data, and data-related services create
utility through structured production pipelines. 
We show that model value in structured AI production should be attributed relative to the predecessor contexts supported by the production process rather than to all possible predecessor contexts. 
When pipelines exhibit provenance dependence, input--operation dependence, or model-state dependence, valuation rules that ignore documented production structure can evaluate contributions in pipeline-unsupported contexts, creating pipeline-context valuation mismatch.
\texttt{ADS} addresses the mismatch by translating auditable production evidence into ordered valuation groups and pipeline-supported predecessor contexts.
We establish an axiomatic characterization of \texttt{ADS} and derive group efficiency and completion invariance. 
We further develop \texttt{MC-ADS} and \texttt{SA-MC-ADS} for scalable estimation when utility evaluation is costly, and expected and credal \texttt{ADS} for
settings in which the pipeline structure remains uncertain.
Our experiments show that ignoring production dependencies shifts value toward redundant derivatives and away from useful upstream data and services, and that decisions guided by \texttt{ADS} improve model performance in multi-stage LLM fine-tuning and derivative data selection.

For AI marketplace operators, the main implication is that a valuation rule cannot be separated from the production context that defines its marginal comparisons. 
A defensible implementation specifies the valuation units and utility, identifies evidence-supported production dependencies, and determines which contributions occupy comparable positions before allocations are observed. 
These choices make the basis of value attribution explicit and auditable. 
When no defensible production-based precedence relation is available, classical \texttt{DS} remains appropriate. 
When documented dependencies determine economically meaningful predecessor contexts, \texttt{ADS} provides a structure-aware allocation aligned with those contexts.
More broadly, our results suggest that value in modern AI production is determined not only by what a contribution contains or performs, but also by where and how it enters the process that creates model utility.

\texttt{ADS} is a value attribution framework rather than a complete market mechanism. 
It takes the marketplace's utility objective as given, does not by itself prevent strategic manipulation of timestamps, provenance, or
valuation-unit boundaries, and depends on a defensible construction of the admissible structure set. 
These limitations motivate future work on inferring
and auditing valuation structures, designing complementary contractual and mechanism-design safeguards, 
and scaling estimation for costly
foundation-model pipelines.

%% file: Online_Appendices.tex
\appendix

\input{appendix_a}
\input{appendix_b}

\input{appendix_c}

\input{appendix_d}

%% file: appendix_a.tex
\section{Proofs for the ADS Framework}
\label{sec:proofs}

\paragraph{Proof of Lemma~\ref{lemma:example1_redundancy}.}
For any coalition excluding \(D_i\) and its copy \(\widetilde D_i\), adding either produces identical training data and utility. Shapley symmetry therefore gives equal values to each pair. Summing over pairs and applying efficiency gives each of the raw and duplicate groups one half of \(v(\mathcal D^{\mathrm{dup}})-v(\varnothing)\). The pooled data of \(\mathcal D^{\mathrm{dup}}\) contain exactly two copies of every training example in \(\mathcal D^{\mathrm{raw}}\). Uniform-duplication invariance therefore implies \(v(\mathcal D^{\mathrm{dup}})=v(\mathcal D^{\mathrm{raw}})\).\hfill\(\square\)

\paragraph{Proof of Proposition~\ref{propo:asv_subset} and the ordered-group derivation.}
Choose a permutation \(\pi_q\in\Pi(G_q)\) for each group \(G_q\). Concatenate these permutations in the order \(G_1,\ldots,G_Q\), written as \(\pi_1\oplus\cdots\oplus\pi_Q\). Each pipeline-consistent permutation corresponds to exactly one combination of within-group permutations. Choosing each \(\pi_q\) independently and uniformly therefore gives a uniform draw from \(\Pi_\sigma(\mathcal U)\). For \(u\in G_q\), the predecessor set \(U_{q-1}\cup S\), with \(S=\pre_{\pi_q}(u)\), depends only on \(\pi_q\). A fixed \(S\subseteq G_q\setminus\{u\}\) occurs in exactly \(|S|!(|G_q|-|S|-1)!\) within-group orders. Dividing by the \(|G_q|!\) possible orders gives
\[
p_u^\sigma(U_{q-1}\cup S)
=\frac{|S|!(|G_q|-|S|-1)!}{|G_q|!}
=\frac{1}{|G_q|}\binom{|G_q|-1}{|S|}^{-1}.
\]
This is Eq.~\eqref{eq:asv_prob}. These probabilities are positive for every \(S\subseteq G_q\setminus\{u\}\) and zero outside \(\mathcal C_\sigma(u)\). Grouping the permutation formula by predecessor sets proves Eq.~\eqref{eq:asv_subset}. Define the normalized induced game
\[
v_q(S):=v(U_{q-1}\cup S)-v(U_{q-1}),\qquad S\subseteq G_q.
\]
For \(u\in G_q\) and \(S\subseteq G_q\setminus\{u\}\), the induced game has marginal contribution \(v_q(S\cup\{u\})-v_q(S)=\Delta(u\mid U_{q-1}\cup S)\). The weights in Eq.~\eqref{eq:asv_subset} are the classical Shapley subset weights for \(|G_q|\) players. Hence that equation gives the Shapley value of \((G_q,v_q)\). Under Eq.~\eqref{eq:state_sufficiency}, subtracting the two induced-game utilities cancels \(\widetilde v_q(\varnothing;\theta_{q-1})\). The resulting marginal contribution is \(\Delta_q(u\mid S;\theta_{q-1})\), which gives Eq.~\eqref{eq:asv_seq}.

For a nonempty \(T\subseteq\mathcal U\), consider the full-domain unanimity game \(u_T(P)=\mathbf 1\{T\subseteq P\}\). A unit contributes one exactly when it is the last member of \(T\) to enter. Under a pipeline-consistent order, this unit belongs to \(T\cap G_k\), where \(k:=\max\{r:T\cap G_r\neq\varnothing\}\). Uniform ordering within \(G_k\) gives each member of \(T\cap G_k\) probability \(1/|T\cap G_k|\) of being last. Thus \texttt{ADS} assigns equal shares to those units and zero to all others. Nonempty unanimity games form a basis for normalized full-domain games. Linearity therefore gives the weighted Shapley value with equal weights within groups and ordered partition \(\sigma\) \citep{kalai1987weighted}. A nonzero baseline can be subtracted without changing any marginal contribution. Definition~\ref{definition:ads} is also a random-order value with fixed group precedence and uniform within-group ordering \citep{weber1988probabilistic,kalai1987weighted,nowak1995axiomatizations}.\hfill\(\square\)

\paragraph{Proof of Theorem~\ref{proposition:ads_characterization}.}
By Proposition~\ref{propo:asv_subset}, \(\phi^\sigma(u;v,\mathcal U)\allowbreak=\phi(u;v_q,G_q)\) for \(u\in G_q\). The induced game \(v_q\) is linear in \(v\). Pipeline nullity and within-group symmetry in \(v\) are exactly nullity and symmetry in \(v_q\). The corresponding properties of the classical Shapley value therefore establish Axioms~\ref{axiom:ads_linearity}--\ref{axiom:ads_symmetry}. Shapley efficiency gives the group total \(v_q(G_q)=v(U_q)-v(U_{q-1})\). Summing these totals over \(q\) gives \(v(\mathcal U)-v(\varnothing)\), establishing pipeline efficiency.

For uniqueness, let \(\psi^\sigma\) satisfy the four axioms. Fix a group \(G_q\) and a normalized game \(w:2^{G_q}\to\mathbb R\), with \(w(\varnothing)=0\). Define its lift to \(\mathcal F_\sigma\) by
\[
(\mathcal L_qw)(P):=w(P\cap G_q),\qquad P\in\mathcal F_\sigma.
\]
Each coalition \(P\) has a unique intersection with \(G_q\), so the lift is well defined. Its values at \(U_{q-1}\) and \(U_q\) are \(0\) and \(w(G_q)\), respectively. Let \(c_v(P)=v(\varnothing)\). For \(P=U_{r-1}\cup S\), with \(S\subseteq G_r\), the lifted games for \(q<r\) sum to \(\sum_{q<r}v_q(G_q)=v(U_{r-1})-v(\varnothing)\). The \(r\)-th lifted game contributes \(v_r(S)=v(P)-v(U_{r-1})\), and those for \(q>r\) contribute zero. Adding these terms gives
\[
v=c_v+\sum_{q=1}^Q\mathcal L_qv_q.
\]
Every unit is pipeline-null in \(c_v\). For \(u\notin G_q\), adding \(u\) leaves \(P\cap G_q\) unchanged, so \(u\) is pipeline-null in \(\mathcal L_qw\). Linearity and nullity therefore imply
\(\psi^\sigma(u;v,\mathcal U)=\psi^\sigma(u;\mathcal L_qv_q,\mathcal U)\) for \(u\in G_q\).
Define \(\chi_q(u;w):=\psi^\sigma(u;\mathcal L_qw,\mathcal U)\) for \(u\in G_q\). The lift is linear in \(w\). Within \(G_q\), the supported marginal contributions and symmetry comparisons under \(\mathcal L_qw\) are exactly those under \(w\). Thus \(\chi_q\) satisfies linearity, nullity, and symmetry on normalized games over \(G_q\). Pipeline efficiency and nullity outside \(G_q\) give \(\sum_{u\in G_q}\chi_q(u;w)=w(G_q)\). 
The classical Shapley characterization yields \(\chi_q(u;w)=\phi(u;w,G_q)\). Taking \(w=v_q\) and applying Proposition~\ref{propo:asv_subset} proves \(\psi^\sigma(u;v,\mathcal U)=\phi^\sigma(u;v,\mathcal U)\) for every unit.\hfill\(\square\)

\paragraph{Proof of Proposition~\ref{proposition:sv_as_special_asv}.}
When \(\sigma=(\mathcal U)\), there is one group \(G_1=\mathcal U\) and \(U_0=\varnothing\). Hence \(\Pi_\sigma(\mathcal U)=\Pi(\mathcal U)\), \(\mathcal F_\sigma=2^{\mathcal U}\), and \(\mathcal C_\sigma(u)=2^{\mathcal U\setminus\{u\}}\). Definition~\ref{definition:ads} then averages over all permutations, exactly as in the classical Shapley formula. The weights in Eq.~\eqref{eq:asv_subset} become \(|S|!(|\mathcal U|-|S|-1)!/|\mathcal U|!\). For \(\mathcal U=\mathcal D^{\mathrm{raw}}\), this gives \texttt{DS} in Eq.~\eqref{eq:ds_marginal}.\hfill\(\square\)

\paragraph{Proof of Proposition~\ref{propo:efficiency_asv_intra_group}.}
By Proposition~\ref{propo:asv_subset}, \(\phi^\sigma(u;v,\mathcal U)=\phi(u;v_q,G_q)\) for \(u\in G_q\).
Shapley efficiency in \((G_q,v_q)\) gives
\(\sum_{u\in G_q}\phi^\sigma(u;v,\mathcal U)=v_q(G_q)-v_q(\varnothing)=v(U_q)-v(U_{q-1})\).
Summing the group totals cancels the intermediate utility terms. The remaining total is \(v(\mathcal U)-v(\varnothing)\).
For \(q=1\), we have \(U_0=\varnothing\) and \(v_1(S)=v(S)-v(\varnothing)\). The restricted game \(v|_{2^{G_1}}\) is well defined because every subset of \(G_1\) belongs to \(\mathcal F_\sigma\). Subtracting \(v(\varnothing)\) leaves every marginal contribution unchanged. Hence \(\phi^\sigma(u;v,\mathcal U)=\phi(u;v|_{2^{G_1}},G_1)\) for \(u\in G_1\).
Thus first-group allocations equal their Shapley values in the first-group subgame and are invariant to appending later groups while that subgame is fixed.\hfill\(\square\)

\paragraph{Proof of Proposition~\ref{proposition:ads_completion_invariance}.}
For \(u\in G_q\) and \(\pi=\pi_1\oplus\cdots\oplus\pi_Q\in\Pi_\sigma(\mathcal U)\), the predecessor set is \(\pre_\pi(u)=U_{q-1}\cup\pre_{\pi_q}(u)\in\mathcal C_\sigma(u)\). Both \(\pre_\pi(u)\) and \(\pre_\pi(u)\cup\{u\}\) therefore belong to \(\mathcal F_\sigma\). Any two full-domain completions that agree with \(v\) on this domain give the same marginal contribution for every supported predecessor set. Averaging these contributions in Definition~\ref{definition:ads} gives the same \texttt{ADS} allocation under either completion. Classical \texttt{DS} need not have this property. If \(Q\geq2\) and \(u\in G_2\), then \(\{u\}\notin\mathcal F_\sigma\) because \(G_1\) is nonempty. For a full-domain completion \(\bar v\), the Shapley formula for \(u\) includes \(\bar v(\{u\})\) with coefficient \(1/|\mathcal U|\). Changing only this unsupported utility therefore changes the classical allocation, as illustrated in Example~\ref{ex:broker_service}.\hfill\(\square\)

%% file: appendix_b.tex
\section{Proofs and Computational Details for Scalable ADS Estimation}
\label{app:technical-details}

\subsection{Finite-Sample Properties and Cost of \texttt{MC-ADS}}
\label{app:mc_accuracy}

\paragraph{Proof of Theorem~\ref{thm:mc_ads_properties}.}
For \(u\in G_q\), define \(Z_u(\pi_q)=\Delta(u\mid U_{q-1}\cup\pre_{\pi_q}(u))\), where \(\pi_q\) is uniform on \(\Pi(G_q)\). Proposition~\ref{propo:asv_subset} gives \(\mathbb EZ_u=\ads{u}{\mathcal U}\). Since the estimator averages copies of \(Z_u\), linearity of expectation proves unbiasedness. For non-singleton groups, these copies are independent and have range length at most \(\rho\). Hoeffding's inequality and a union bound over at most \(N=|\mathcal U|\) units give
\[
\Pr\!\left(\max_u|\widehat\phi_{\mathrm{MC}}^\sigma(u)-\phi^\sigma(u)|>\epsilon\right)
\leq 2N\exp\!\left(-2R_{\mathrm{MC}}\epsilon^2/\rho^2\right).
\]
The stated sample size makes this bound at most \(\delta\). The union bound allows dependence between unit estimates. Singleton groups are exact. On each complete within-group path, the intermediate utility terms cancel, giving \(\sum_{u\in G_q}Z_u(\pi_q)=v(U_q)-v(U_{q-1})\). Averaging the same paths for all units preserves that group total at every finite sample size.\hfill\(\square\)

\paragraph{Computational cost.}
Each group requires one baseline utility evaluation and at most \(|G_q|\) additional evaluations per path. Since \(R_q\leq R_{\mathrm{MC}}\), the total is at most \(Q+\sum_qR_q|G_q|\leq Q+R_{\mathrm{MC}}N\), proving Remark~\ref{remark:mc_ads_cost}. Reusing cached utilities can reduce this count.

\subsection{Finite-Sample Properties and Variance of \texttt{SA-MC-ADS}}
\label{app:sa-mcads-proof}

\paragraph{Proof of Theorem~\ref{thm:sa-mcads-main}.}
For a uniform within-group order, define \(X_u^{\mathrm L}=\Delta^{\mathrm L}(u\mid U_{q-1}\cup\pre_{\pi_q}(u))\) and \(X_u^{\mathrm{res}}=Z_u-X_u^{\mathrm L}\). All low-only and paired permutations are drawn independently from this uniform distribution. Since \(X_u^{\mathrm L}+X_u^{\mathrm{res}}=Z_u\), the two sample means have expectations that sum to \(\ads{u}{\mathcal U}\), proving unbiasedness.

The low-only mean has \(R_{\mathrm L}\) terms, each with range length at most \(\rho_{\mathrm L}/R_{\mathrm L}\). The residual mean has \(R_{\mathrm H}\) terms, each with range length at most \(\rho_{\mathrm{res}}/R_{\mathrm H}\). The sum of their squared range lengths is at most \(a:=\rho_{\mathrm L}^2/R_{\mathrm L}+\rho_{\mathrm{res}}^2/R_{\mathrm H}\). If \(a>0\), Hoeffding's inequality and a union bound give
\[
\Pr\!\left(\max_u|\widehat\phi_{\mathrm{SA}}^\sigma(u)-\phi^\sigma(u)|>\epsilon\right)
\leq 2N\exp(-2\epsilon^2/a).
\]
The condition in Eq.~\eqref{eq:sa-mcads-accuracy-main} makes this bound at most \(\delta\). If \(a=0\), the low-fidelity and residual contributions are constant over the within-group permutations, so the estimator is exact. Singleton groups are also evaluated exactly.

For group efficiency, let \(b_q=v(U_q)-v(U_{q-1})\) and \(b_q^{\mathrm L}=v^{\mathrm L}(U_q)-v^{\mathrm L}(U_{q-1})\). Each sampled path is shared by all units in \(G_q\). On a complete low-only path, the utility increments sum to \(b_q^{\mathrm L}\). On a complete paired path, the high- and low-fidelity totals are \(b_q\) and \(b_q^{\mathrm L}\), respectively. Thus the residual total is \(b_q-b_q^{\mathrm L}\). Averaging preserves both group totals. Their sum is \(b_q\).\hfill\(\square\)

\paragraph{Variance decomposition.}
The finite permutation spaces ensure finite second moments. Independence of the two samples gives \(\operatorname{Var}(\widehat\phi_{\mathrm{SA}}^\sigma(u))=\allowbreak\operatorname{Var}(X_u^{\mathrm L})/R_{\mathrm L}+\operatorname{Var}(X_u^{\mathrm{res}})/R_{\mathrm H}\). This is Eq.~\eqref{eq:sa-mcads-variance-main}, proving Remark~\ref{remark:sa-mcads-variance}. Both terms vanish for singleton groups.
Expanding \(X_u^{\mathrm{res}}=Z_u-X_u^{\mathrm L}\) gives \(\operatorname{Var}(X_u^{\mathrm{res}})=\operatorname{Var}(Z_u)-2\operatorname{Cov}(Z_u,X_u^{\mathrm L})+\operatorname{Var}(X_u^{\mathrm L})\). At \(R_{\mathrm{MC}}=R_{\mathrm H}\), \texttt{MC-ADS} has variance \(\operatorname{Var}(Z_u)/R_{\mathrm H}\). The residual component has strictly smaller variance exactly when \(2\operatorname{Cov}(Z_u,X_u^{\mathrm L})>\operatorname{Var}(X_u^{\mathrm L})\).

\subsection{Exact-Control Variant}
\label{app:sa-mcads-algorithm}
\label{app:sa-coefficient}
Let \(\phi_{q,u}^{\mathrm L}:=\mathbb EX_u^{\mathrm L}\). Proposition~\ref{propo:asv_subset} applied to \(v^{\mathrm L}\) gives \(\phi_{q,u}^{\mathrm L}=\phi^\sigma(u;v^{\mathrm L},\mathcal U)\). If this control mean is known exactly, replace the low-only sample mean by it:
\begin{equation}
\widehat\phi_{\mathrm{SA,exact}}^\sigma(u)
=\phi_{q,u}^{\mathrm L}+\frac{1}{R_{\mathrm H}}\sum_{r=1}^{R_{\mathrm H}}X_u^{\mathrm{res}}(\pi_q^{(r)}).
\label{eq:sa-exact-control-mean}
\end{equation}
Since \(\mathbb EX_u^{\mathrm{res}}=\phi^\sigma(u)-\phi_{q,u}^{\mathrm L}\), the estimator is unbiased. Its variance is the residual term \(\operatorname{Var}(X_u^{\mathrm{res}})/R_{\mathrm H}\) in Eq.~\eqref{eq:sa-mcads-variance-main}. By Proposition~\ref{propo:efficiency_asv_intra_group} applied to \(v^{\mathrm L}\), the exact control means sum to \(b_q^{\mathrm L}\). Averaging complete residual paths shared by all units gives the group total \(b_q-b_q^{\mathrm L}\). Adding these totals preserves \(b_q\). The concentration argument now uses only the residual term, giving the sufficient sample size
\[
R_{\mathrm H}\geq\max\!\left\{1,\left\lceil\frac{\rho_{\mathrm{res}}^2}{2\epsilon^2}\log\frac{2N}{\delta}\right\rceil\right\}.
\]

%% file: appendix_c.tex
\section{Additional Experimental Evidence}
\label{sec:allocation_robustness}

\subsection{Completion Sensitivity}
\label{app:completion_sensitivity}
Table~\ref{tab:exp1b_completion_sensitivity} varies unsupported utilities while fixing all pipeline-supported utilities in Section~\ref{subsubsec:exp1b_broker_pipeline}. Direct evaluation is the main completion shown in the main text.

\begin{table}[!htbp]
    \centering
    \fontsize{10}{12}\selectfont
    \setlength{\tabcolsep}{5pt}
    \begin{tabular}{@{}llccc@{}}
        \toprule
        Rule & Completion
        & \shortstack{Raw\\contributors}
        & \shortstack{Deduplication\\service}
        & \shortstack{Redundant\\bundle} \\
        \midrule
        \texttt{DS} & Direct evaluation
        & 0.2346 (0.0033) & 0.0022 (0.0010) & 0.1436 (0.0024) \\
        & Prefix projection
        & 0.3757 (0.0045) & 0.0047 (0.0015) & 0.0000 (0.0000) \\
        & Prerequisite closure
        & 0.1696 (0.0028) & 0.1054 (0.0020) & 0.1054 (0.0020) \\
        & Constant baseline
        & 0.4232 (0.0052) & 0.0420 (0.0014) & $-$0.0848 (0.0014) \\
        \addlinespace
        \texttt{ADS} & All four completions
        & 0.3662 (0.0054) & 0.0142 (0.0044) & 0.0000 (0.0000) \\
        \bottomrule
    \end{tabular}
    \caption[Full-market allocations under alternative utility completions.]{Full-market allocations under alternative utility completions. Means across 100 independently resampled markets. Allocations are measured in balanced-accuracy surplus above \(v(\varnothing)=0.5\). Parentheses give 95\% Student-\(t\) confidence-interval half-widths. Raw-data allocations sum the values assigned to both contributors. Prefix projection discards active units after the first incomplete group. Prerequisite closure adds missing predecessors required by the latest active contribution. Constant baseline assigns \(v(\varnothing)=0.5\) to unsupported coalitions.}
    \label{tab:exp1b_completion_sensitivity}
\end{table}

The \texttt{DS} allocation changes across completions and can assign the redundant bundle positive or negative value. \texttt{ADS} is invariant, as Proposition~\ref{proposition:ads_completion_invariance} predicts.

\subsection{Estimator Precision and Total Computational Cost}\label{appC:computational cost}
\begin{figure}[!htbp]
\centering
\captionsetup[subfigure]{font=footnotesize,labelfont=bf,justification=centering}
\begin{subfigure}[t]{0.245\textwidth}
\centering
\includegraphics[width=\linewidth]{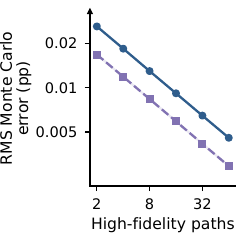}
\caption{Adult: by path count}
\label{fig:sa_adult_paths}
\end{subfigure}\hfill%
\begin{subfigure}[t]{0.245\textwidth}
\centering
\includegraphics[width=\linewidth]{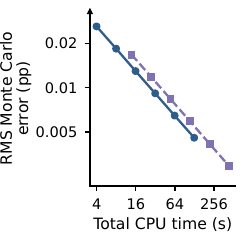}
\caption{Adult: by CPU time}
\label{fig:sa_adult_time}
\end{subfigure}\hfill%
\begin{subfigure}[t]{0.245\textwidth}
\centering
\includegraphics[width=\linewidth]{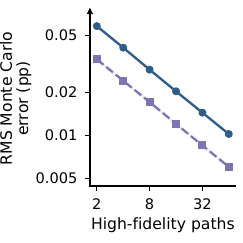}
\caption{EMNIST: by path count}
\label{fig:sa_emnist_paths}
\end{subfigure}\hfill%
\begin{subfigure}[t]{0.245\textwidth}
\centering
\includegraphics[width=\linewidth]{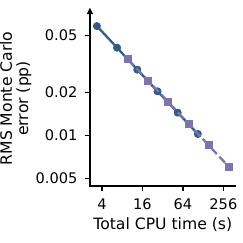}
\caption{EMNIST: by CPU time}
\label{fig:sa_emnist_time}
\end{subfigure}\par
\vspace{0.3em}
\includegraphics[width=0.45\textwidth]{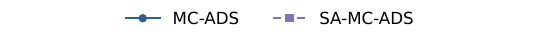}
\caption[Estimator precision and total CPU time, with logarithmic axes.]{Estimator precision and total CPU time, with logarithmic axes. Pointwise 95\% market-bootstrap intervals show RMS Monte Carlo error (bands and vertical bars) and mean CPU time (horizontal bars). All derivative records in 20 randomly drawn markets per dataset (Section~\ref{subsubsec:derivative_procurement}). Both methods use \(R_{\mathrm H}=R\in\{2,4,8,16,32,64\}\) high-fidelity paths. \texttt{SA-MC-ADS} adds \(R_{\mathrm L}=2R\) independent low-only paths (Theorem~\ref{thm:sa-mcads-main}). Surrogates use the same logistic learner, with at most six iterations and accuracy scale 0.635 on Adult, and 24 iterations and scale 0.654 on EMNIST Letters. The iteration limits and scaling coefficients were selected during preliminary development and held fixed for this comparison. RMS error (percentage points) is the square root of the mean sample variance from eight independent estimates per record, averaged first within markets and then across markets. CPU time includes all surrogate computation.}
\label{fig:sa_precision_runtime}
\end{figure}

At fixed high-fidelity budgets, \texttt{SA-MC-ADS} reduces estimated mean squared Monte Carlo error by approximately 58\% on Adult and 65\% on EMNIST Letters, while requiring 3.4 and 2.9 times as much CPU time as \texttt{MC-ADS}, respectively (Figure~\ref{fig:sa_precision_runtime}). Over the observed overlapping CPU-time ranges, \texttt{MC-ADS} is more precise on Adult, while the two methods attain similar precision on EMNIST Letters. Computational savings therefore also require a sufficiently cheap surrogate.

\FloatBarrier

\subsection{Dataset Procurement across Additional Base Models}\label{app:llm_selection_additional}
\begin{figure}[!htbp]
\centering
\captionsetup[subfigure]{font=footnotesize,labelfont=bf,justification=centering}
\begin{subfigure}[t]{0.32\textwidth}\centering\includegraphics[width=\linewidth]{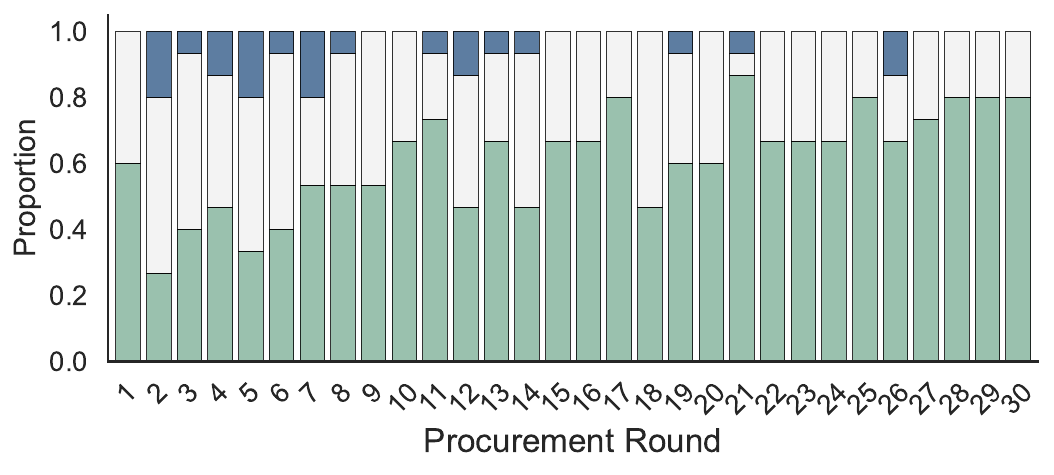}\caption{\texttt{LOO}: Llama 3.2-3B}\end{subfigure}\hfill
\begin{subfigure}[t]{0.32\textwidth}\centering\includegraphics[width=\linewidth]{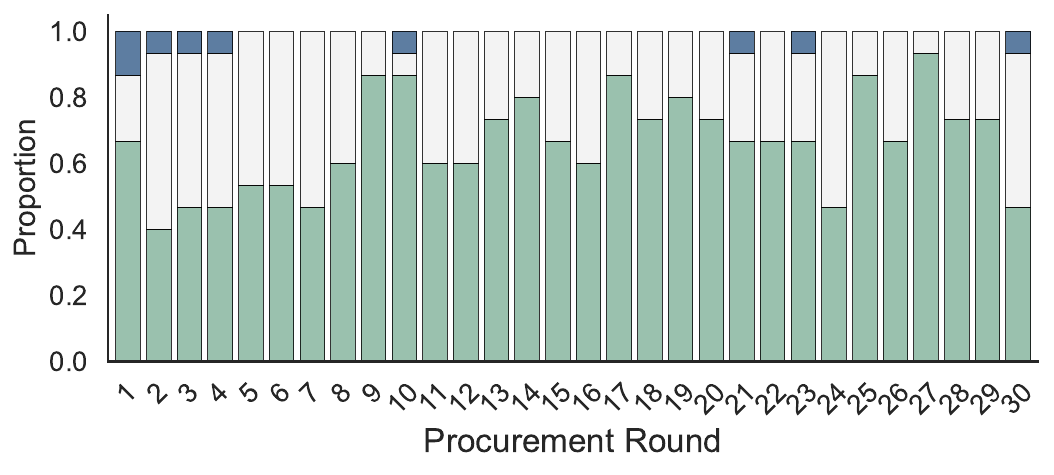}\caption{\texttt{LOO}: Qwen 3-0.6B}\end{subfigure}\hfill
\begin{subfigure}[t]{0.32\textwidth}\centering\includegraphics[width=\linewidth]{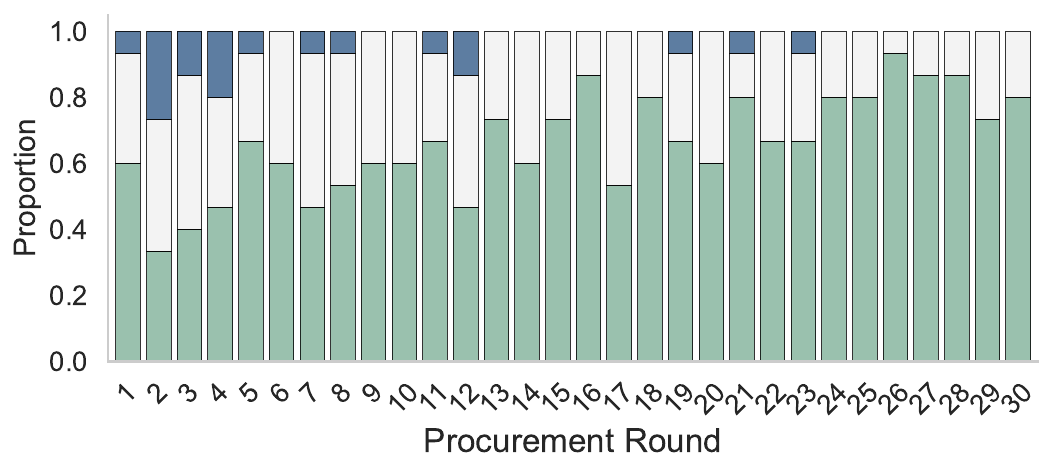}\caption{\texttt{LOO}: Qwen 3-8B}\end{subfigure}\par\vspace{0.4em}
\begin{subfigure}[t]{0.32\textwidth}\centering\includegraphics[width=\linewidth]{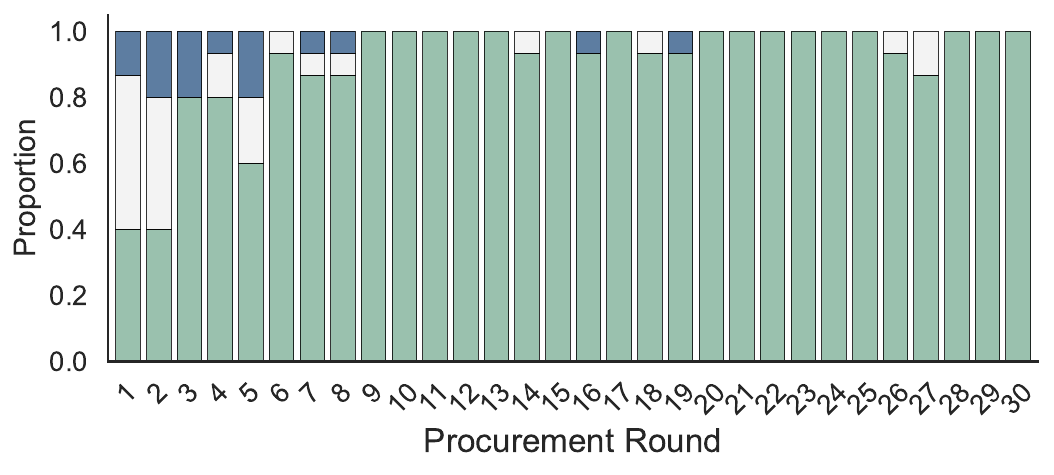}\caption{\texttt{ADS}: Llama 3.2-3B}\end{subfigure}\hfill
\begin{subfigure}[t]{0.32\textwidth}\centering\includegraphics[width=\linewidth]{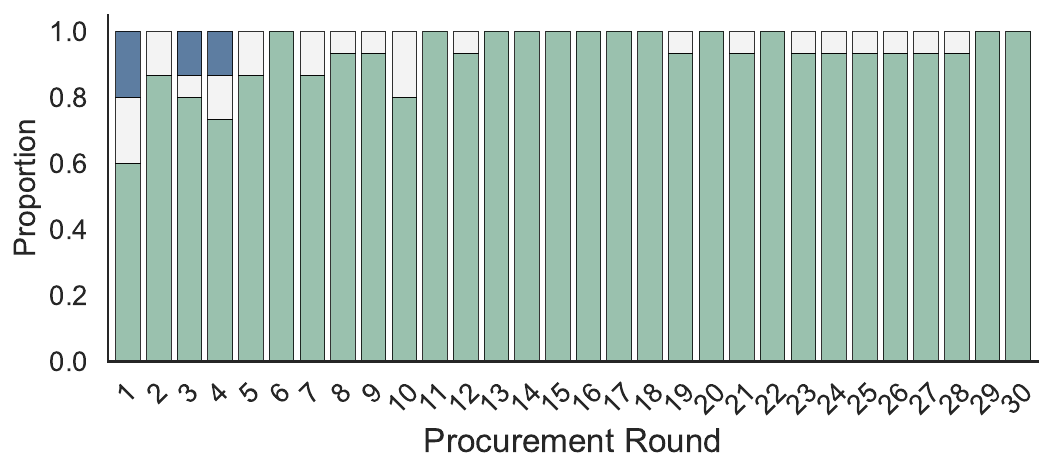}\caption{\texttt{ADS}: Qwen 3-0.6B}\end{subfigure}\hfill
\begin{subfigure}[t]{0.32\textwidth}\centering\includegraphics[width=\linewidth]{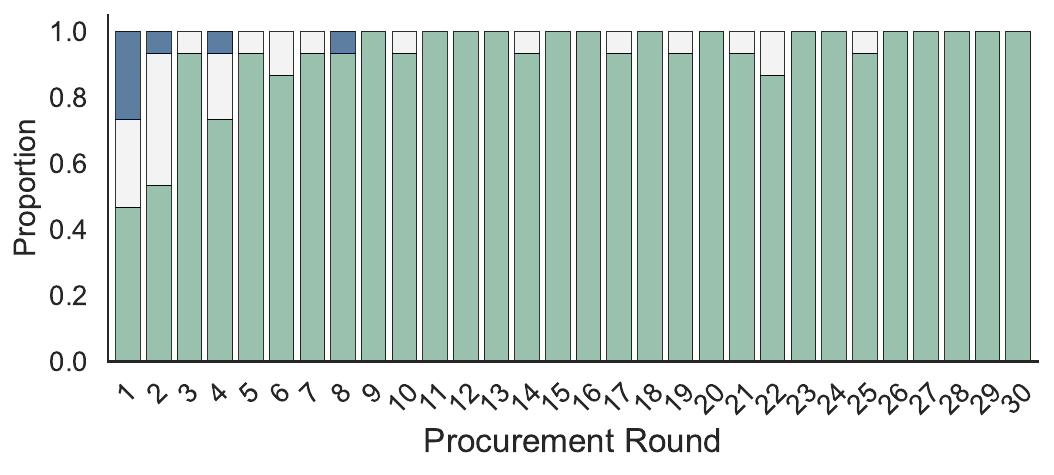}\caption{\texttt{ADS}: Qwen 3-8B}\end{subfigure}\par\vspace{0.4em}
\includegraphics[width=0.75\textwidth]{figures/llm_fig/legend_bar.pdf}
\caption{Dataset-selection shares for the three additional base LLMs. Each bar represents one procurement round. Columns correspond to Llama 3.2-3B, Qwen 3-0.6B, and Qwen 3-8B. The upper row shows \texttt{LOO} and the lower row \texttt{ADS}. Colors denote the 30\%, 60\%, and 90\% off-target candidates, as in Figure~\ref{fig:selection_llm_experiment}.}
\label{fig:llm_selection_additional}
\end{figure}
\FloatBarrier

%% file: appendix_d.tex
\section{Expected and Credal ADS under Uncertain Pipeline Structure}
\label{app:credal_ads}

\subsection{Credal Bounds and Market Allocation}
\label{app:credal_bounds}
Let \(c_u=(\phi^\sigma(u;v,\mathcal U))_{\sigma\in\Sigma}\). For nonempty \(\mathcal M=\{\mu\geq0:\mathbf1^\top\mu=1,\ A\mu\leq b\}\), Eq.~\eqref{eq:credal_ads_bounds} requires \(2|\mathcal U|\) linear programs with objective \(c_u^\top\mu\) and a shared feasible region. Compactness ensures attained extrema. These programs take exact fixed-structure allocations as inputs.

Let \(B=v(\mathcal U)-v(\varnothing)\). Allocating the lower values leaves the reserve
\begin{equation}
R_{\mathcal M}:=B-\sum_uL_{\mathcal M}(u)
=\sum_u[\widetilde\phi^\mu(u)-L_{\mathcal M}(u)]\geq0,
\quad\mu\in\mathcal M.
\label{eq:credal-reserve}
\end{equation}
Expected-\texttt{ADS} efficiency gives the equality, and each summand is nonnegative. The reserve covers adjustments to any allocation under a common \(\mu\in\mathcal M\). With \(\Sigma,v\) fixed, lower bounds increase weakly and upper bounds and the reserve decrease weakly as the nonempty credal set shrinks. This monotonicity need not hold if \(\Sigma\) or \(v\) changes.

\subsection{Adjacent-Order and Unit-Level Sensitivity}
\label{app:adjacent_order_sensitivity}
\paragraph{Proof of Proposition~\ref{prop:order_sensitivity}.}
Write \(P=U_{q-1}\) and \(T=U_{q+1}\). Group efficiency gives the two \(G_q\) allocations as \(v(P\cup G_q)-v(P)\) and \(v(T)-v(P\cup G_{q+1})\). Their difference is \(I_q\), and both orders allocate \(v(T)-v(P)\) to the two groups jointly. Mixing gives Eq.~\eqref{eq:order_transfer}. Compactness and convexity of \(\mathcal M\) make the attainable probabilities \(\mu(\sigma)\) an interval \([l,h]\), with endpoints equal to their minimum and maximum. The group totals have slopes \(I_q\) and \(-I_q\), so their widths are \((h-l)|I_q|\), proving Eq.~\eqref{eq:order_range_width}.\hfill\(\square\)

\paragraph{Unit-level sensitivity and reserve.}
In the same two-structure setting, let \(l=\min_{\mu\in\mathcal M}\mu(\sigma)\), \(h=\max_{\mu\in\mathcal M}\mu(\sigma)\), and \(d_u=\phi^\sigma(u)-\phi^{\sigma'}(u)\). For units outside \(G_q\cup G_{q+1}\), the unchanged group and prefix give \(d_u=0\). Since \(\widetilde\phi^\mu(u)=\phi^{\sigma'}(u)+\mu(\sigma)d_u\), endpoint evaluation gives
\begin{equation}
H_{\mathcal M}(u)-L_{\mathcal M}(u)=(h-l)|d_u|.
\label{eq:credal-unit-width}
\end{equation}
The group sums \(\sum_{u\in G_q}d_u=I_q\) and \(\sum_{u\in G_{q+1}}d_u=-I_q\) allow offsetting unit effects, so group totals can be invariant while individual widths remain positive. The admissible midpoint \(\mu_0(\sigma)=(l+h)/2\) gives every unit interval's midpoint. Efficiency and Eq.~\eqref{eq:credal-reserve} then yield
\begin{equation}
R_{\mathcal M}=\tfrac12\sum_u[H_{\mathcal M}(u)-L_{\mathcal M}(u)]
=\tfrac{h-l}{2}\sum_u|d_u|.
\label{eq:credal-reserve-adjacent}
\end{equation}

\subsection{Illustration with Incomplete Service-Order Information}
\label{app:credal_illustration}
We examine how service-order records reduce uncertainty about expected \texttt{ADS} allocations. Label correction, deduplication, and feature restoration follow the raw-data group. For any fixed subset of services, utility on the same raw data is order-independent, satisfying the common-utility assumption in Section~\ref{sec:robustness_strategy}. Marginal contributions can still vary with predecessors.

\begin{figure}[!htbp]
\centering
\captionsetup[subfigure]{font=small,labelfont=bf,justification=centering}
\begin{subfigure}[t]{0.485\textwidth}
\centering
\includegraphics[width=\linewidth]{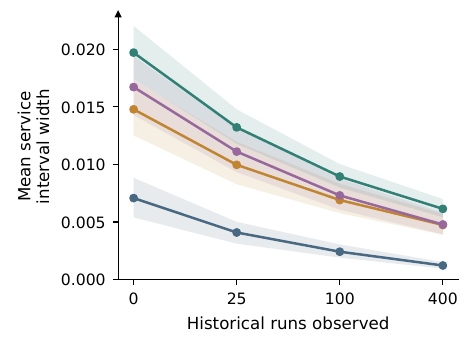}
\caption{Allocation intervals}
\label{fig:appE_intervals}
\end{subfigure}\hfill%
\begin{subfigure}[t]{0.485\textwidth}
\centering
\includegraphics[width=\linewidth]{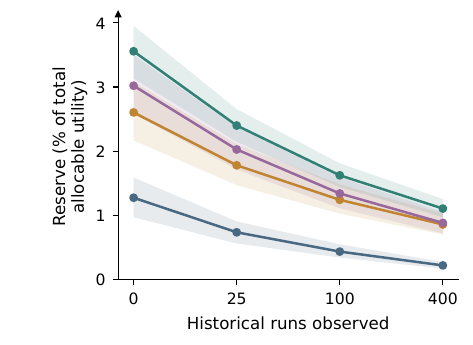}
\caption{Reserve}
\label{fig:appE_reserve}
\end{subfigure}\par
\vspace{0.3em}
\includegraphics[width=0.85\textwidth]{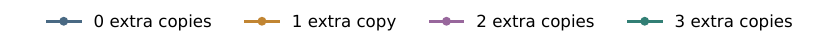}
\caption[Allocation uncertainty with increasing service-order information.]{Allocation uncertainty with increasing service-order information. Panel (a) averages the three services' allocation-interval widths. Panel (b) reports the reserve as a percentage of total allocable utility. Across 30 paired optical-digits seeds, we add zero to three copies per image for a selected quarter of training images. Utility is the balanced accuracy of a ridge classifier. Fixed-order allocations are computed exactly. Credal sets intersect Hoeffding bounds on three pairwise-order frequencies from 25, 100, and 400 independent historical runs of the same stationary scheduler (at least 95\% simultaneous distribution coverage). With three extra copies, 0--400 logs reduce width from 0.0197 to 0.0061 and reserve share from 3.557\% to 1.105\%. Shading shows 95\% percentile bootstrap intervals across seeds (2,000 resamples), conditional on the dataset.}
\label{fig:appE_evidence_refinement}
\end{figure}

In Figure~\ref{fig:appE_evidence_refinement}, each curve fixes the number of extra copies, while the horizontal axis counts historical runs. More logs restrict possible order probabilities. The nested credal sets therefore narrow allocation intervals and reduce the reserve. Higher duplication yields greater uncertainty here, consistent with greater sensitivity to service order. These bounds concern expected allocations under a representative stationary scheduler, not allocations in each realized order.

%% file: refs.bib
@inproceedings{chen2020simple,
  title={A simple framework for contrastive learning of visual representations},
  author={Chen, Ting and Kornblith, Simon and Norouzi, Mohammad and Hinton, Geoffrey},
  booktitle={Proceedings of the 37th {International Conference on Machine Learning}},
  pages={1597--1607},
  year={2020},
  volume={119},
  series={Proceedings of Machine Learning Research},
  editor={Daum{\'e}, III, Hal and Singh, Aarti},
  publisher={PMLR},
  url={https://proceedings.mlr.press/v119/chen20j.html}
}

@article{trofimov2023multi,
  title={Multi-fidelity neural architecture search with knowledge distillation},
  author={Trofimov, Ilya and Klyuchnikov, Nikita and Salnikov, Mikhail and Filippov, Alexander and Burnaev, Evgeny},
  journal={IEEE Access},
  volume={11},
  pages={59217--59225},
  year={2023},
  publisher={IEEE},
  doi={10.1109/ACCESS.2023.3234810}
}

@inproceedings{wu2020practical,
  title={Practical multi-fidelity {Bayesian} optimization for hyperparameter tuning},
  author={Wu, Jian and Toscano-Palmerin, Saul and Frazier, Peter I and Wilson, Andrew Gordon},
  booktitle={Proceedings of the 35th {Uncertainty in Artificial Intelligence Conference}},
  pages={788--798},
  year={2020},
  volume={115},
  series={Proceedings of Machine Learning Research},
  editor={Adams, Ryan P. and Gogate, Vibhav},
  publisher={PMLR},
  url={https://proceedings.mlr.press/v115/wu20a.html}
}

@inproceedings{goldwasser2024stabilizing,
  title={Stabilizing estimates of {Shapley} values with control variates},
  author={Goldwasser, Jeremy and Hooker, Giles},
  booktitle={Explainable artificial intelligence},
  pages={416--439},
  year={2024},
  editor={Longo, Luca and Lapuschkin, Sebastian and Seifert, Christin},
  series={Communications in Computer and Information Science},
  volume={2154},
  publisher={Springer},
  doi={10.1007/978-3-031-63797-1_21}
}

@incollection{derakhshan2022materialization,
  title={Materialization and reuse optimizations for production data science pipelines},
  author={Derakhshan, Behrouz and Rezaei Mahdiraji, Alireza and Kaoudi, Zoi and Rabl, Tilmann and Markl, Volker},
  booktitle={Proceedings of the 2022 {International Conference on Management of Data}},
  pages={1962--1976},
  year={2022},
  doi={10.1145/3514221.3526186},
  publisher={Association for Computing Machinery}
}

@article{han2026data,
  title={Data valuation for vertical federated learning: A model-free and privacy-preserving method},
  author={Han, Xiao and Wang, Leye and Wu, Junjie and Fang, Xiao},
  journal={MIS Quarterly},
  volume={50},
  number={1},
  pages={177--210},
  year={2026},
  publisher={Management Information Systems Research Center, University of Minnesota},
  doi={10.25300/MISQ/2025/19161}
}

@article{pei2023data,
  title={Data and {AI} model markets: Opportunities for data and model sharing, discovery, and integration},
  author={Pei, Jian and Fernandez, Raul Castro and Yu, Xiaohui},
  journal={Proceedings of the VLDB Endowment},
  volume={16},
  number={12},
  pages={3872--3873},
  year={2023},
  publisher={VLDB Endowment},
  doi={10.14778/3611540.3611573}
}

@article{bi2024understanding,
  title={Understanding partnership formation and repeated contributions in federated learning: An analytical investigation},
  author={Bi, Xuan and Gupta, Alok and Yang, Mochen},
  journal={Management Science},
  volume={70},
  number={8},
  pages={4974--4994},
  year={2024},
  publisher={INFORMS},
  doi={10.1287/mnsc.2023.00611}
}

@inproceedings{wang2023data,
  title={Data {Banzhaf}: A robust data valuation framework for machine learning},
  author={Wang, Jiachen T and Jia, Ruoxi},
  booktitle={Proceedings of the 26th {International Conference on Artificial Intelligence and Statistics}},
  pages={6388--6421},
  year={2023},
  volume={206},
  series={Proceedings of Machine Learning Research},
  editor={Ruiz, Francisco and Dy, Jennifer and van de Meent, Jan-Willem},
  publisher={PMLR},
  url={https://proceedings.mlr.press/v206/wang23e.html}
}

@article{dubey1981value,
  title={Value theory without efficiency},
  author={Dubey, Pradeep and Neyman, Abraham and Weber, Robert James},
  journal={Mathematics of Operations Research},
  volume={6},
  number={1},
  pages={122--128},
  year={1981},
  publisher={INFORMS},
  doi={10.1287/moor.6.1.122}
}

@article{hue2026measuring,
  title={Measuring the driving forces of predictive performance: Application to credit scoring},
  author={Hu{\'e}, Sullivan and Hurlin, Christophe and P{\'e}rignon, Christophe and Saurin, S{\'e}bastien},
  journal={Management Science},
  year={2026},
  publisher={INFORMS},
  doi={10.1287/mnsc.2023.02025},
  howpublished={Advance online publication}
}

@inproceedings{lee2026priority,
  title={Priority-aware {Shapley} value},
  author={Lee, Kiljae and Liu, Ziqi and Tang, Weijing and Zhang, Yuan},
  year={2026},
  booktitle={Proceedings of the 43rd {International Conference on Machine Learning}},
  pages={64673--64710},
  volume={306},
  series={Proceedings of Machine Learning Research},
  editor={Zhang, Tong and Dudik, Miroslav and Jaggi, Martin and Agarwal, Alekh and Li, Sharon and Schuurmans, Dale and Zhu, Jerry and Berkenkamp, Felix and Dong, Hanze and Bietti, Alberto},
  publisher={PMLR},
  url={https://proceedings.mlr.press/v306/lee26at.html}
}

@online{lee2026generalized,
  title={Generalized priority-aware {Shapley} value},
  author={Lee, Kiljae and Liu, Ziqi and Tang, Weijing and Zhang, Yuan},
  year={2026},
  organization={arXiv},
  doi={10.48550/arXiv.2605.15018},
  note={Accepted at {NeurIPS} 2026}
}

@article{faigle1992shapley,
  title={The {Shapley} value for cooperative games under precedence constraints},
  author={Faigle, Ulrich and Kern, Walter},
  journal={International Journal of Game Theory},
  volume={21},
  number={3},
  pages={249--266},
  year={1992},
  publisher={Springer},
  doi={10.1007/BF01258278}
}

@article{cook1977detection,
  author  = {Cook, R. Dennis},
  title   = {Detection of Influential Observation in Linear Regression},
  journal = {Technometrics},
  volume  = {19},
  number  = {1},
  pages   = {15--18},
  year    = {1977},
  doi={10.1080/00401706.1977.10489493}
}

@incollection{wang2020principled,
  title={A principled approach to data valuation for federated learning},
  author={Wang, Tianhao and Rausch, Johannes and Zhang, Ce and Jia, Ruoxi and Song, Dawn},
  pages={153--167},
  year={2020},
  publisher={Springer},
  booktitle={Federated learning: Privacy and incentive},
  editor={Yang, Qiang and Fan, Lixin and Yu, Han},
  series={Lecture Notes in Computer Science},
  volume={12500},
  doi={10.1007/978-3-030-63076-8_11}
}

@inproceedings{ghorbani2019data,
  title={Data {S}hapley: {E}quitable valuation of data for machine learning},
  author={Ghorbani, Amirata and Zou, James},
  booktitle={Proceedings of the 36th {International Conference on Machine Learning}},
  pages={2242--2251},
  year={2019},
  volume={97},
  series={Proceedings of Machine Learning Research},
  editor={Chaudhuri, Kamalika and Salakhutdinov, Ruslan},
  publisher={PMLR},
  url={https://proceedings.mlr.press/v97/ghorbani19c.html}
}

@incollection{shapley1953value,
  title={A value for n-person games},
  author={Shapley, Lloyd S},
  volume={2},
  pages={307--317},
  year={1953},
  publisher={Princeton University Press},
  booktitle={Contributions to the theory of games},
  editor={Kuhn, Harold W. and Tucker, Albert W.},
  doi={10.1515/9781400881970-018}
}

@article{lecun1998gradient,
  title={Gradient-based learning applied to document recognition},
  author={LeCun, Yann and Bottou, L{\'e}on and Bengio, Yoshua and Haffner, Patrick},
  journal={Proceedings of the IEEE},
  volume={86},
  number={11},
  pages={2278--2324},
  year={1998},
  publisher={IEEE},
  doi={10.1109/5.726791}
}

@article{shorten2019survey,
  title={A survey on image data augmentation for deep learning},
  author={Shorten, Connor and Khoshgoftaar, Taghi M},
  journal={Journal of Big Data},
  volume={6},
  number={1},
  year={2019},
  publisher={Springer},
  doi={10.1186/s40537-019-0197-0},
  eid={60}
}

@inproceedings{ghorbani2020distributional,
  title={A distributional framework for data valuation},
  author={Ghorbani, Amirata and Kim, Michael and Zou, James},
  booktitle={Proceedings of the 37th {International Conference on Machine Learning}},
  pages={3535--3544},
  year={2020},
  volume={119},
  series={Proceedings of Machine Learning Research},
  editor={Daum{\'e}, III, Hal and Singh, Aarti},
  publisher={PMLR},
  url={https://proceedings.mlr.press/v119/ghorbani20a.html}
}

@article{lundberg2020local,
  title={From local explanations to global understanding with explainable {AI} for trees},
  author={Lundberg, Scott M and Erion, Gabriel and Chen, Hugh and DeGrave, Alex and Prutkin, Jordan M and Nair, Bala and Katz, Ronit and Himmelfarb, Jonathan and Bansal, Nisha and Lee, Su-In},
  journal={Nature Machine Intelligence},
  volume={2},
  number={1},
  pages={56--67},
  year={2020},
  publisher={Nature Publishing Group},
  doi={10.1038/s42256-019-0138-9}
}

@article{covert2020understanding,
  title={Understanding global feature contributions with additive importance measures},
  author={Covert, Ian and Lundberg, Scott M and Lee, Su-In},
  volume={33},
  pages={17212--17223},
  year={2020},
  journaltitle={{Advances in Neural Information Processing Systems}},
  editor={H. Larochelle and M. Ranzato and R. Hadsell and M. F. Balcan and H. Lin},
  publisher={Curran Associates, Inc.},
  url={https://proceedings.neurips.cc/paper_files/paper/2020/file/c7bf0b7c1a86d5eb3be2c722cf2cf746-Paper.pdf}
}

@article{frye2020asymmetric,
  title={Asymmetric {S}hapley Values: Incorporating Causal Knowledge into Model-Agnostic Explainability},
  author={Frye, Christopher and Rowat, Colin and Feige, Ilya},
  journaltitle={{Advances in Neural Information Processing Systems}},
  volume={33},
  pages={1229--1239},
  year={2020},
  editor={H. Larochelle and M. Ranzato and R. Hadsell and M. F. Balcan and H. Lin},
  publisher={Curran Associates, Inc.},
  url={https://proceedings.neurips.cc/paper_files/paper/2020/file/0d770c496aa3da6d2c3f2bd19e7b9d6b-Paper.pdf}
}

@article{lundberg2017unified,
  title={A unified approach to interpreting model predictions},
  author={Lundberg, Scott M and Lee, Su-In},
  volume={30},
  year={2017},
  journaltitle={{Advances in Neural Information Processing Systems}},
  editor={I. Guyon and U. von Luxburg and S. Bengio and H. Wallach and R. Fergus and S. Vishwanathan and R. Garnett},
  pages={4765--4774},
  publisher={Curran Associates, Inc.},
  url={https://papers.nips.cc/paper/2017/hash/8a20a8621978632d76c43dfd28b67767-Abstract.html}
}

@inproceedings{kwon2022beta,
  title        = {{Beta Shapley}: A Unified and Noise-Reduced Data Valuation
                  Framework for Machine Learning},
  author       = {Kwon, Yongchan and Zou, James},
  booktitle    = {Proceedings of the 25th {International Conference on Artificial Intelligence and Statistics}},
  series       = {Proceedings of Machine Learning Research},
  volume       = {151},
  pages        = {8780--8802},
  year         = {2022},
  publisher    = {PMLR},
  editor={Camps-Valls, Gustau and Ruiz, Francisco J. R. and Valera, Isabel},
  url={https://proceedings.mlr.press/v151/kwon22a.html}
}

@dataset{misc_adult_2,
  author       = {Becker,Barry and Kohavi,Ronny},
  title        = {{Adult}},
  year         = {1996},
  entrysubtype={Data set},
  publisher={UCI Machine Learning Repository},
  doi={10.24432/C5XW20}
}

@article{nowak1995axiomatizations,
  title={On axiomatizations of the weighted {S}hapley values},
  author={Nowak, Andrzej S and Radzik, T},
  journal={Games and Economic Behavior},
  volume={8},
  number={2},
  pages={389--405},
  year={1995},
  publisher={Elsevier},
  doi={10.1016/S0899-8256(05)80008-9}
}

@article{jia2019knn,
  title     = {Efficient Task-Specific Data Valuation for Nearest Neighbor Algorithms},
  author    = {Jia, Ruoxi and Dao, David and Wang, Boxin and Hubis, Frances Ann and
               Gurel, Nezihe Merve and Li, Bo and Zhang, Ce and Spanos, Costas and Song, Dawn},
  journal   = {Proceedings of the VLDB Endowment},
  volume    = {12}, number = {11}, pages = {1610--1623}, year = {2019},
  doi={10.14778/3342263.3342637}
}

@article{kalai1987weighted,
  title={On weighted {S}hapley values},
  author={Kalai, Ehud and Samet, Dov},
  journal={International Journal of Game Theory},
  volume={16},
  number={3},
  pages={205--222},
  year={1987},
  publisher={Springer},
  doi={10.1007/BF01756292}
}

@online{wang2024economic,
  title={An economic solution to copyright challenges of generative {AI}},
  author={Wang, Jiachen T and Deng, Zhun and Chiba-Okabe, Hiroaki and Barak, Boaz and Su, Weijie J},
  year={2024},
  organization={arXiv},
  doi={10.48550/arXiv.2404.13964}
}

@incollection{chen2019towards,
  title={Towards model-based pricing for machine learning in a data marketplace},
  author={Chen, Lingjiao and Koutris, Paraschos and Kumar, Arun},
  booktitle={Proceedings of the 2019 {International Conference on Management of Data}},
  pages={1535--1552},
  year={2019},
  doi={10.1145/3299869.3300078},
  publisher={Association for Computing Machinery}
}

@article{qu2026shapley,
  author  = {Qu, Xinxue and Darku, Francis Bilson and Guo, Hong},
  title   = {{Shapley} value-based feature attribution for data masking},
  journal = {MIS Quarterly},
  year    = {2026},
  volume  = {50},
  number  = {1},
  pages   = {145--176},
  doi     = {10.25300/MISQ/2025/18502}
}

@article{xia2023equitable,
  title={Equitable data valuation meets the right to be forgotten in model markets},
  author={Xia, Haocheng and Liu, Jinfei and Lou, Jian and Qin, Zhan and Ren, Kui and Cao, Yang and Xiong, Li},
  journal={Proceedings of the VLDB Endowment},
  volume={16},
  number={11},
  pages={3349--3362},
  year={2023},
  publisher={VLDB Endowment},
  doi={10.14778/3611479.3611531}
}

@article{sculley2015hidden,
  title     = {Hidden Technical Debt in Machine Learning Systems},
  author    = {Sculley, D. and Holt, Gary and Golovin, Daniel and Davydov, Eugene and Phillips, Todd and Ebner, Dietmar and Chaudhary, Vinay and Young, Michael and Crespo, Jean-Fran{\c{c}}ois and Dennison, Dan},
  journaltitle = {{Advances in Neural Information Processing Systems}},
  volume    = {28},
  pages     = {2503--2511},
  year      = {2015},
  editor={C. Cortes and N. Lawrence and D. Lee and M. Sugiyama and R. Garnett},
  publisher={Curran Associates, Inc.},
  url={https://proceedings.neurips.cc/paper_files/paper/2015/file/86df7dcfd896fcaf2674f757a2463eba-Paper.pdf}
}

@incollection{baylor2017tfx,
  author    = {Baylor, Denis and Breck, Eric and Cheng, Heng-Tze and Fiedel, Noah and Foo, Chuan Yu and Haque, Zakaria and Haykal, Salem and Ispir, Mustafa and Jain, Vihan and Koc, Levent and Koo, Chiu Yuen and Lew, Lukasz and Mewald, Clemens and Modi, Akshay Naresh and Polyzotis, Neoklis and Ramesh, Sukriti and Roy, Sudip and Whang, Steven Euijong and Wicke, Martin and Wilkiewicz, Jarek and Zhang, Xin and Zinkevich, Martin},
  title     = {{TFX}: A {TensorFlow}-Based Production-Scale Machine Learning Platform},
  booktitle = {Proceedings of the 23rd {ACM SIGKDD International Conference on Knowledge Discovery and Data Mining}},
  pages     = {1387--1395},
  year      = {2017},
  publisher = {Association for Computing Machinery},
  doi       = {10.1145/3097983.3098021},
}

@article{polyzotis2018data,
  author  = {Polyzotis, Neoklis and Roy, Sudip and Whang, Steven Euijong and Zinkevich, Martin},
  title   = {Data Lifecycle Challenges in Production Machine Learning: A Survey},
  journal = {ACM SIGMOD Record},
  volume  = {47},
  number  = {2},
  pages   = {17--28},
  year    = {2018},
  month   = jun,
  doi     = {10.1145/3299887.3299891},
  url     = {https://doi.org/10.1145/3299887.3299891}
}

@inproceedings{han2023optimal,
  title={Optimal pricing for data-augmented {AutoML} marketplaces},
  author={Han, Minbiao and Xia, Steven and Light, Jonathan and Fernandez, Raul Castro and Xu, Haifeng and Galhotra, Sainyam},
  pages={40133--40149},
  year={2026},
  booktitle={Proceedings of the 43rd {International Conference on Machine Learning}},
  volume={306},
  series={Proceedings of Machine Learning Research},
  editor={Zhang, Tong and Dudik, Miroslav and Jaggi, Martin and Agarwal, Alekh and Li, Sharon and Schuurmans, Dale and Zhu, Jerry and Berkenkamp, Felix and Dong, Hanze and Bietti, Alberto},
  publisher={PMLR},
  url={https://proceedings.mlr.press/v306/han26w.html}
}

@article{mahmood2024pricing,
  title={Pricing and competition for generative {AI}},
  author={Mahmood, Rafid},
  volume={37},
  pages={75727--75748},
  year={2024},
  journaltitle={{Advances in Neural Information Processing Systems}},
  editor={A. Globerson and L. Mackey and D. Belgrave and A. Fan and U. Paquet and J. Tomczak and C. Zhang},
  publisher={Curran Associates, Inc.},
  url={https://proceedings.neurips.cc/paper_files/paper/2024/file/8a6eb62e87a287f91734ca33e1585e6e-Paper-Conference.pdf},
  doi={10.52202/079017-2411}
}

@article{zhang2025fairshare,
  title={Fairshare data pricing via data valuation for large language models},
  author={Zhang, Luyang and Jiao, Cathy and Li, Beibei and Xiong, Chenyan},
  pages={33306--33345},
  year={2025},
  journaltitle={{Advances in Neural Information Processing Systems}},
  volume={38},
  editor={D. Belgrave and C. Zhang and H. Lin and R. Pascanu and P. Koniusz and M. Ghassemi and N. Chen},
  publisher={Curran Associates, Inc.},
  url={https://proceedings.neurips.cc/paper_files/paper/2025/file/2fed719c7af739e5a8615170629ed7fc-Paper-Conference.pdf},
  doi={10.52202/085713-1119}
}

@incollection{agarwal2019marketplace,
  title     = {A Marketplace for Data: An Algorithmic Solution},
  author    = {Agarwal, Anish and Dahleh, Munther and Sarkar, Tuhin},
  booktitle = {Proceedings of the 2019 {ACM Conference on Economics and Computation} ({EC} '19)},
  pages     = {701--726},
  year      = {2019},
  publisher = {Association for Computing Machinery},
  address   = {New York, NY, USA},
  doi       = {10.1145/3328526.3329589}
}

@online{ohrimenko2019collaborative,
  title         = {Collaborative machine learning markets with data-replication-robust payments},
  author        = {Ohrimenko, Olga and Tople, Shruti and Tschiatschek, Sebastian},
  year          = {2019},
  organization={arXiv},
  doi={10.48550/arXiv.1911.09052}
}

@article{balazinska2011data,
  title     = {Data Markets in the Cloud: An Opportunity for the Database Community},
  author    = {Balazinska, Magdalena and Howe, Bill and Suciu, Dan},
  journal   = {Proceedings of the VLDB Endowment},
  volume    = {4},
  number    = {12},
  pages     = {1482--1485},
  year      = {2011},
  doi       = {10.14778/3402755.3402801}
}

@incollection{koutris2012query,
  title     = {Query-Based Data Pricing},
  author    = {Koutris, Paraschos and Upadhyaya, Prasang and Balazinska, Magdalena and Howe, Bill and Suciu, Dan},
  booktitle = {Proceedings of the 31st {ACM SIGMOD-SIGACT-SIGAI Symposium on Principles of Database Systems} ({PODS} '12)},
  pages     = {167--178},
  year      = {2012},
  publisher = {Association for Computing Machinery},
  doi       = {10.1145/2213556.2213582}
}

@article{linkifer2014arbitrage,
  title     = {On Arbitrage-Free Pricing for General Data Queries},
  author    = {Lin, Bing-Rong and Kifer, Daniel},
  journal   = {Proceedings of the VLDB Endowment},
  volume    = {7},
  number    = {9},
  pages     = {757--768},
  year      = {2014},
  doi       = {10.14778/2732939.2732948}
}

@incollection{xin2021production,
  title={Production machine learning pipelines: Empirical analysis and optimization opportunities},
  author={Xin, Doris and Miao, Hui and Parameswaran, Aditya and Polyzotis, Neoklis},
  booktitle={Proceedings of the 2021 {International Conference on Management of Data}},
  pages={2639--2652},
  year={2021},
  doi={10.1145/3448016.3457566},
  publisher={Association for Computing Machinery}
}

@article{liu2021dealer,
  title={Dealer: An end-to-end model marketplace with differential privacy},
  author={Liu, Jinfei and Lou, Jian and Liu, Junxu and Xiong, Li and Pei, Jian and Sun, Jimeng},
  journal={Proceedings of the VLDB Endowment},
  volume={14},
  number={6},
  year={2021},
  doi={10.14778/3447689.3447700},
  pages={957--969}
}

@article{gebru2021datasheets,
  title={Datasheets for Datasets},
  author={Gebru, Timnit and Morgenstern, Jamie and Vecchione, Briana and Vaughan, Jennifer Wortman and Wallach, Hanna and Daum{\'e}, III, Hal and Crawford, Kate},
  journal={Communications of the ACM},
  volume={64},
  number={12},
  pages={86--92},
  year={2021},
  doi={10.1145/3458723}
}

@incollection{weber1988probabilistic,
  title={Probabilistic Values for Games},
  author={Weber, Robert J.},
  booktitle={The {Shapley} value: Essays in honor of {Lloyd S. Shapley}},
  editor={Roth, Alvin E.},
  pages={101--120},
  year={1988},
  publisher={Cambridge University Press},
  doi={10.1017/CBO9780511528446.008}
}

@incollection{cohen2017emnist,
  author = {Cohen, Gregory and Afshar, Saeed and Tapson, Jonathan and van Schaik, Andre},
  title = {{EMNIST}: Extending {MNIST} to handwritten letters},
  year = {2017},
  booktitle={2017 {International Joint Conference on Neural Networks} ({IJCNN})},
  publisher={IEEE},
  pages={2921--2926},
  doi={10.1109/IJCNN.2017.7966217}
}

@article{liu2023cost,
  author  = {Liu, Xiaoping and Li, Xiao-Bai and Sarkar, Sumit},
  title   = {Cost-restricted feature selection for data acquisition},
  journal = {Management Science},
  year    = {2023},
  volume  = {69},
  number  = {7},
  pages   = {3976--3992},
  doi     = {10.1287/mnsc.2022.4551}
}

@article{singal2022shapley,
  author  = {Singal, Raghav and Besbes, Omar and Desir, Antoine and Goyal, Vineet and Iyengar, Garud},
  title   = {{Shapley} meets uniform: An axiomatic framework for attribution in online advertising},
  journal = {Management Science},
  year    = {2022},
  volume  = {68},
  number  = {10},
  pages   = {7457--7479},
  doi     = {10.1287/mnsc.2021.4263}
}

@article{pujol2024commoning,
  author = {Pujol Priego, Laia and Wareham, Jonathan},
  title = {Data commoning in the life sciences},
  journal = {MIS Quarterly},
  year = {2024},
  volume = {48},
  number = {2},
  pages = {491--520},
  doi = {10.25300/MISQ/2023/17439}
}

@article{pine2026myth,
  author = {Pine, Kathleen H. and Mazmanian, Melissa},
  title = {The myth of good data: An ecosystem perspective on data integrity breakdown and rehabilitation},
  journal = {MIS Quarterly},
  year = {2026},
  volume = {50},
  number = {2},
  pages = {615--640},
  doi = {10.25300/MISQ/2025/17587}
}

@article{parsons2026repurposing,
  author = {Parsons, Jeffrey and Lukyanenko, Roman and Greenwood, Brad N. and Cooper, Caren B.},
  title = {Understanding and improving data repurposing},
  journal = {MIS Quarterly},
  year = {2026},
  volume = {50},
  number = {1},
  pages = {35--58},
  doi = {10.25300/MISQ/2025/18361}
}

@article{heinrich2025different,
  author = {Heinrich, Bernd and Klier, Mathias and Obermeier, Andreas and Schiller, Alexander},
  title = {Different but the same? {An} event-driven approach to determine probabilities of data duplication},
  journal = {MIS Quarterly},
  year = {2025},
  volume = {49},
  number = {4},
  pages = {1539--1566},
  doi = {10.25300/MISQ/2025/18178}
}

@article{vandenbroek2021machine,
  author = {van den Broek, Elmira and Sergeeva, Anastasia and Huysman, Marleen},
  title = {When the machine meets the expert: An ethnography of developing {AI} for hiring},
  journal = {MIS Quarterly},
  year = {2021},
  volume = {45},
  number = {3},
  pages = {1557--1580},
  doi = {10.25300/MISQ/2021/16559}
}
